\documentclass[lettersize,journal]{IEEEtran}
\usepackage{amsmath,amsfonts, amssymb}
\usepackage{algorithmic}
\usepackage{algorithm}
\usepackage{array}
\usepackage[caption=false,font=normalsize,labelfont=sf,textfont=sf]{subfig}
\usepackage{textcomp}
\usepackage{stfloats}
\usepackage{url}
\usepackage{verbatim}
\usepackage{graphicx}
\usepackage{gensymb}
\usepackage{cite}
\usepackage{textcomp}
\usepackage{xcolor}
\usepackage{adjustbox,xspace}
\usepackage{nomencl}
\usepackage{multirow}
\usepackage{makecell}
\usepackage{ulem}
\usepackage{booktabs}

\makenomenclature
\newcommand{\todo}[1]{{\color{black}{#1}}}
\newcommand{\review}[1]{{\color{black}{#1}}}
\newcommand{\reviewII}[1]{{\color{black}{#1}}}
\usepackage{etoolbox}

\begin{document}

\title{A Survey of mmWave-based Human Sensing: Technology, Platforms and Applications}

\author{Jia~Zhang,~\IEEEmembership{Student~Member,~IEEE,}
        Rui~Xi,~\IEEEmembership{Member,~IEEE,}
        Yuan~He,~\IEEEmembership{Senior~Member,~IEEE,}
        Yimiao~Sun,~\IEEEmembership{Student~Member,~IEEE,}
        Xiuzhen~Guo,~\IEEEmembership{Member,~IEEE,}
        Weiguo~Wang,~\IEEEmembership{Student~Member,~IEEE,}
        Xin~Na,~\IEEEmembership{Student~Member,~IEEE,}
        Yunhao~Liu,~\IEEEmembership{Fellow,~IEEE,}
        Zhenguo~Shi,~\IEEEmembership{Member,~IEEE,}
        Tao~Gu,~\IEEEmembership{Senior~Member,~IEEE}
        % <-this % stops a space
\thanks{This paper was produced by the IEEE Publication Technology Group. They are in Piscataway, NJ.}% <-this % stops a space
\thanks{Manuscript received April 19, 2021; revised August 16, 2021.}}

% The paper headers
\markboth{Journal of \LaTeX\ Class Files,~Vol.~14, No.~8, August~2021}%
{Shell \MakeLowercase{\textit{et al.}}: A Sample Article Using IEEEtran.cls for IEEE Journals}

% \IEEEpubid{0000--0000/00\$00.00~\copyright~2021 IEEE}
% Remember, if you use this you must call \IEEEpubidadjcol in the second
% column for its text to clear the IEEEpubid mark.

\maketitle

\begin{abstract}
With the rapid development of the Internet of Things (IoT) and the rise of 5G communication networks and automatic driving, millimeter wave (mmWave) sensing is emerging and starts impacting our life and workspace. mmWave sensing can sense humans and objects in a contactless way, providing fine-grained sensing ability. In the past few years, many mmWave sensing techniques have been proposed and applied in various human sensing applications (e.g., human localization, gesture recognition, and vital monitoring). We discover the need of a comprehensive survey to summarize the technology, platforms and applications of mmWave-based human sensing. In this survey, we first present the mmWave hardware platforms and some key techniques of mmWave sensing. We then provide a comprehensive review of existing mmWave-based human sensing works. Specifically, we divide existing works into four categories according to the sensing granularity: human tracking and localization, motion recognition, biometric measurement and human imaging. Finally, we discuss the potential research challenges and present future directions in this area.
\end{abstract}

\begin{IEEEkeywords}
Millimeter wave, human sensing, mmWave sensing, mmWave radar.
\end{IEEEkeywords}

%intro, hardware, tech 20%; sensing target 50%; challenge & future 30%.
\section{Introduction}

Wireless sensing (e.g., Wi-Fi, Zigbee, RFID, and acoustic) offers contactless sensing, and it has become a new human sensing paradigm in the last decade. As the propagation of wireless signals is affected by the sensing target, the received signals contain target-related information. By analyzing the characteristic changes of wireless signals (i.e., phase, amplitude, and frequency), the target-related information, such as human gesture or respiration, can be captured and analyzed.

In recent years, millimeter wave (mmWave) sensors have been developed rapidly and attracted extensive attention, which catalyze the emergence of mmWave sensing technology. mmWave signals operate at a high frequency (30-300 GHz) with a large bandwidth. Hence it can provide high sensing sensitivity and precision. The short wavelength of mmWave signals further enables the antennas to be highly integrated, enabling beamforming and other techniques that support directional sensing capabilities. As a result, mmWave sensing has great advantages in human sensing over low-frequency sensing technologies such as Wi-Fi, UWB, and LoRa.

In addition, mmWave sensing offers many advantages over sensor-based techniques (i.e., camera, Lidar, and ultrasonic sensor), including a wide sensing range, fine-grained and directional sensing capability, and resistance to weather and illumination conditions. Camera-based sensing techniques provide high-resolution imaging results under good illumination conditions, thereby supporting the sensing of human status such as gesture and gait. However, camera-based sensing techniques are limited by illumination conditions and concerns about privacy. In addition, a low-resolution camera may not be able to capture micro motions. Lidar-based sensing can measure micro motions but it has a narrow sensing range. They are likely to be affected by weather conditions so that their application scenarios are limited. Ultrasonic-based sensing can resist different weather conditions but may suffer from severe signal attenuation and non-directional sensing, hence it may perform poorly in multipath scenarios.

\review{mmWave sensing has been widely explored to ``see, listen, inquire and touch'' the human body to provide accurate and ubiquitous human sensing capability. It can enable various human sensing tasks such as human tracking and localization, activity recognition, vital sign monitoring, sound recovery, and human imaging. For example, in 2016 Google \cite{soli_home, soli} designed a tiny mmWave sensing module Soli that supports finger gesture recognition, which has been integrated into its Nest Hub and Pixel 4 smartphone. Texas Instruments (TI) \cite{TI1443, TI6843, IWR1642, AWR1443, AWR1642} has developed a series of mmWave radar products for human sensing. In the field of autonomous driving, mmWave radars have been widely used in object detection and classification and are being explored in human imaging. Furthermore, considering the rapid large-scale deployment of 5G and the indispensable application of mmWave signals in 5G, mmWave sensing in 5G scenarios can play a vital role in Integration Sensing and Communication (ISAC), digital twins and so on. mmWave-based human sensing is being continuously explored and deployed by researchers and application developers, and its applications range from human-computer interaction, smart medical, smart home to smart city.} 

\review{mmWave-based human sensing presents unique technical challenges compared to non-human sensing. Firstly, the mobility of the human body needs to be considered. Compared with other static objects, the human's position and posture have greater variation and uncertainties, which are important factors that affect the human sensing result. Secondly, The human vital signal usually has a low signal-to-noise ratio (SNR) and is uncertain, resulting in higher noise cancellation requirements. Compared with the micro-motion of non-human objects, the human vital signal is more complex and easily buried by the movement-related noise. Furthermore, human vital signals are affected by various factors such as physiology, psychology, and the surrounding environment, making it difficult to analyze. Last but not least, the differences among individuals are significant. The contours, gestures, vital signals, and vocal habits of different individuals are very diverse. How to overcome these differences between individuals to achieve accurate human sensing has become a critical problem. To achieve accurate human sensing, people must carefully address these unique challenges. It is worth noting that many human sensing techniques can be extended to non-human sensing scenarios due to their attractive sensing capabilities, but this is not the focus of this article.}
%\begin{figure}
%    \centering
%    \includegraphics[width=\linewidth]{figure/mmwave.png}
%    \caption{Comparison of different sensing modalities.}
%    \label{fig:my_label}
%\end{figure}

\begin{figure*}[!tb]
\centering
\includegraphics[width=0.9\linewidth]{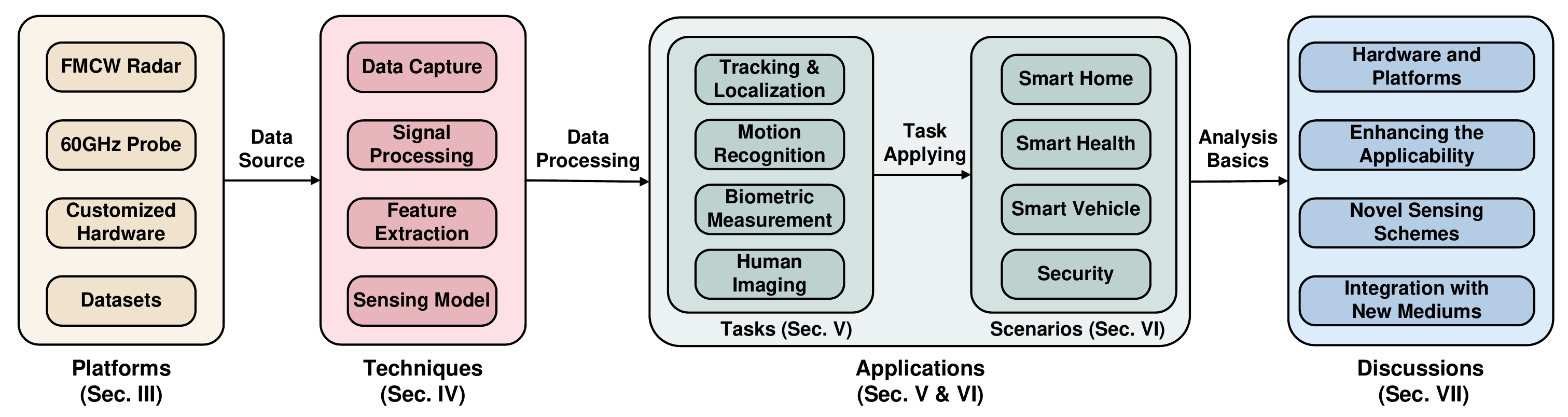}
\caption{\review{Article organization.}}
\label{fig:structure}
\end{figure*}

%With the rapid advancement of mmWave sensing and to further boost innovative research in this promising area, we discover the need of a comprehensive survey in mmWave-based human sensing technology, platforms, and applications. 

%Although there have been many mmWave-based human sensing applications, we find that the actual deployment is still very limited. As mmWave-based human sensing is reported to have excellent performance in various sensing tasks, we wonder what factors hinder its widespread deployment. promote its widespread deployment in practical scenarios

\review{Existing mmWave-based human sensing works have attempted to overcome these technical challenges. However, there is still a significant gap between the performance of these works and the accurate and ubiquitous sensing capability we expect, in terms of sensing range, sensing granularity, sensing accuracy, human body constraints, etc. We wonder what factors prevent mmWave-based human sensing from this vision. To this end, we need to study the existing works so that we can summarize the common core challenges and figure out their limitations. Moreover, it is critical to point out the key challenges and development trends of mmWave-based human sensing technology so as to illuminate potential directions and improve sensing capability. Therefore, we find that there is a strong need for a comprehensive survey in mmWave-based human sensing technology, platforms, and applications.}

%As mmWave-based human sensing is increasingly researched and deployed in the industry and academia, a comprehensive survey is necessary and important for researchers and application developers.

\review{This paper fills the gap by providing a comprehensive review of mmWave-based human sensing. The main contributions of this paper are summarized as follows:

\textbullet\ This paper presents a complete picture of the literature in the area of mmWave-based human sensing.

\textbullet\ The hardware platform and key techniques of mmWave-based human sensing are summarized in detail. Specifically, we summarize the human sensing pipeline. Then the related techniques in each module are compared and summarized.

\textbullet\ We comprehensively compare and summarize existing mmWave-based human sensing works into four categories according to the sensing granularity. Furthermore, we point out the core challenges and alternative solutions for each category of sensing tasks. Finally, the lessons learned about mmWave-based human sensing are given to provide readers with our suggestions.

\textbullet\ Following an elaborated discussion of the literature, we comprehensively discuss the key challenges and future directions related to mmWave-based human sensing, including hardware and platforms, enhancing the applicability, novel sensing schemes, and integration with new mediums. We also give a detailed discussion on the potential sensing techniques, including ISAC, THz sensing and so on.

\review{There have been some related surveys that provide a summary of particular scopes, such as mmWave-based communication \cite{busari2017millimeter, zhou2018ieee}, application scenarios \cite{singh2020multi, kebe2020human} and sensing techniques \cite{device-based-survey, deep-learning-survey, venon2022millimeter}. We compare and summarize them in detail in the next section. We find that none of the current surveys focuses on summarizing the existing mmWave-based human sensing works and analyzing their key challenges. As we mentioned, a comprehensive survey is necessary and important for researchers and application developers.} This survey is expected to inspire researchers and developers to conduct further research in mmWave sensing and build various applications to realize accurate, ubiquitous and stable human sensing in real life.
}

%This survey fills the gap by providing a comprehensive review of mmWave-based human sensing. We classify existing mmWave-based human sensing works into four categories according to the sensing granularity: human tracking and localization, motion recognition, biometric measurement and human imaging. Following elaborated discussion of the literature, we further discuss the potential research challenges of mmWave-based human sensing and discuss the future directions. This survey may inspire researchers and developers to conduct further research in mmWave sensing and build various applications to realize accurate, ubiquitous and stable human sensing in real life. 

\begin{table*}[h]
\renewcommand{\arraystretch}{1.3}
  \centering
  %\captionof{table}{THE SIMULATION PARAMETERS.}
\caption{\review{Summary of Related Surveys on mmWave-based Human Sensing}}

\begin{tabular}{lcccc} 
\toprule
\textbf{Reference} & \textbf{Signal Sources} & \textbf{Application Scenarios} & \textbf{Sensing Tasks} & \textbf{Taxonomy} \\ \hline
Wang \textit{et al.} \cite{cross-sensing-survey}& \begin{tabular}[c]{@{}c@{}}RF, Acoustic, Motion,\\ Biometric, Optical\end{tabular} & Wearable Application & \begin{tabular}[c]{@{}c@{}}Movement Tracking,\\ Human-machine-interface, \\ Vital Monitoring, Smart Hearth, \\ Mental State Monitoring,\\ User Authentication\end{tabular} & \begin{tabular}[c]{@{}c@{}}Wearables' Usages within\\  the Scope of Cross-sensing\end{tabular} \\ \hline
Abdu \textit{et al.} \cite{deep-learning-survey}& mmWave Radar Signals & Autonomous Driving & Object Detection, Classification & \begin{tabular}[c]{@{}c@{}}Learning-based Radar Data Processing\\ for Object Detection and Classification\end{tabular} \\ \hline
Shastri \textit{et al.} \cite{device-based-survey} & \begin{tabular}[c]{@{}c@{}}mmWave AP Signals, \\ mmWave Radar Signals\end{tabular} & \begin{tabular}[c]{@{}c@{}}Indoor Localization\\ and Sensing\end{tabular} & \begin{tabular}[c]{@{}c@{}}Localization,\\ Human Activity Recognition,\\ Object Detection, Health Monitoring\end{tabular} & \begin{tabular}[c]{@{}c@{}}Device-based Localization \\ and Device-free Sensing\end{tabular} \\ \hline
Singh \textit{et al.} \cite{singh2020multi}& Radar Signals & \begin{tabular}[c]{@{}c@{}}Non-Contact Vital \\ Sign Monitoring\end{tabular} & \makecell[c]{Heart Rate and Respiration\\ Rate Measurement} & \begin{tabular}[c]{@{}c@{}}Research Challenges Associated with \\ Hardware and Signal Processing\end{tabular} \\ \hline
Venon \textit{et al.} \cite{venon2022millimeter}& mmWave Radar Signals & Automotive & \begin{tabular}[c]{@{}c@{}}Motion Estimation, SLAM, \\ Place Recognition,\\ Detection and Recognition,\\  Semantic Segmentation\end{tabular} & Algorithms for Automotive \\ \hline
Kebe \textit{et al.} \cite{kebe2020human}& Radar Signals & Vital Sign Detection & \begin{tabular}[c]{@{}c@{}}Cardio-pulmonary Signals'\\ Detection and Monitoring\end{tabular} & \begin{tabular}[c]{@{}c@{}}Radar Categories Used\\  in Vital Signs Detection\end{tabular} \\ \hline
This Survey & mmWave Radar Signals & Human Sensing & \begin{tabular}[c]{@{}c@{}}Tracking and Localization,\\ Activity Recognition,\\  Gesture Recognition, \\Handwriting Tracking,\\ Gait Recognition, Vital Sensing,\\ Sound Recognition, Human Imaging\end{tabular} & \begin{tabular}[c]{@{}c@{}}Human Sensing Applications \\ w.r.t. the Sensing Granularity\end{tabular} \\

\bottomrule
\end{tabular}
\label{table:related_survey}
\end{table*}

\review{As shown in Fig.~\ref{fig:structure}, the structure of the survey is organized as follows. Sec. \ref{sec:related work} summarizes the related surveys and points out the salient novelties of our paper. Sec.\ref{sec:platform} discusses the mmWave hardware platform and datasets, which provide reliable data sources for subsequent sensing techniques and tasks. Sec.\ref{sec:techniques} briefly introduces the key techniques of mmWave sensing from data capture to sensing model. These techniques further process the obtained data and extract corresponding human-related information for task-related analysis. Sec.\ref{sec:sensing_task} elaborates on the various mmWave-based sensing tasks from tracking and localization to human imaging. These various sensing tasks can be applied to many practical application scenarios, which are introduced in Sec.\ref{sec:application scenarios}. Sec.\ref{sec:sensing_task} and Sec.\ref{sec:application scenarios} together form a detailed discussion of mmWave-based human sensing applications. Based on the analysis and summary of these works, Sec.\ref{sec:challenge} discusses the potential challenges and future directions. We conclude this survey in Sec.\ref{sec:conclusion}.}

%As shown in Fig.~\ref{fig:structure}, the structure of the survey is organized as follows. Sec.\ref{sec:platform} discusses the mmWave hardware platform and datasets. Sec.\ref{sec:techniques} briefly introduces the key techniques of mmWave sensing. Sec.\ref{sec:sensing_task} elaborates on the various mmWave-based sensing tasks. Sec.\ref{sec:challenge} discusses the potential challenge and future directions. We conclude this survey in Sec.\ref{sec:conclusion}.

% IEEE survey & tutorial
%调研list前20篇文章，两篇页数较多分别为58和62，除去这两篇外，平均页数35，最大42，最小20。规定多于30页每页多收220刀，最多收8页，页数再多也不收了。
%25页左右

\review{
\section{Related Work}
\label{sec:related work}

There have been some related surveys that provide a mmWave-related summary of particular scopes, such as mmWave-based communication, application and sensing techniques. In this section, we first introduce these related surveys and summarize them in Table.~\ref{table:related_survey}. Then we discuss the differences between the most related work and our work in detail. Finally, we point out the salient novelties of our work.

Wang \textit{et al.} \cite{cross-sensing-survey} provided a review on the cross-sensing technologies towards wearable applications. It includes sensing techniques based on multiple wireless signals, such as WiFi, mmWave and RFID, but the analysis of mmWave sensing is not enough. Abdu \textit{et al.} \cite{deep-learning-survey} wrote a survey of deep learning approaches processing mmWave radar signals in autonomous driving applications, but their focus is on deep learning rather than on mmWave sensing. Busari \textit{et al.} \cite{busari2017millimeter} summarized the research on mmWave massive MIMO communication and discussed research issues and future directions on mmWave massive MIMO.
Although its discussions about mmWave massive MIMO communication, such as channel model and channel estimation, may be further used for mmWave sensing, the related analysis is not presented. Zhou \textit{et al.} \cite{zhou2018ieee} conducted a review on the IEEE 802.11ay-based medium access control (MAC) layer related issues. Similarly, it mainly discussed the technical challenges, design issues and beamforming training in the IEEE 802.11ay, rather than the mmWave-based sensing issues.

The most related to our paper is \cite{device-based-survey}, which summarized the state-of-the-art in device-based localization and device-free sensing using mmWave communication and radar devices. In the device-free sensing section, that paper concentrates on the relevant signal processing techniques and learning techniques. Then the application and performance of these techniques are described in terms of some sensing applications including human activity recognition, object detection and health monitoring. Compared with that work, we analyze the key techniques related to human sensing and summarize them based on the human sensing pipeline. We focus on mmWave-based human sensing tasks and the corresponding application scenarios. Furthermore, we discuss the lessons learned, key challenges and future directions of mmWave-based human sensing in detail.

There are also some related surveys that provide an analysis of particular narrow scopes, such as application scenarios \cite{singh2020multi, kebe2020human} and sensing techniques \cite{deep-learning-survey, venon2022millimeter}. We summarize these surveys in Table.~\ref{table:related_survey}. \reviewII{Specifically, Wang \textit{et al.} \cite{cross-sensing-survey} provided a review on the cross-sensing techniques towards wearable applications and summarized the applied signal processing and machine learning algorithms. Abdu \textit{et al.} \cite{deep-learning-survey} presented a survey of deep learning approaches processing radar signals to accomplish some sensing tasks in autonomous driving applications. Shastri \textit{et al.} \cite{device-based-survey} focused on indoor mmWave device-based localization and device-free sensing, and provided a review of indoor localization approaches, technologies, schemes and algorithms. Singh \textit{et al.} \cite{singh2020multi} focused on the research challenges of non-contact vital sign monitoring in multi-resident environments. Venon \textit{et al.} \cite{venon2022millimeter} introduced the perception, recognition and localization techniques in automotive applications and described algorithms and applications adapted or developed for mmWave radars. Kebe \textit{et al.} \cite{kebe2020human} presented a review on contactless radar-based vital signs detection and highlighted the challenges in biomedical applications.} However, none of the current surveys particularly summarizes and compares existing mmWave-based human sensing works. As mmWave-based human sensing is increasingly researched in the industry and academia, a comprehensive survey is necessary and important for researchers and application developers.

\begin{table*}[h]
\renewcommand{\arraystretch}{1.3}
  \centering
  %\captionof{table}{THE SIMULATION PARAMETERS.}
\caption{Comparison of mmWave hardware devices}

\begin{tabular}{llllllll} 
\toprule
   \makecell[c]{\textbf{Products}}  & \makecell[c]{\textbf{Freq.}} & \makecell[c]{\textbf{Antenna size} \\ (Tx $\times$ Rx)} & \makecell[c]{\textbf{ADC sampling} \\ \textbf{rate (max)} \\ \textbf{(MSPS)}} & \makecell[c]{\textbf{TX power}\\{(dBm)}}& \makecell[c]{\textbf{Package Size}\\$mm^2$:(W $\times$ L)}  & \makecell[c]{\textbf{Manufacturer}} & \makecell[c]{\textbf{Image}} \\  \hline

    \makecell[c]{AWR1xx \\ IWR1xx }     &    \makecell[c]{76 GHz \\$\sim$ \\81 GHz}       &    \makecell[c]{3 $\times$ 4;\\2 $\times$ 4\\(xWR1642)}      &     \makecell[c]{12.5~(xWR162,\\AWR1443);\\25~(xWR1843);\\37.5~(AWR1243,\\IWR1443)}  &  \makecell[c]{12 } &     \makecell[c]{10.4 $\times$ 10.4;\\15 $\times$ 15\\(AWR1843AOP)}   & \makecell[c]{TI} & \begin{minipage}[b]{0.3\columnwidth}
		\centering
		\raisebox{-.5\height}{\includegraphics[width=\linewidth]{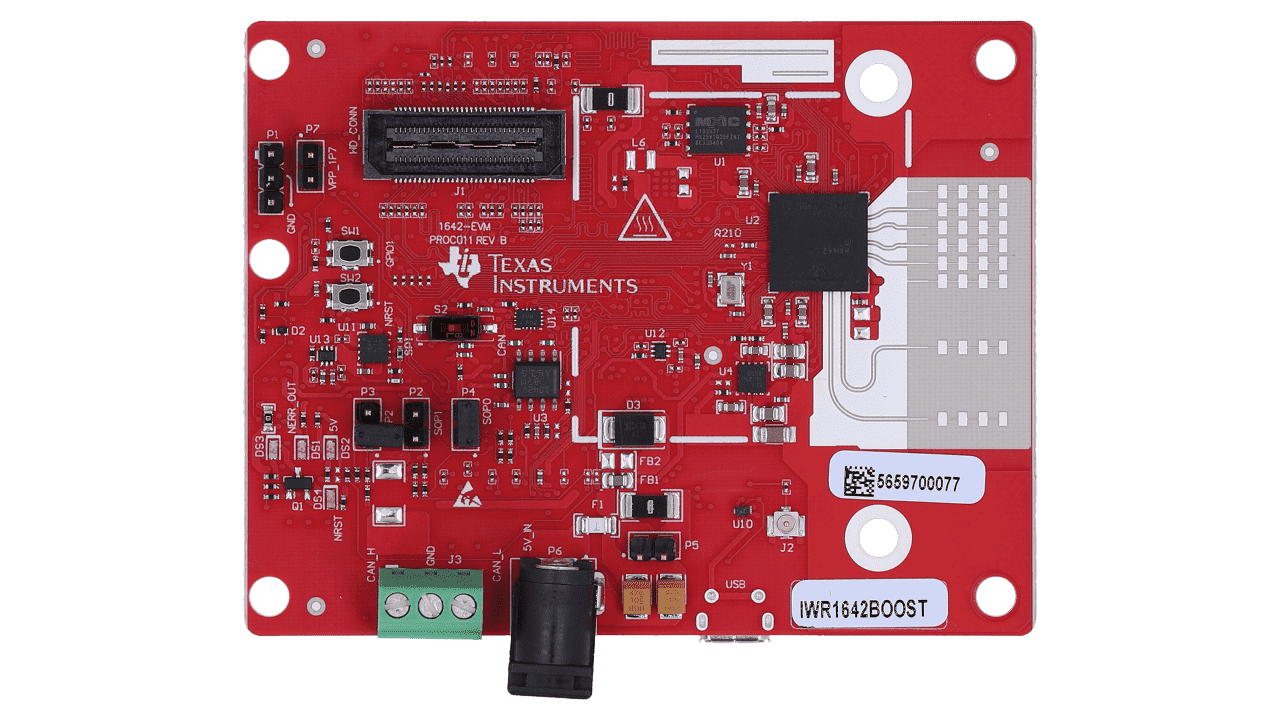}}
	\end{minipage}
 \\ \hline

    \makecell[c]{AWR6xx \\ IWR6xx }     &    \makecell[c]{60 GHz \\$\sim$ \\64 GHz}       &    \makecell[c]{3 $\times$ 4}      &     \makecell[c]{25 }  &  \makecell[c]{12;\\10\\(A/IWR6843AOP) } &     \makecell[c]{10.4 $\times$ 10.4;\\15 $\times$ 15\\(A/IWR6843AOP)}        & \makecell[c]{TI} & 
    \begin{minipage}[b]{0.3\columnwidth}
		\centering
		\raisebox{-.5\height}{\includegraphics[width=0.6\linewidth]{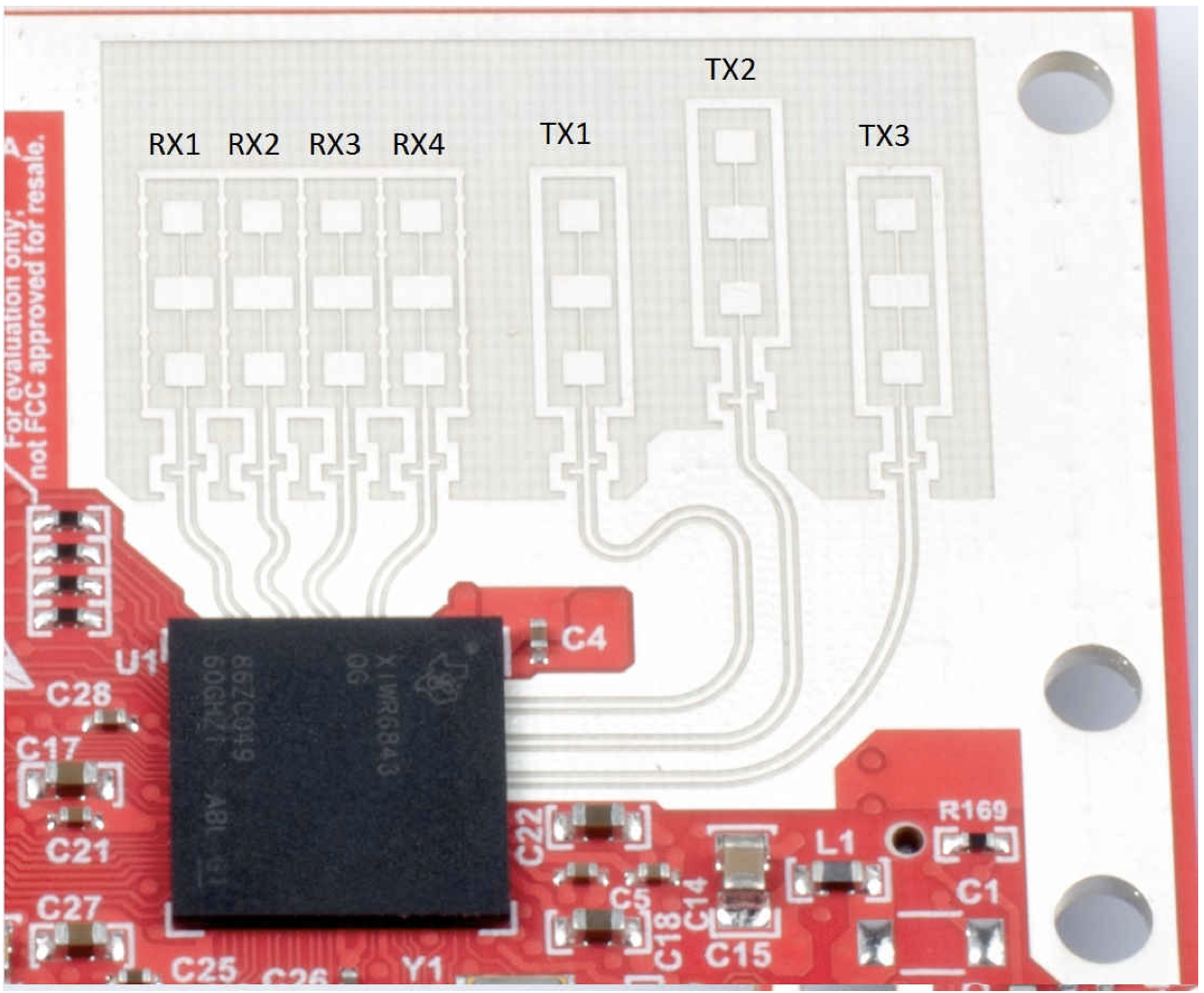}}
	\end{minipage}
 \\ \hline

    \makecell[c]{AWR2xx }     &    \makecell[c]{76 GHz \\$\sim$ \\81 GHz}       &    \makecell[c]{3 $\times$ 4;\\4 $\times$ 4\\(AWR2944)}      &     \makecell[c]{45~(AWR2243);\\37.5~(AWR29xx) }  &  \makecell[c]{13(AWR2243);\\12(AWR29xx)} &     \makecell[c]{10.4 $\times$ 10.4\\(AWR2243);\\12 $\times$ 12\\(AWR29xx)}   & \makecell[c]{TI} &
    \begin{minipage}[b]{0.3\columnwidth}
		\centering
		\raisebox{-.5\height}{\includegraphics[width=0.6\linewidth]{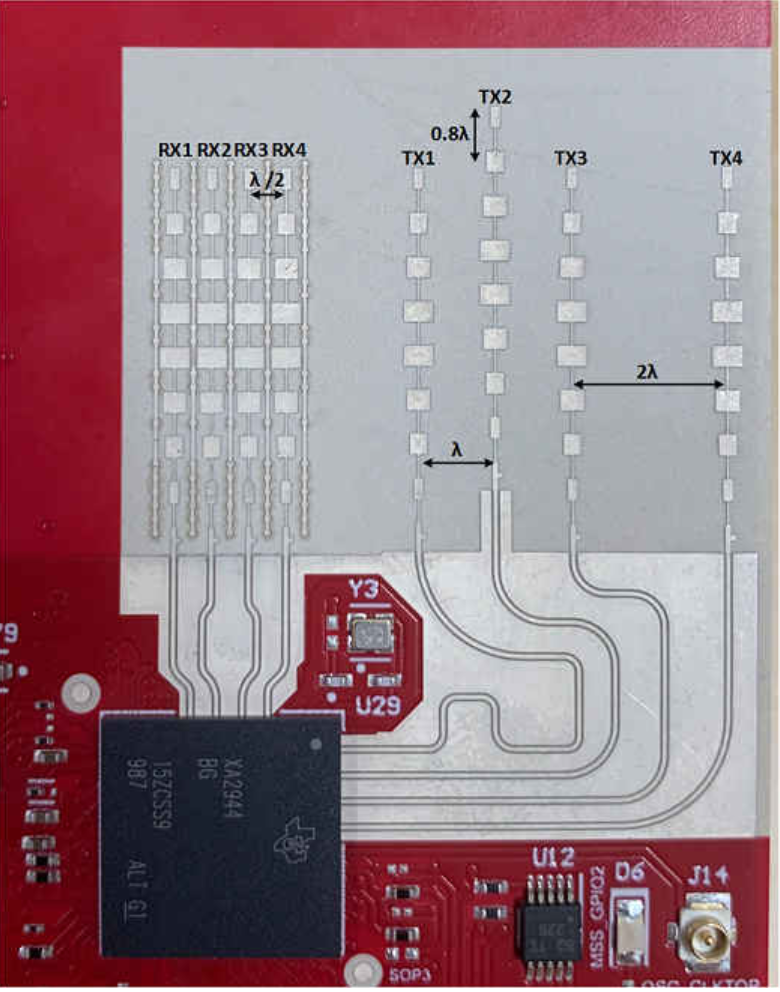}}
	\end{minipage}
 \\ \hline
    
    \makecell[c]{TINYRAD24G}     &   \makecell[c]{24 GHz \\$\sim$ \\24.25 GHz}        &     \makecell[c]{2 $\times$ 4}     &    &   \makecell[c]{8}    &      \makecell[c]{85 $\times$ 85}  & \makecell[c]{Analog Device} &
    \begin{minipage}[b]{0.3\columnwidth}
		\centering
		\raisebox{-.5\height}{\includegraphics[width=0.6\linewidth]{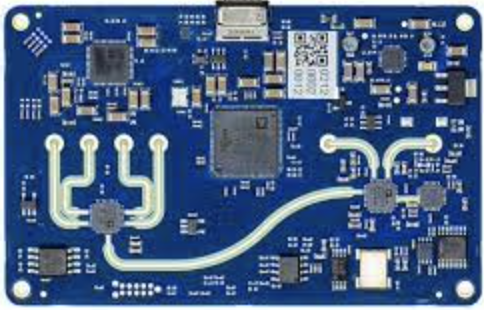}}
	\end{minipage}
 \\ \hline
    
    \makecell[c]{TRX\_024}     &  \makecell[c]{21.5 GHz \\$\sim$ \\28.7 GHz}         &   \makecell[c]{1 $\times$ 1}       &   \makecell[c]{2.5}     & \makecell[c]{6}  &  \makecell[c]{3 $\times$ 3}  & \makecell[c]{Silicon Radar} & 
    \begin{minipage}[b]{0.3\columnwidth}
		\centering
		\raisebox{-.5\height}{\includegraphics[width=0.6\linewidth]{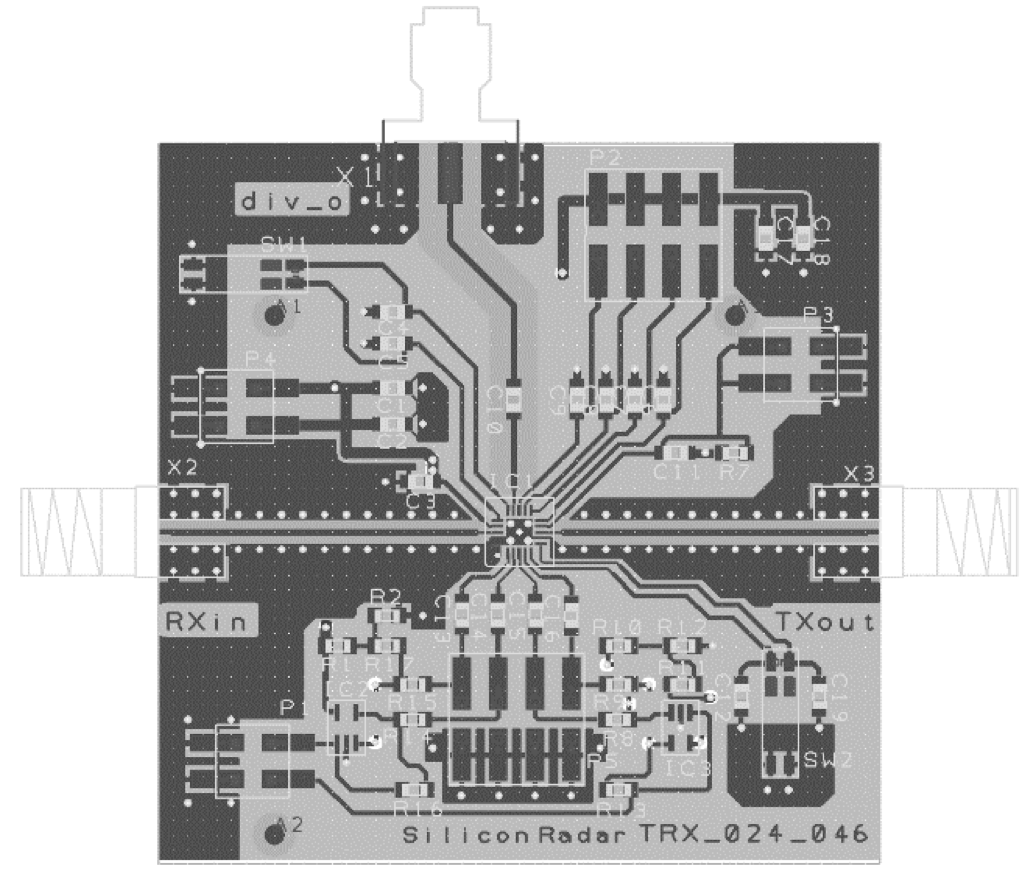}}
	\end{minipage}
 \\ \hline
    
    \makecell[c]{TRX\_120\\TRA\_120}     & \makecell[c]{119.3 GHz \\$\sim$  \\125.8 GHz  }       &  \makecell[c]{4 $\times$ 4}   &  \makecell[c]{2.5}   &      \makecell[c]{-3}     &   \makecell[c]{8 $\times$ 8\\5 $\times$ 5}  & \makecell[c]{Silicon Radar} & 
    \begin{minipage}[b]{0.3\columnwidth}
		\centering
		\raisebox{-.5\height}{\includegraphics[width=0.6\linewidth]{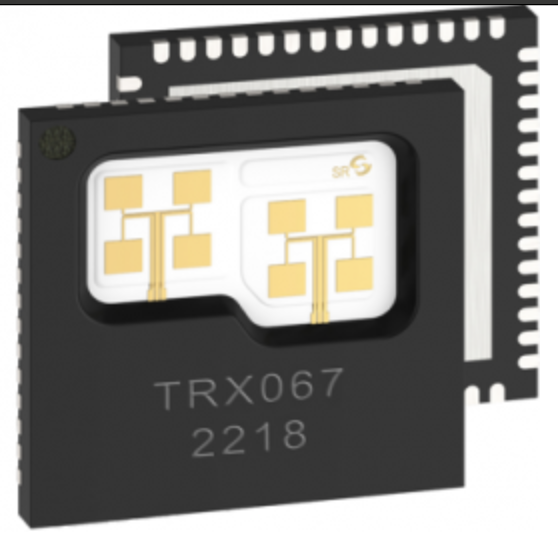}}
	\end{minipage}
 \\ \hline
    
    \makecell[c]{BGT60LTR11AIP}     &    \makecell[c]{61 GHz \\$\sim$  \\61.5 GHz}       &   \makecell[c]{1 $\times$ 1}       &     \makecell[c]{-}      &  \makecell[c]{10}  &  \makecell[c]{3.3 $\times$ 6.7}   & \makecell[c]{Infineon} & 
    \begin{minipage}[b]{0.3\columnwidth}
		\centering
		\raisebox{-.5\height}{\includegraphics[width=0.6\linewidth]{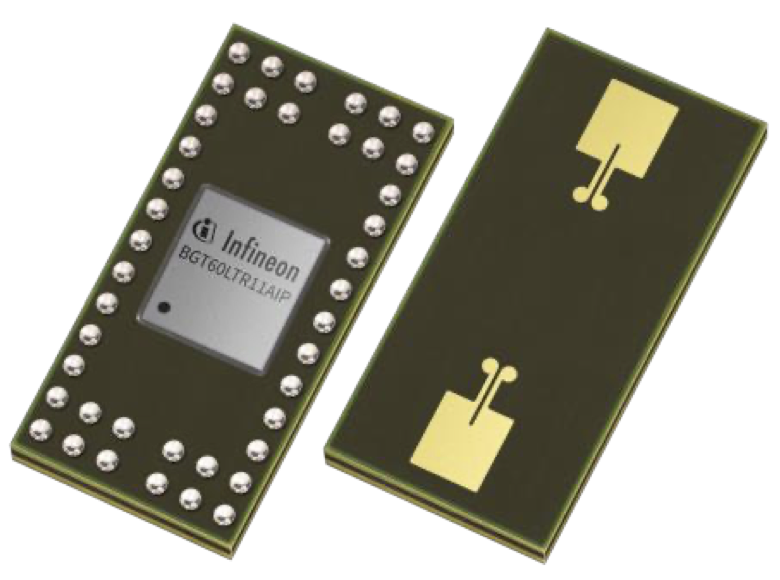}}
	\end{minipage}
 \\ \hline
    
    \makecell[c]{TEF82xx}     &   \makecell[c]{76 GHz \\$\sim$ \\81 GHz}        &     \makecell[c]{3 $\times$ 4}      &    \makecell[c]{40}       &     \makecell[c]{13.5}  &  \makecell[c]{7.5 $\times$ 7.5} & \makecell[c]{NXP} &
    \begin{minipage}[b]{0.3\columnwidth}
		\centering
		\raisebox{-.5\height}{\includegraphics[width=0.6\linewidth]{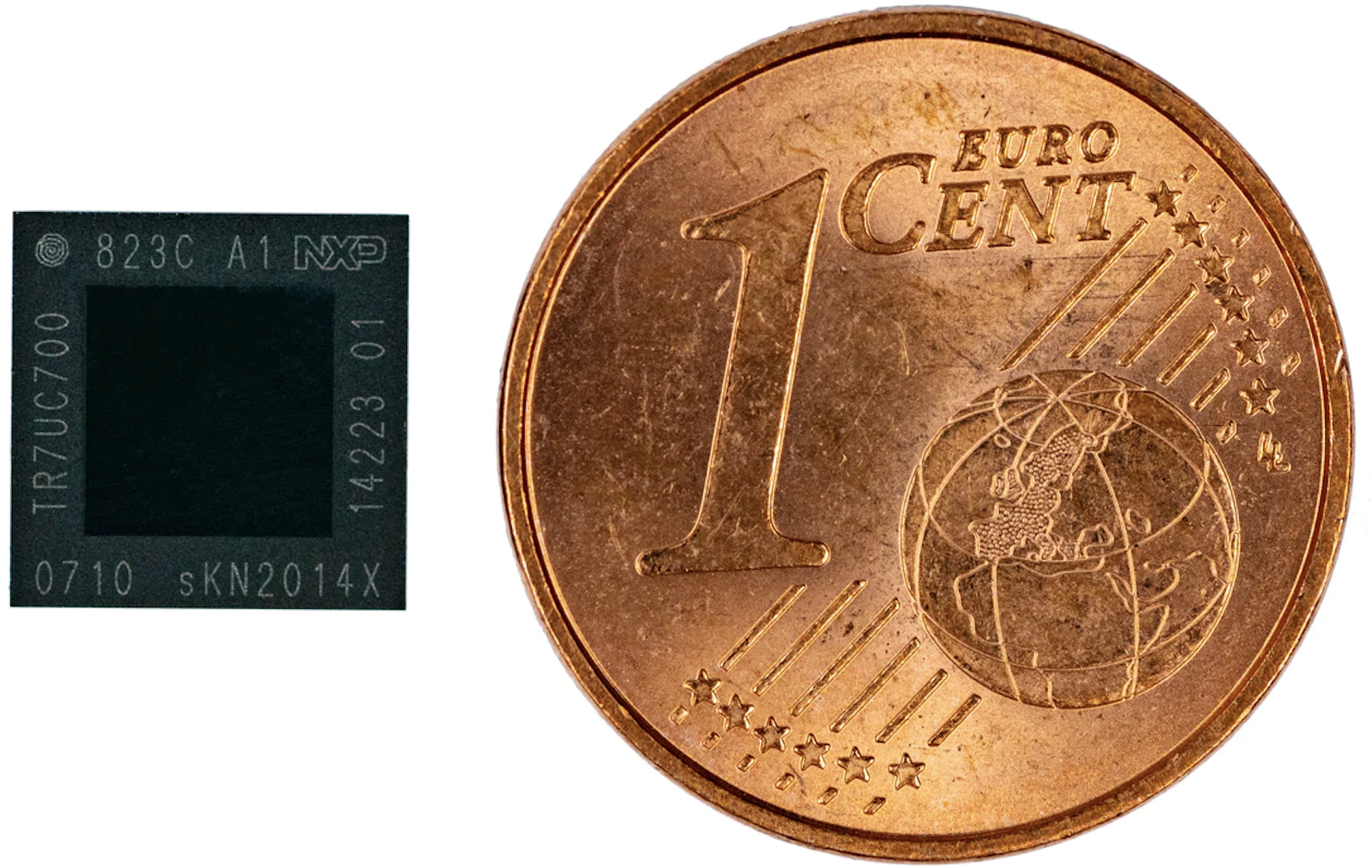}}
	\end{minipage}\\

\bottomrule
\end{tabular}
\label{table:platforms}
\end{table*}

Compared with these works, our paper presents new results in the following aspects: (1) Our paper is more inclusive and contains the latest advances in the field of mmWave-based human sensing. (2) We propose a comprehensive taxonomy of mmWave-based human sensing. According to the sensing granularity, our paper classifies the existing works into four categories: human tracking and localization, motion recognition, biometric measurement and human imaging. (3) We conduct a comprehensive comparison and summary of the existing works. Specifically, the hardware platform, signal form, and performance of these works are compared in detail. (4) The scope of study in our paper is broader. Besides introducing the specific mmWave-based human sensing works, we also discuss the key challenges and future directions, which may motivate more follow-up research in mmWave sensing.

}
\section{Platform \& Datasets}
\label{sec:platform}

To meet the needs of different sensing tasks, various kinds of hardware devices have been set up to perform mmWave sensing experiments, such as commercial FMCW radars, 60GHz probes, and customized devices. In the following, we briefly introduce three hardware platforms and compare different products. Furthermore, we list some public datasets containing mmWave signals for sensing different human statuses, such as activity, vital, and 3D pose.

\subsection{FMCW Radar}

\begin{table*}[h]
\renewcommand{\arraystretch}{1.3}
  \centering
  %\captionof{table}{THE SIMULATION PARAMETERS.}
\caption{Comparison of the customized hardware.}

\begin{tabular}{lccccc} 
\toprule
  Reference  & \makecell[c]{Openmili \cite{openmili}} & \makecell[c]{$M^3$ \cite{mcube}} & \makecell[c]{mm-Flex \cite{mmflex}} & \makecell[c]{\cite{eeye}} & \makecell[c]{\textit{Soli} \cite{soli}}  \\ \hline 
  \makecell[l]{Image}
   & \begin{minipage}[b]{0.3\columnwidth}
		\centering
		\raisebox{-.5\height}{\includegraphics[width=\linewidth]{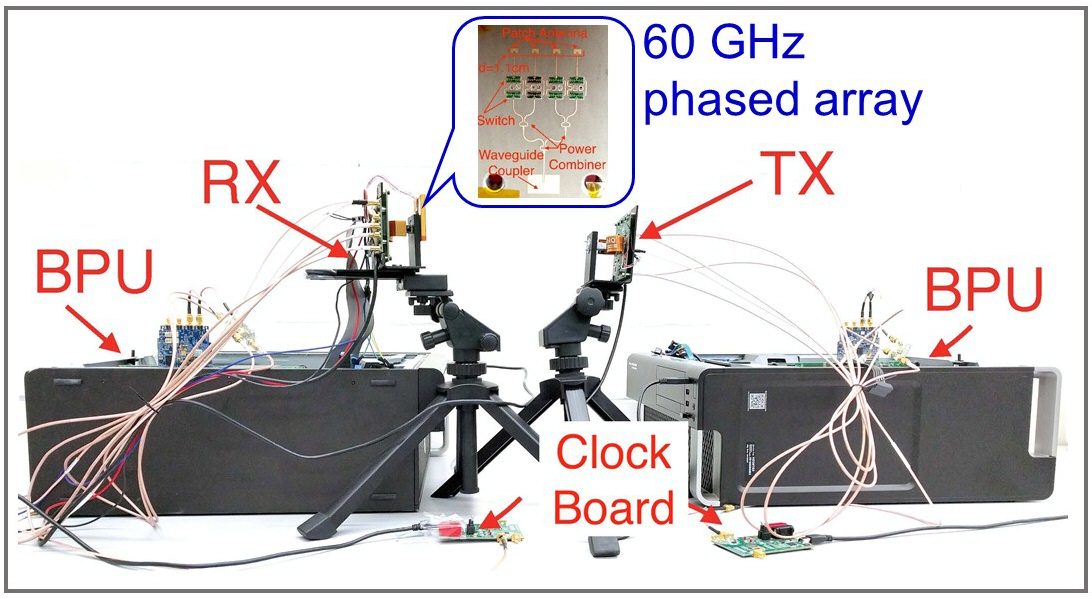}}
	\end{minipage}
    & \begin{minipage}[b]{0.3\columnwidth}
		\centering
		\raisebox{-.5\height}{\includegraphics[width=\linewidth]{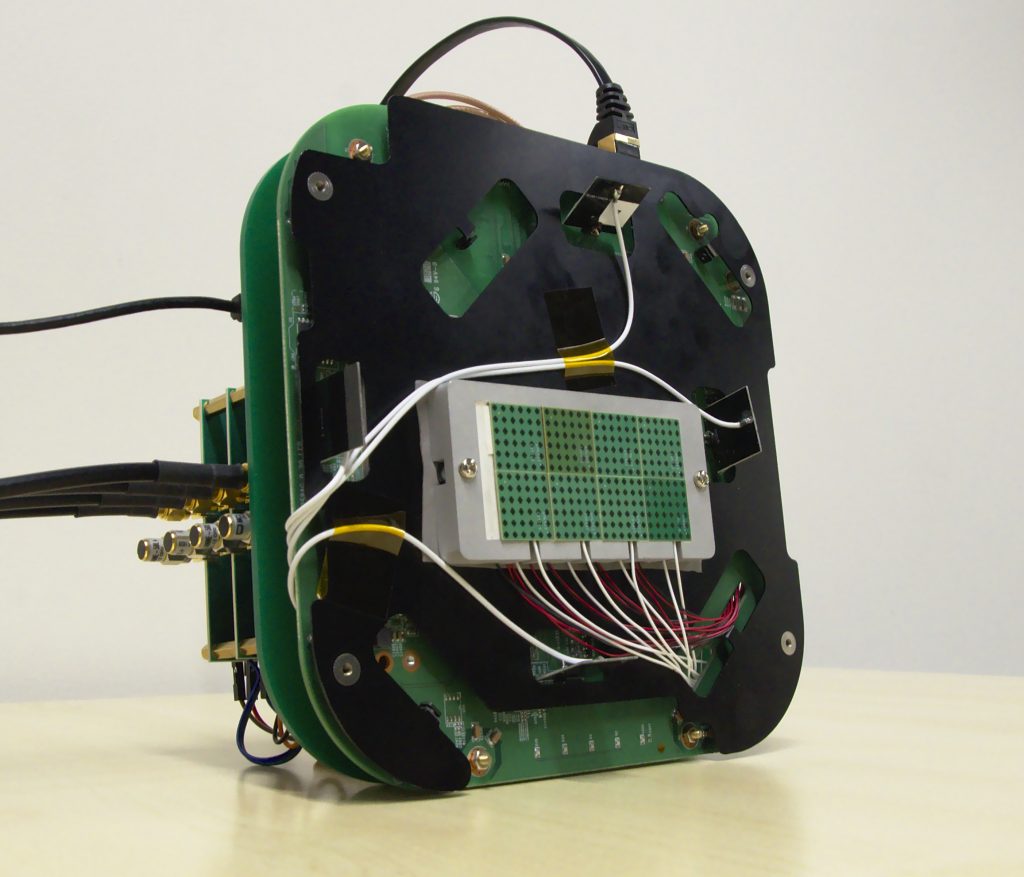}}
	\end{minipage}
    & \begin{minipage}[b]{0.3\columnwidth}
		\centering
		\raisebox{-.5\height}{\includegraphics[width=\linewidth]{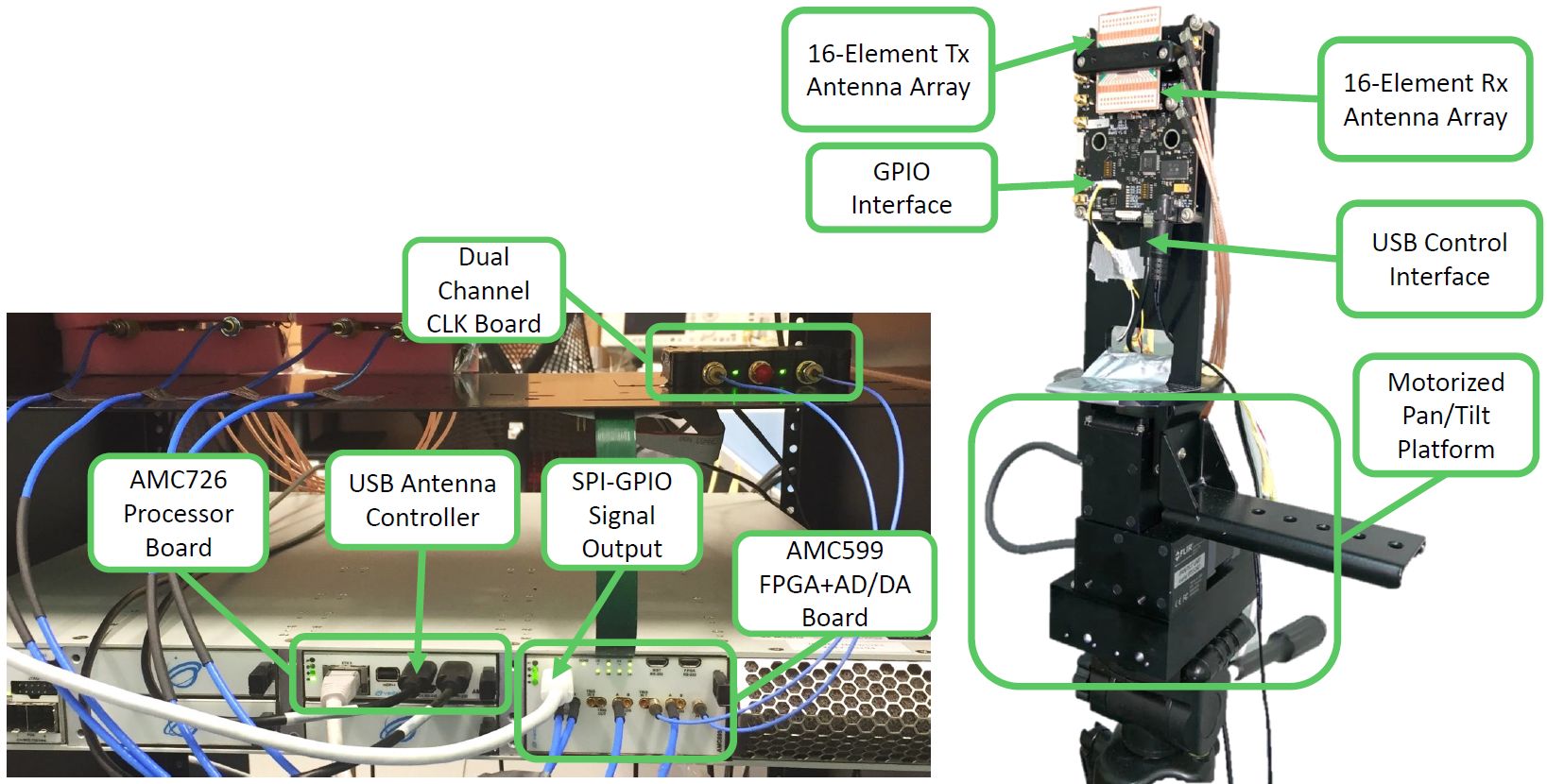}}
	\end{minipage}
    & \begin{minipage}[b]{0.3\columnwidth}
		\centering
		\raisebox{-.5\height}{\includegraphics[width=\linewidth]{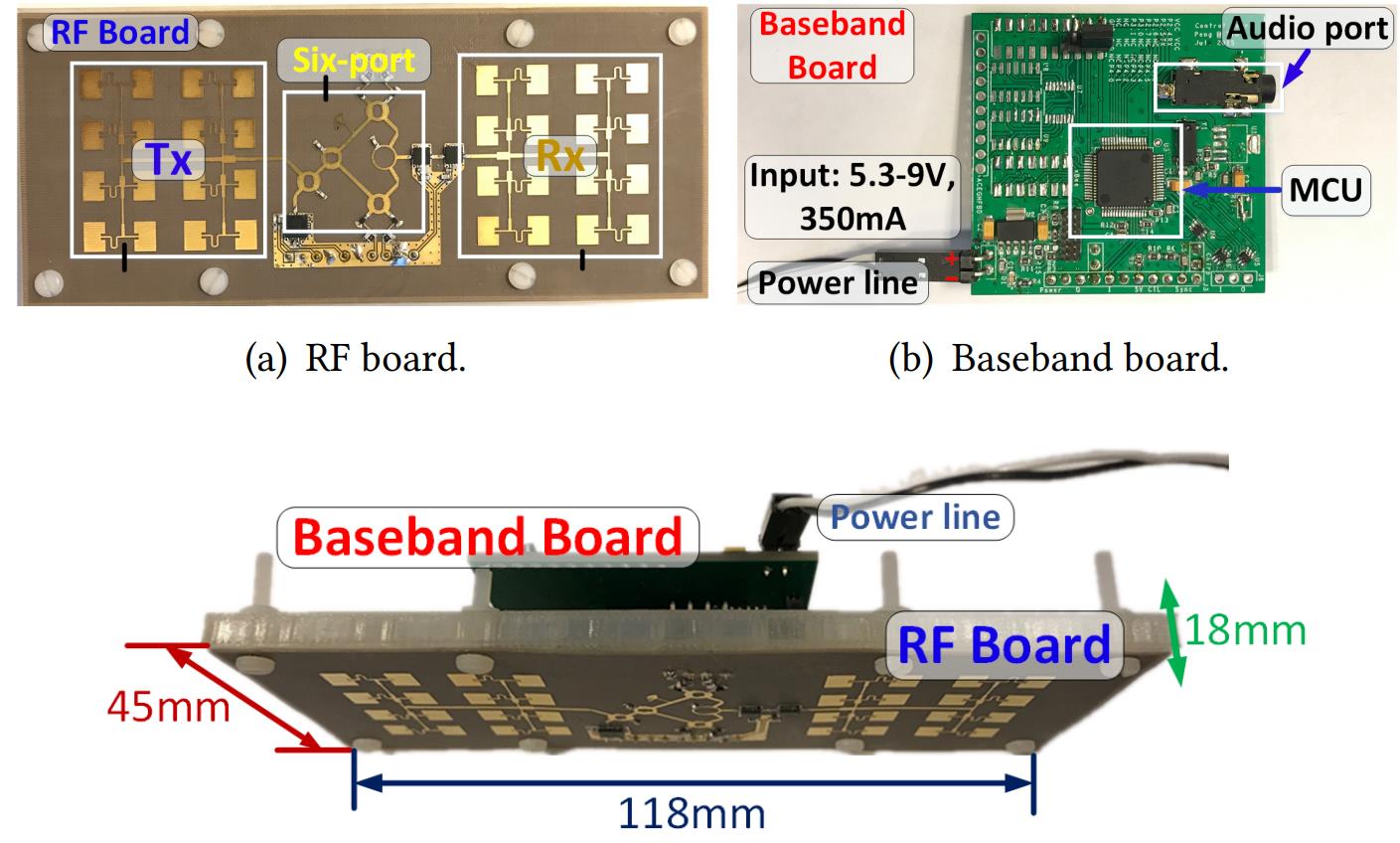}}
	\end{minipage}
    & \begin{minipage}[b]{0.3\columnwidth}
		\centering
		\raisebox{-.5\height}{\includegraphics[width=\linewidth]{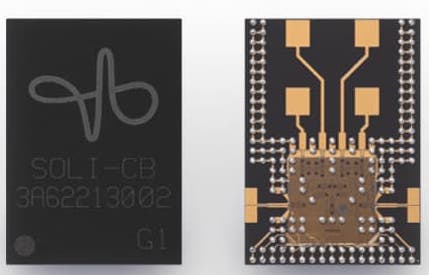}}
	\end{minipage}

    \\ %\hline
  Baseband BW  & 1 GHz   &  4 GHz &   2 GHz  &  450 MHz   & 7 GHz   \\ %\hline
  
  Carries Freq. & 57 $\sim$ 64 GHz  &  60 GHz &  60 GHz &  24 GHz & 60 GHz   \\ %\hline
  
   Antenna &  Horn/Phased-array  & Phased-array  & Phased-antenna  & Array  & Array  \\ %\hline
   
   Cost & \$15K  & below \$15K  &  \$40K & below \$100  & -   \\

\bottomrule
\end{tabular}
\label{tab:customized}
\end{table*}

\begin{figure}[tbp]
\centering
\includegraphics[width=0.9\linewidth]{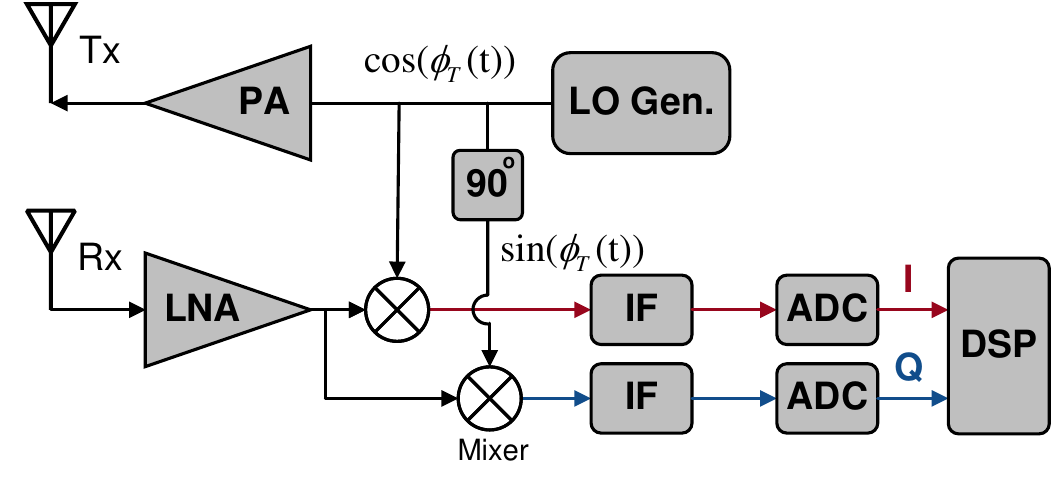}
\caption{The block diagram of mmWave radar system.}
\label{fig:radarblok}
\end{figure}

As Fig.~\ref{fig:radarblok} illustrates, a mmWave radar system is mainly consisting of four components: transmit (Tx)/receive (Rx), radio frequency (RF), analog components, and digital components. The mmWave signal is generated by a voltage-controlled oscillator (like LO Gen. in Fig.~\ref{fig:radarblok}), a part of which is additionally amplified by the power amplifier (PA) and fed to the transmitting antenna, and the other part is coupled to the mixer, mixed with the received echo. The transmitter can use different types of waveforms, such as pulsed, FSK (Frequency Shift Keying), CW (Continuous Wave), and FMCW (Frequency Modulated Continuous Waveform). In this paper, we mainly focus on the FMCW radar. The transmission signal is modulated in frequency, linear changing over a defined time. The received signals by Rx are processed by LNA (Low noise amplifier) and then mixed with the transmitted signals to obtain the \review{intermediate frequency (IF) signals, whose frequencies are equal to the frequency differences between the transmitted signals and the received signals.} After analog-to-digital conversion (ADC), they are sent to the signal processor for further processing. \review{Radar systems can either generate complex signals (i.e., I/Q signals) using I/Q quadrature mixing, or real ones (i.e., I-channel signals) by I single-channel mixing. As Fig.~\ref{fig:radarblok} illustrates, the complex signals contain I and Q components that have the same amplitude and frequency but are shifted a quarter cycle relative to each other, which can be generated by adding a 90$\degree$ phase shift difference between the two mixers. Contrarily, the I single-channel mixing only uses a mixer that provides the absolute frequency shift.}

% \begin{table*}[h]
%   \centering
%     \caption{comparison of the customized hardwares}
%   \begin{tabular}{ |l| c | c | c |c|c| }
%     \hline
%     & \makecell[c]{Openmili} & \makecell[c]{$M^3$} & \makecell[c]{mm-Flex} & \makecell[c]{\cite{eeye}} & \makecell[c]{\textit{Soli}}  \\ \hline \makecell[c]{Image}
%   & \begin{minipage}[b]{0.3\columnwidth}
% 		\centering
% 		\raisebox{-.5\height}{\includegraphics[width=\linewidth]{figure/openmilli.jpeg}}
% 	\end{minipage}
%     & \begin{minipage}[b]{0.3\columnwidth}
% 		\centering
% 		\raisebox{-.5\height}{\includegraphics[width=\linewidth]{figure/m3.jpg}}
% 	\end{minipage}
%     & \begin{minipage}[b]{0.3\columnwidth}
% 		\centering
% 		\raisebox{-.5\height}{\includegraphics[width=\linewidth]{figure/mmflex.jpeg}}
% 	\end{minipage}
%     & \begin{minipage}[b]{0.3\columnwidth}
% 		\centering
% 		\raisebox{-.5\height}{\includegraphics[width=\linewidth]{figure/eeye.jpg}}
% 	\end{minipage}
%     & \begin{minipage}[b]{0.3\columnwidth}
% 		\centering
% 		\raisebox{-.5\height}{\includegraphics[width=\linewidth]{figure/soli.jpeg}}
% 	\end{minipage}

%     \\ \hline
%   Baseband BW  & 1GHz   &  4GHz &   2 GHz  &  450MHz   & 7GHz   \\ \hline
  
%   Carries Freq. & 57 $\sim$ 64 GHz  &  60GHz &  60GHz &  24GHz & 60GHz   \\ \hline
  
%   Antenna &  horn/phased-array  & phased-array  & phased-antenna  & array  & array  \\ \hline
   
%   Cost & \$15K  & below \$15K  &  \$40K & below \$100  & -   \\ 
%   \end{tabular}
% \end{table*}

Actually, a complete radar system is much more complicated and highly integrated. Besides the above components, the manufacturer will integrate some additional subsystems in an extremely small form-factor by leveraging some advanced techniques, such as Monolithic Microwave Integrated Circuits (MMICs), Antenna in Package (AiP), Silicon-Germanium-Technology (SiGe), RFCMOS process, etc. For example, TI IWR1843 is a self-contained solution that includes a monolithic implementation of a 3TX, 4RX system with built-in PLL and ADC. It also integrates a DSP subsystem, an ARM Cortex-R4F-based processor subsystem, and a hardware accelerator (HWA), built with the low-power 45nm RFCOMS process. More details of the commercial mmWave radars can be found in TABLE~\ref{table:platforms}.

\subsection{60GHz Probe}

Benefiting from the high bandwidth, high directivity, and low latency of mmWave signals, mmWave networks have enabled new applications for businesses and consumers, including enhanced telehealth and education, industrial automation, VR/AR, etc. At the 2019 World Radiocommunication Conference (WRC-19), delegates identified several mmWave frequency bands that could be used for 5G networks, including 24.25-27.5 GHz, 37-43.5 GHz, 45.5-47 GHz, 47.2-48.2 and 66-71 GHz.
On the other hand, WiGig, alternatively known as 60 GHz Wi-Fi, has been promoted and released by Wireless Gigabit Alliance to support wireless communication at multi-gigabit speed. It includes the current IEEE 802.11ad standard and 802.11ay standard. These network protocols have promoted the development of mmWave-based communication probes \cite{80211ad, ad7200}. These probes can not only achieve high throughput communication with mmWave signals but also have the potential of mmWave sensing.

%\begin{figure}[!tb]
%\centering
%\includegraphics[width=0.6\linewidth]{figure/msense_new.pdf}
%\caption{The Qualcomm 802.11ad chipsets.}
%\label{fig:msense}
%\end{figure}

There are a few probe-based mmWave sensing works in the current literature. Most of them  \cite{vimo, mmtrack, mmeye} utilize the Qualcomm chipset to obtain the sensing information. Qualcomm has provided a series of 60GHz WiFi solutions with the IEEE 802.11ad standard \cite{80211ad}. %As shown in Fig.~\ref{fig:msense},
The commodity Qualcomm 802.11ad chipsets have 32 antennas assembled in a $6 \times 6$ layout with a size factor of 1.8 cm $\times$ 1.8 cm for both the transmitter and the receiver. The receiver hardware further implements a Golay correlator to obtain the sensing information, i.e., \todo{channel impulse response (CIR)}. With this information, many human sensing applications can be achieved, such as human tracking and localization, vital sensing, human imaging, etc.

\subsection{Customized Hardware}

\begin{table*}[h]
\renewcommand{\arraystretch}{2.5}
  \centering
  %\captionof{table}{THE SIMULATION PARAMETERS.}
\caption{Comparisons of the open mmWave datasets.}

\begin{tabular}{llccccrrr} 
\toprule

\textbf{Dataset} & \textbf{Sensors} & \textbf{Data Format} & \textbf{Setups} & \textbf{Types of sensing} & \textbf{Scenario} & \makecell[l]{\# of \\ \textbf{Subjects}} & \makecell[l]{\# of \\ \textbf{Activities}} & \makecell[l]{\# of \\ \textbf{Frames}}  \\

\midrule

    \textbf{mRI}~\cite{mri} & \makecell[l]{RGB,Depth,IMU,\\TI IWR1443} &  \makecell[c]{Images,\\Point Cloud,\\ Inertial Signals} &  \makecell[l]{RGB:30 Hz,\\Depth:30 Hz,\\IMU:50 Hz,\\mmWave:10 Hz} &  \makecell[c]{Rehabilitation} &  Single  &  20 & 12 & 160K\\   \hline
    
    \textbf{MARS}~\cite{mars} & \makecell[l]{Microsoft Kinect,\\TI IWR1443} &   Point Cloud &  \makecell[l]{mmWave:3.2 GHz\\ BW, 10 Hz\\Camera: 30 Hz} &  Rehabilitation &  Single & 4  & 10 & 40K\\  \hline
    
    \textbf{mmPose}~\cite{mmpose} & \makecell[l]{TI AWR1642,\\Microsoft Kinect} &  Point Cloud  & \makecell[l]{mmWave: 3.072 GHz \\BW, 10 Hz}  &  Walking, Swing  &  Single   &  2 & 4 & 38K\\  \hline
    
    \textbf{mmBody}~\cite{mmbody} & \makecell[l]{ Azure Kinect,\\mmWave radar }  &  \makecell[c]{ Point Clouds,\\RGB(D) Images} & \makecell[l]{Radar: 10-30 fps,\\Camera: 30 fps}  &  \makecell[c]{Skeleton, Mesh} &  - & 20  & 100 & 200K\\ \hline
    
    \textbf{MMActivity}~\cite{radhar} & TI IWR1443  &   Point Cloud & \makecell[l]{3.19 GHz BW,\\30 fps}  & \makecell[c]{Boxing, Jumping,\\Squats, Walking}  & Single & 2  & 5 & 16k\\   \hline
    
    \textbf{mmMesh}~\cite{mmmesh} & \makecell[l]{TI AWR1843,\\VICON}  &  Point Cloud  & \makecell[l]{VICON:10 fps\\mmWave:3.9 GHz \\BW, 10 fps}  &  Skeleton &  Single  &  20 & 8 & 480K\\  \hline
    
    \textbf{M-gesture}~\cite{mgesture} &  TI IWR1443  & \makecell[c]{Raw signals,\\ Point Cloud, \\Raw RDIs}  &  \makecell[l]{4 GHz BW,\\18.18 fps} &  Gestures &  Single  &  144 & 5 & 56K \\   \hline
    
    \textbf{mmGait}~\cite{mmgaitnet} &  \makecell[l]{TI IWR6843,\\TI IWR 1443} &   Point Cloud  & \makecell[l]{IWR6843: 3.75 GHz\\ BW, 10 fps,\\IWR1443:4 GHz\\ BW, 10 fps}  & Gaits  & up to 5 & 95  & - & -\\  \hline
    
    \textbf{mHomeGes}~\cite{mhomeges} & TI IWR1443 &  Doppler Profiles  & \makecell[l]{3.19 GHz BW,\\10 fps}  & Arm Gestures  & Single  & 25  & 10 & 22K\\   \hline
    
    \textbf{Pantomime}~\cite{pantomime} &  TI IWR1443  &  Point Cloud  & \makecell[l]{3.19 GHz BW,\\ 30 fps} &  \makecell[c]{Mid-air Gestures} & Single &  45 & 21 & 23K\\   

\bottomrule
\end{tabular}
\label{tab:datasets}
\vspace{-0.6cm}
\end{table*}

Furthermore, many researchers intend to open up new directions for mmWave sensing and protocol development and support system-oriented research by building customized experimental platforms, such as mmWave software radio, mmWave radar, etc. We list them in TABLE \ref{tab:customized}.

\textbf{WiMi} \cite{mtrack} is the first 60GHz testbed with a reconfigurable RF front-end and software-defined baseband processing modules. Based on this platform, \textbf{Openmili} \cite{openmili}, a software-defined mmWave network stack, is designed to span PHY layer signal processing to applications. OpenMili’s most outstanding feature is a reconfigurable phased-array antenna that can switch between 16 beam patterns at microsecond granularity. The designed phased-array specifically fits the WR-15 waveguide (a standard antenna interface on 60GHz radios), ensuring it can retrofit both OpenMili’s RF front-end and other commercial mmWave radios that are typically equipped with WR-15 horn antennas RF front-end. 

${M}^3$ \cite{mcube} is the first mmWave massive MIMO software radio. It provides up to eight 32-element phased-arrays, at a cost that is an order of magnitude lower than existing commercial mmWave SDR solutions. The baseband module (BM or 11ad NIC) and phased array module (PM) are commercial multi-array 802.11ad nodes. The IF bridge board is a customized PCB. The baseband processing unit (BPU) can be an FPGA or an SDR, such as a WARP board or a USRP. The control FPGA is a Cmod A7 module with an adapter board. It can hijack commercial mmWave radios for any waveform transmission and real-time phased array reconfiguration.

\textbf{mm-FLEX} \cite{mmflex} supports a bandwidth of 2 GHz and is compatible with mmwave standard requirements. Moreover, it integrates a powerful FPGA-based baseband processor with full-duplex capabilities together with mm-wave RF front-ends and phased antenna arrays that are fully configurable from the processor in real time.

In \cite{eeye}, a customized light mmWave platform consisting of the RF board, the baseband board, and the MCU board, is also proposed. For the RF board, Rogers RT/duroid 5880, which can provide a thickness lower to 0.0096 in, is used as the substrate. Furthermore, both Tx and Rx consist of 16 antennas following the 4 $\times$ 4 layouts and the size of the microstrip patch antenna is 5 mm $\times$ 3.7 mm. The distance between Tx and Rx of the probe is 40 mm. Besides, it also integrates a voltage regulator, difference amplifier, on-chip 12-bit ADC and MCU together. The total size reaches 4.65 in $\times$ 4.65 in $\times$ 0.59 in with 45.4 g weights, and its power consumption is only 1.23 W.

Moreover, Google ATAP (Advanced Technologies and Projects) also designs a radar chip, termed \textit{Soli} \cite{soli}, implemented at mmWave RF frequencies. It can be integrated into consumer electronic devices with minimal effort. The antenna, RF front-end, baseband processing, VCO and serial communication are integrated on the chip. Similar to the commercial mmWave radar, it also operates in the 60 GHz band with 7 GHz bandwidth and requires only minor hardware modifications to be compatible with 802.11ad and Wi-Gig standards~\cite{wigig}. The chip uses an antenna-in-package (AiP) patch array to form a wide 150$^\circ$ beam. At the same time, beamforming is realized at the receiver. It uses digital beamforming technology in SiGe-based FMCW chips instead of the analog phase shifters used on CMOS chips.

\subsection{Datasets}
\label{sec:dataset}

To advance mmwave-based research efforts, some researchers have published their datasets covering several different types, such as gait, gesture, and rehabilitation. In the following section, we will give a detailed description and list them in TABLE~\ref{tab:datasets}.

\textbf{MARS}~\cite{mars} is the first rehabilitation movement dataset using mmWave point cloud with well-labeled joints. It uses a TI IWR1443 Boost mmWave radar to collect mmWave data and a Microsoft Kinect V2 sensor to generate its respective labels. During collections, a subject is required to perform 10 distinct rehabilitation movements. Each frame or image will be processed to compute the 3D positions of 19 joints covering the upper and lower body. Besides, \textbf{mRI}~\cite{mri} also builds a high-quality and large-scale dataset, consisting of many complementary modalities, including mmWave point clouds, RGB frames, depth frames and inertial signals.%, as Figure \ref{fig:mri_setup} shows.

\textbf{MMActivity} dataset~\cite{radhar} is the first dataset used for human activity recognition through mmWave radars. TI IWR1443BOOST, which works in the 76 GHz to 81 GHz frequency range, is used for data collection. This dataset consists of five different activities collected from two users, including walking, jumping, jumping jacks, squats and boxing. The captured point clouds contain spatial coordinates, velocities, distances, intensities and bearing angles. The sampling rate of the radar is 30 frames per second. In addition, \textbf{mm-Pose}~\cite{mmpose} uses two TI IWR1642, where one is rotated 90$^{\circ}$ counter-clockwise with respect to the another, to collect mmWave samples from four actions. The ground truth is obtained by a Microsoft Kinect. 

Furthermore, \textbf{mmBody}~\cite{mmbody} builds a data acquisition platform using the Phoenix mmWave radar by Arbe Robotics, and two depth cameras, where the master one is right under the mmWave radar and the slave one is located on one side of the radar-body line. The subject is at a distance of 3-5 m away from the radar and the slave one is 1.5-2.3 m from the body. Furthermore, it requires 20 volunteers with different weights and heights to perform 100 motions. Additionally, it conducts the experiments under different conditions, such as poor lighting, rain, smoke, and occlusion with different materials. Similarly, \textbf{mmMesh}~\cite{mmmesh} chooses TI AWR1843BOOST to collect mmWave data reflected from 8 daily activities. VICON motion capture system~\cite{vicon} with a sampling rate of 10 fps is used to obtain high-precision dynamic pose information of the subject, which can be utilized to generate the ground truth human mesh.

\textbf{M-gesture}~\cite{mgesture} builds the first dataset to collect mmWave radar data for gesture recognition. A total of 1357 minutes of data are collected, involving 144 people (including 64 males and 80 females). The dataset contains not merely direct sensing, but also sensing with certain blockages (e.g., paper, corrugated paper, metal board). Moreover, \textbf{mHomeGes}~\cite{mhomeges} and \textbf{Pantomime}~\cite{pantomime} also use TI-IWR1443 to collect gesture datasets in smart home scenarios.

\textbf{mmGait}~\cite{mmgaitnet} builds the first mmWave dataset, collecting gait data from 95 volunteers (including 45 males and 50 females) using two mmWave radars, TI IWR6843 and IWR1443. their locations are 1 m away from each other. During collections, volunteers are required to walk in two modes, fixed route and freely, and it is up to 5 people for the multiple scenarios. In
data processing, it further converts the point clouds, collected by two devices with different rectangular coordinate systems, into the same coordinate system.

%\textbf{mHomeGes}~\cite{mhomeges} uses TI-IWR1443 to collect 10 self-defined arm gestures for human-computer interaction in smart home scenarios, and it will collect the data on 13 anchor positions for each gesture. Additionally, \textbf{mHomeGes} recruits 25 volunteers, including 11 males and 14 females with different heights, weights and ages. Everyone repeats each predefined gesture 30 times and the experiment includes 7 indoor scenes, covering bedrooms, studies, and parlors. 

%\textbf{Pantomime}~\cite{pantomime} uses a TI IWR1443 to collect the reflected mmWave signals from 21 mid-air gestures, performed by 45 participants consisting of 13 females and 32 males aged 18–55, weight 50–110 kg, and height 1.55–2 m. The configurations of radar are with a frame rate of 30 fps, a range resolution of 0.047 m, a velocity resolution of 0.87 m/s, and a maximum velocity of 6.9 m/s up to a maximum range of 5 m. During the collection, participants are required to move their body parts with a relatively large radar cross-section to enable detection up to 5 m of distance. \textbf{Pantomime} chooses five indoor environments (\textit{Open}, \textit{Office}, \textit{Restaurant}, \textit{Factory} and \textit{Through-wall}), in addition, it conducts experiments for single and multiple people scenarios.

\review{
\subsection{Summary}
To sum up, there are already many COTS mmWave devices or customized platforms to provide reliable data sources to support various sensing tasks. While different mmWave hardware and platforms have different signal bandwidths, carrier frequencies, antenna arrangements and transmission powers, which can provide different sensing ranges, sensing granularities and sensing dimensions, how to choose a suitable platform is a prerequisite for achieving a particular task. For example, vital-related motion is very tiny and its sensing requires sensing accuracy to be even as high as sub-millimeters. Therefore, the vital sensing task should use mmWave devices with high carrier frequency and large bandwidth. Moreover, human imaging requires the two-dimensional spatial information of the target, which requires the radar to provide corresponding two-dimensional sensing capabilities. In this case, these radars with the linear arrangement of the receiving antenna cannot meet the requirements. 

\reviewII{In terms of FMCW radar, although various radars have been used for different sensing tasks, most of them are limited in the number of antennas. This leads to limited angular resolution and insufficient point cloud outputs, resulting in fragile robustness for practical applications. Therefore, developers should pay attention to these indicators in the design of future radars. More details about hardware development are discussed in Sec. \ref{sec:harddev}.

In terms of 60GHz probe, the 60GHz devices used for sensing are still limited. With the continuous development of ISAC in 5G communication, 60GHz probes may become popular for sensing. The designers may need to pay more attention to the sensing range of the 60GHz probe due to its directivity and high attenuation. The design of the interplay between communication and sensing also needs to be considered. Sec. \ref{sec: sensing with communication} further discusses the challenges and opportunities of sensing combined with communication.

Moreover, customized hardware potentially provides more tailored designs that COTS devices generally do not provide, including antennas with adjustable polarization directions, customized antenna array arrangement, scene-adapted denoising hardware, etc. These customized designs can greatly help researchers to explore novel mmWave-based human sensing techniques.
}

Furthermore, there are some public datasets listed in this section, and different parameters of each dataset are shown. Researchers can find appropriate datasets quickly, so as to propose advanced information processing techniques and models, and promote the development of the field.

}

%To sum up, there are already many COTS mmWave devices or customized platforms

%\reviewII{The successful development of miniaturized millimeter-wave radar can be attributed to advancements in CMOS technology and single-chip IC. These innovations have enabled the creation of single-chip radar ICs operating in various frequency bands, including 24, 60, 76-77, and 94 GHz. These radar ICs have found applications in automotive safety and convenience systems such as adaptive cruise control, collision mitigation with automatic braking, lane-departure sensing, and object detection for backup assistance. The integration of all RF components on a single chip is particularly crucial at millimeter-wave frequencies, offering advantages such as improved performance reliability, cost savings, smaller form factor, and excellent transmit-to-receive isolation. CMOS technology further enhances these benefits by enabling high-level integration, lower production costs, and facilitating the development of multifunction RF transceivers and RF system-on-chip designs. It also provides flexibility for signal processing integration and custom IP integration, allowing designers to create differentiated systems. More details about hardware development are discussed in Section \ref{sec:harddev}.
%}

%\input{chapter/2_hardware_platform}
\section{Key Techniques}
\label{sec:techniques}
%\begin{figure*}[t]
%\centering
%\includegraphics[width=0.8\linewidth]{figure/sensing pipeline.pdf}
%\caption{The general flowchart of mmWave-based human sensing}
%\label{fig:pipeline}
%\end{figure*}

In this section, we briefly introduce some key techniques related to mmWave-based human sensing. We first introduce the general working pipeline of human sensing, including data capture, signal preprocessing, feature extraction, sensing model and sensing task. Then we state the key techniques in each module separately. The sensing task will be elaborated on in the next section.

\subsection{Human Sensing Pipeline}
The general flowchart of mmWave-based human sensing is as follows: The mmWave transceiver always periodically transmits the mmWave signal and continuously receives the reflected signal. The received signal is downconverted by the mixer and sampled by the ADC. Then these samples can be captured and converted to the raw data \textbf{(Data capture)}. The raw data is further processed into various signal forms to complete different sensing tasks. Some denoising techniques are performed to resist environmental interference \textbf{(Signal preprocessing)}. To obtain human-related information from mmWave signals, various features can be extracted and analyzed \textbf{(Feature extraction)}. Furthermore, according to different sensing tasks, the task-related features need to be fed into different models for accurate sensing \textbf{(Sensing model)}. Finally, the sensing task can be achieved \textbf{(Sensing task)}.

We first introduce some details of data capture in Sec.~\ref{sec:data_capture}. Then we discuss the key techniques of signal preprocessing in Sec.~\ref{sec:signal_preprocessing}. A brief introduction to feature extraction is presented in Sec.~\ref{sec:feature_extraction}. Sec.~\ref{sec:sensing_model} described the classification of the sensing model. Finally, we elaborate on the details of the sensing task in Sec.~\ref{sec:sensing_task}.

\subsection{Data Capture}
\label{sec:data_capture}
To achieve accurate mmWave-based human sensing, the raw data must be carefully captured and processed. According to the differences in data sources, we divide data capture into two categories, namely data capture based on FMCW radars and data capture based on 24/60GHz probes. We introduce them separately in the following.

\subsubsection{Data capture based on FMCW radar}
\label{sec:data capture}

\begin{figure}[!tb]
\centering
\includegraphics[width=\linewidth]{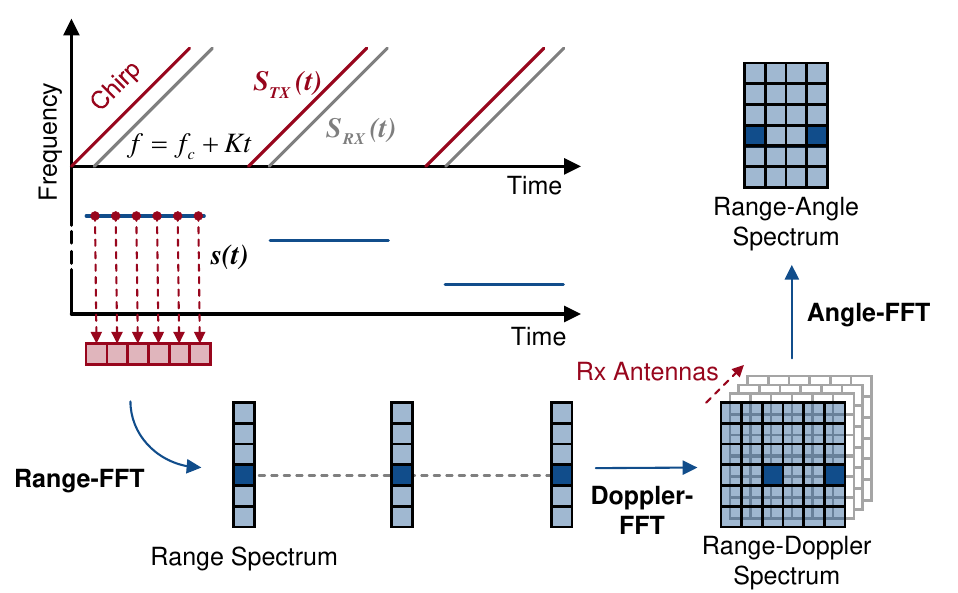}
\caption{Illustration of data capture based on FMCW radar.}
\label{fig:mmvib}
\end{figure}

The mmWave radar usually periodically sends FMCW chirp signals for distance and velocity measurement. As shown in Fig.~\ref{fig:mmvib}, the frequency difference between the transmitted signal and the received signal corresponds to the signal propagation time and can be utilized to determine the object distance. Denoting the distance between the radar and the sensing target by $R(t)$, the transmitted and the received signal can be represented as:

\begin{equation}
    \begin{aligned}
        & S_{Tx}(t) = \exp [j(2\pi f_c t+\pi Kt^2)] \\
        & S_{Rx}(t) = \alpha S_{Tx}[t-2R(t)/c] \\
    \end{aligned}
\end{equation}
where $\alpha$ is the path loss. $f_c$ and $K$ are the chirp start frequency and the \review{chirp} slope of the FMCW signal, respectively. By mixing the transmitted signal and the received signal, the \textit{beat frequency signal} $s(t)$, \review{namely IF signals,} can be obtained as:

\begin{equation}
    \begin{aligned}
        s(t) = S_{Tx}(t)S_{Rx}(t)^* \approx \alpha \exp [j4\pi (f_c + Kt)R(t)/c]
    \end{aligned}
\end{equation}
whose phase values indicate the distance information $R(t)$. To separate received signal components reflected from different ranges, a \textit{Range-FFT} \cite{TI-FFT} operation is performed on the samples of $s(t)$ within a chirp for signal separation. As shown in Fig.~\ref{fig:mmvib}, this operation maps the frequency spectrum of $s(t)$ to the range spectrum. To further measure the velocity of the target, the samples in the corresponding range bin are selected from the Range-FFT results and combined to form the \textit{reflected signal}, which can be obtained as:

\begin{equation}
    \begin{aligned}
        s(t) \xrightarrow[\text{at object range bin}]{\text{Range-FFT}} S(t) = \alpha \exp [j4\pi f_c R(t)/c]
    \end{aligned}
\end{equation}

Then the velocity of the object can be estimated by performing another FFT operation called \textit{Doppler-FFT} \cite{TI-FFT} on $S(t)$. With Range-Doppler-FFT, the radar can obtain the Range-Doppler spectrum and detect the existence of the object.

To further depict the position of the object, multiple receive antennas of the mmWave radar are utilized to derive the angle between the radar and the object. By performing the third FFT, Angle-FFT \cite{TI-FFT}, on these received signals, the radar can obtain the Range-Angle spectrum and detect the exact position of the object. To further improve the angular resolution, the beamforming technology \cite{capon, beamforming} instead of Angle-FFT can be utilized. By calculating the optimal aggregation weights of the transmit/receive antennas, they can be concentrated in the desired direction to obtain a much higher resolution.

%\paragraph{FMCW signal}
%\paragraph{Range FFT, Angle FFT, Doppler FFT and beamforming}
%\paragraph{beam steering, digital beamforming}

\subsubsection{Data capture based on 24/60GHz probe}

Unlike the mmWave radar, which utilizes the beat frequency signal to locate sensing targets, the mmWave probe exploits the \todo{channel impulse responses (CIRs)} between transmit antennas and receive antennas to determine the target's location. The time delay of the CIR indicates the signal propagation time and can be used to determine the object distance. Denoting the traveling distance of the mmWave signal reflected by the target as $d(t)$, the CIR between transmit antenna $m$ and receive antenna $n$ can be expressed as:

\begin{equation}
    \begin{aligned}
        h_{m,n}(t) = \alpha_{m,n} \exp (-j2\pi d_{m,n}(t)/\lambda_c)
    \end{aligned}
\end{equation}
where $\alpha_{m,n}$ and $\lambda_c$ represent the complex channel gain and the carrier wavelength, respectively. If the reflected signal falls into a certain CIR tap, the distance between the probe and the target can be determined by mapping the CIR spectrum into the range spectrum. Furthermore, by adjusting the coefficients of the transmitter/receiver antennas in sequence, the mmWave probe can perform beamforming at both transmitter and receiver to obtain the reflected signal from each angle and determine the exact position of the object.  

%\paragraph{CIR measurement}
%RSSI, CIR intro
%\paragraph{CIR-based range, angle, doppler extraction}

\subsection{Signal Preprocessing}
\label{sec:signal_preprocessing}
After the data capture module, the raw data needs to be processed into various signal forms to complete different sensing tasks. We first introduce each signal form for further human sensing. Then some denoising techniques are discussed to resist environmental interference.

\subsubsection{Signal form}
The signal forms of the mmWave data include the Range-Angle spectrum, Range-Doppler spectrum, phase waveform, point cloud, etc. People can choose the suitable signal form according to their sensing tasks. For example, the Range-Angle spectrum is suitable for human tracking while the phase waveform can be utilized for vital sensing. Following, we will introduce each signal form separately.

\paragraph{Range-Angle spectrum \& Range-Doppler spectrum}
As we introduced in Sec.~\ref{sec:data_capture}, the Range-Angle spectrum can be obtained by either Range-Angle-FFT from the FMCW radar or CIR mapping and beamforming from the mmWave probe. The complex value in each Range-Angle bin in the Range-Angle spectrum indicates the reflected signal from the corresponding spatial position. The intensities can be utilized to detect the existence of objects, and the phases can further represent the micro displacement of the object.

Similarly, the Range-Doppler spectrum can be obtained by either Range-Doppler FFT from the FMCW radar or CIR mapping and phase differencing from the mmWave probe. The value in each Range-Doppler bin indicates the moving objects and their corresponding velocities. Therefore, the Range-Doppler spectrum can be utilized to separate these objects with different velocities in the same range.

\paragraph{Phase waveform}
To further characterize the micro displacement of the human body, a series of phase values can be extracted from the Range-Angle spectrum or directly from the Range-FFT result to form the phase waveform. As the phases of the reflected signal indicate the reflector's micro displacement, the phase waveform can be utilized for fine-grained human sensing, such as vital sensing.

\paragraph{Point cloud}

\begin{figure}[!tb]
\centering
    \includegraphics[width=\linewidth]{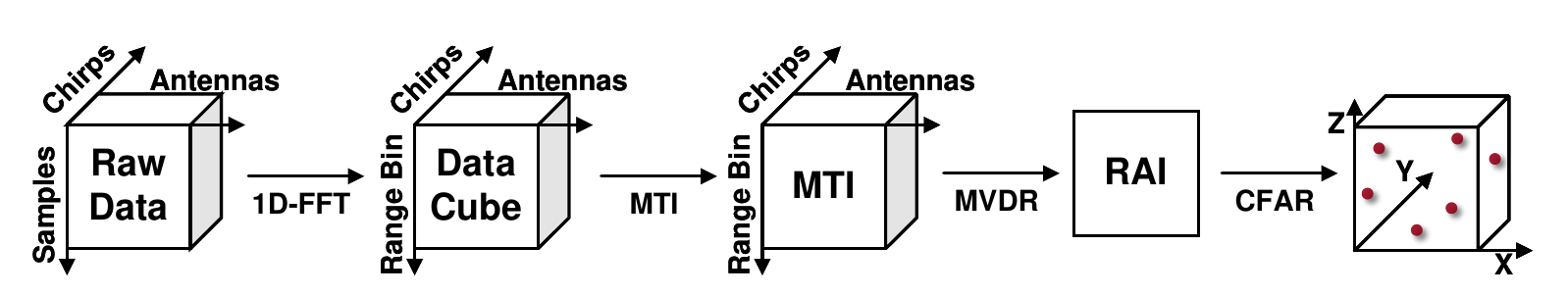}
    \caption{The flowchart of point cloud generation.}
    \label{fig:pcgen}
\end{figure}

A point cloud refers to a collection of the reflection points representing the object's surface, which has been widely used in various sensing tasks such as human imaging. Similar to LiDAR and vision, mmWave radars can obtain sparse point clouds from the raw data through a series of operations. Recently, the point cloud generation method proposed by Texas Instrument Technology has been widely used. As shown in Fig.~\ref{fig:pcgen}, the generation pipeline includes Range-FFT, Moving Target Indication (MTI) method \cite{MTI} and Minimum Variance Distortionless Response algorithm (MVDR) \cite{mvdr}. Then the \todo{Range Angle Image (RAI)} can be obtained, and the peak points can be further detected by Constant False Alarm Rate algorithm (CFAR) \cite{cfar}. It is worth noting that due to the limited hardware capability, the point number in each point cloud is usually limited.

\subsubsection{Signal denoise}
In this section, we discuss some signal denoising techniques for each signal form mentioned before to resist environmental interference.
\paragraph{CFAR \& Spectrum subtraction}
For the Range-Angle spectrum and the Range-Doppler spectrum, there is a lot of background noise from static objects and multipath interference existing in the area without humans. These noise confuses the localization of the human body and needs to be properly removed. There are many background noise elimination techniques, such as CFAR and spectrum subtraction. CFAR is a classical adaptive algorithm used to detect targets against environmental noise. It adaptively selects a noise threshold level to detect the bins with objects. Spectrum subtraction algorithm \cite{spectrum_sub} is another effective method to eliminate the background noise. The main idea is to subtract the estimation of the average background noise from the noisy measurement.   

\paragraph{Fitting \& Filter}
For the phase waveform, as the reflected signal from the area with humans includes not only the human-reflected signal but also the signal reflected from other objects in the same area, the phase values extracted from such reflected signals are distorted. Researchers have proposed some solutions to handle this problem, such as performing the circle fitting algorithm \cite{mmvib, gwaltz} or the line fitting algorithm \cite{radiomic} on the reflected signal, applying the FIR filter or the bandpass filter to the phase waveform, and so on. 
\paragraph{Clustering}
For the point cloud, the sparse point clouds generated from the mmWave radar are dispersed and the noise can be significant. Some data clustering algorithms have been employed to determine which points are caused by reflections from humans. For example, DBSCAN clustering algorithm \cite{dbscan} and K-mean clustering algorithm \cite{kmeans} have been exploited to merge these points into clusters to separate these human-related points from noise.

\review{\subsubsection{Comparison}
As different signal preprocessing techniques are suitable for different sensing tasks, we compare these techniques according to their applicable sensing tasks. In human localization and motion recognition, Range-Angle spectrums and point clouds are commonly used as they indicate the environment reflection, including the position and posture of the human. In this case, CFAR, spectrum subtraction and clustering can help remove background noise from static objects and multipath interference, thereby helping us to focus on the reflected signal from the human body. Whereas in biometric measurement, phase waveforms are chosen due to their unique ability to characterize the micro-motions. Compared with CFAR and clustering, fitting and filtering are more suitable for time series data. In human imaging, point clouds and Range-Angle spectrums can be selected. The former is more visible and the latter contains more spatial information. On the contrary, phase waveforms are rarely considered because there is less attention to tiny movements in human imaging.  
}

\subsection{Feature Extraction}
\label{sec:feature_extraction}
According to the different sensing tasks, the various features need to be extracted from the denoised signals. Time-domain analysis and frequency-domain analysis are two categories of feature extraction methods widely used in mmWave-based human sensing. The former focuses on analyzing the periodicity, duration, amplitude and other features of signals in the time domain, while the latter mainly extracts the frequency features, including spectrum, periodicity, power spectrum, etc.

\subsubsection{Time-domain analysis}
Some time-domain analysis techniques have been exploited to obtain the signal period or separate the superposition of multiple signals. For example, the template matching algorithm has been utilized in sensing human vital to obtain the heart rate and the breathing rate. Some signal decomposition techniques, such as \todo{variational modal decomposition algorithm (VMD)} and \todo{empirical mode decomposition algorithm (EMD)}, have been exploited to separate the vital sign waveforms from the reflected signal. For example, mmHRV \cite{mmhrv} regards the chest motion derived from the reflected mmWave signal as the linear superposition of the respiration signal, heartbeat signals and other motion signals. By taking each signal as an intrinsic mode function (IMF) component, mmHRV utilizes the VMD algorithm to realize signal decomposition and extracts a pure heartbeat signal.

\subsubsection{Frequency-domain analysis}
Compared with time-domain features, frequency-domain features are often more stable and distinguishable. Many feature-domain analysis techniques have been utilized to resist various noises and extract distinguishing features. For example, researchers have utilized the Fourier transform and wavelet transform techniques \cite{wavelet} to precisely extract the vital sign features. Furthermore, \todo{Short-time Fourier transform (STFT)} \cite{stft} and some frequency-domain features, such as \todo{residual phase Cepstrum coefficients (RPCC) \cite{RPCC} and Mel frequency cepstral coefficient (MFCC) \cite{MFCC}}), have been exploited to achieve sound recognition or motion recognition. For example, SPARCS \cite{sparcs} employs the STFT algorithm to extract the micro-Doppler spectrum of human motion. By dividing a longer reflected signal into shorter segments and computing their Fourier transform separately, SPARCS can simultaneously extract the time-frequency features and achieve accurate activity recognition.

\subsection{Sensing Model}
\label{sec:sensing_model}

While the sensing tasks are application-dependent, researchers need to build a suitable model to obtain sensing results with the extracted features or mmWave signals as inputs. For simplicity, the commonly used models can be divided into two categories: domain knowledge-based and deep learning-based. The former is strongly dependent on the expert's experience. On the contrary, the latter is data-intensive and requires a large amount of data. Both of them will be described in detail below.

\subsubsection{Domain knowledge-based model}
For some periodic or simple predefined motions, each exhibits specific physical characteristics. Researchers can easily build task-specific models that cast above-extracted features to different motions by analyzing signals or bringing in domain-specific knowledge, such as anthropology and physics. For example, gait is a repetitive motion whose cycle can be divided into eight phases (e.g., heel strike, foot flat, midstance, mid-swing, and so on). The features of the gait, such as speed, step time, step length, etc., can be computed by peak finding, pattern matching, and other techniques. Furthermore, based on the built model, researchers can exploit traditional machine learning techniques (e.g., \todo{random forest \cite{random_forest}, decision tree \cite{decision_tree} and support vector machine (SVM) \cite{svm}}). However, this method has poor generalization ability and can only be used for simple or specific tasks. Moreover, its performance is limited by the researcher's domain knowledge. 

\subsubsection{Deep learning-based model}
As opposed to the domain knowledge-based model relying on handcrafted features, due to the powerful feature representation capability, the deep learning-based model can automatically extract features directly from mmWave signals (e.g., spectrogram, point clouds) for different tasks. Amongst, \todo{Convolutional Neural Networks (CNNs) \cite{cnn}} and Recurrent Neural Networks (RNNs)~\cite{rnn} are usually exploited to learn the spatial relationship and temporal dependencies, respectively. For example, RF-SCG~\cite{rf-scg} applies a CNN-based architecture to recover SCG waveform from the reflected signals. However, these methods require a large amount of high-quality data to obtain an accurate model. To tackle it, many works choose Generative Adversarial Networks (GANs)~\cite{gan} to enhance the samples by relying on the signal propagation model or task-related physical models. For example, MILLIEAR~\cite{milliear} directly employs a conditional generative adversarial network (cGAN) to enhance the audio components and reduce noise. Furthermore, MilliPose~\cite{millipose} leverages a cGAN architecture to generate a high-resolution 2D full-body silhouette image from the low-resolution 3D mmWave signals.

\review{\subsubsection{Comparison}
We compare these two approaches as follows: (1) Domain knowledge-based methods can use domain knowledge and mathematical principles to derive and implement signal representations. Furthermore, they can also predict the performance limitations and robustness of the processing techniques. However, they may have difficulty dealing with non-linear, noisy, or uncertain signals that require advanced mathematical tools or assumptions, resulting in less flexibility or scalability. (2) Deep learning-based methods can learn from data and adapt to changing situations, optimize system performance, and intelligently filter signals. Moreover, they can handle complex and high-dimensional data that may be difficult for the former methods. Even though, deep learning requires large amounts of labeled data, computational resources, and human expertise to train and deploy effective models. Besides, they may also suffer from overfitting, underfitting, or bias issues if the data is not representative or sufficient.

}

\review{
\subsection{Summary}
As mentioned in this section, there are already many well-performing signal processing techniques from data capture to sensing models. Therefore, how to choose the appropriate techniques for each step in the sensing pipeline becomes critical to achieve a preset sensing task. Roughly speaking, point clouds and deep learning-based sensing models are often chosen when performing contour-related sensing tasks such as human imaging and gesture recognition. The reason is that deep learning-based sensing models are often used to achieve super-resolution detection or accurate multi-class classification. Furthermore, point clouds are suitable for imaging tasks due to their 3D discrete point characteristics. When performing micro-motion-related sensing tasks such as vital sensing and sound recognition, phase waveform and frequency-domain analysis are often considered. The reason lies in the ability of phase waveform to represent tiny displacements and the requirement for frequency domain features in these sensing tasks. In summary, people are recommended to take advantage of the most suitable techniques to obtain better sensing results.

\reviewII{Moreover, existing sensing techniques may not meet the requirements of specific sensing tasks due to the diversity and complexity of these tasks. This prompts researchers to continuously explore novel sensing techniques, including customized data capture and signal preprocessing techniques, novel feature extraction techniques and compound sensing models. For example, mHomeGes \cite{mhomeges} proposes a novel signal form namely Concentrated Position-Doppler Profile (CPDP) to avoid both signal noise and ambiguous angular resolution, thereby enabling accurate arm gesture recognition. mmFace \cite{mmface} reconstructs the mmWave signals into 3D facial images and then extracts distance-resistant facial structure features for accurate face authentication. Such spatial features are more suitable for human imaging, compared with time-domain features or frequency-domain features. AmbiEar \cite{ambiear} proposes an indirect voice sensing model from surrounding objects and incorporates a deep learning model to achieve voice recognition in non-line-of-sight (NLoS) scenarios. This can inspire researchers to combine the domain knowledge-based model and deep learning-based model in a complementary manner, which may improve the interpretability and accuracy of the sensing pipeline simultaneously. These techniques can help researchers overcome specific challenges and achieve better sensing performance.

}
}

\section{Sensing Task}
\label{sec:sensing_task}

\begin{figure}[!tb]
\centering
\includegraphics[width=0.85\linewidth]{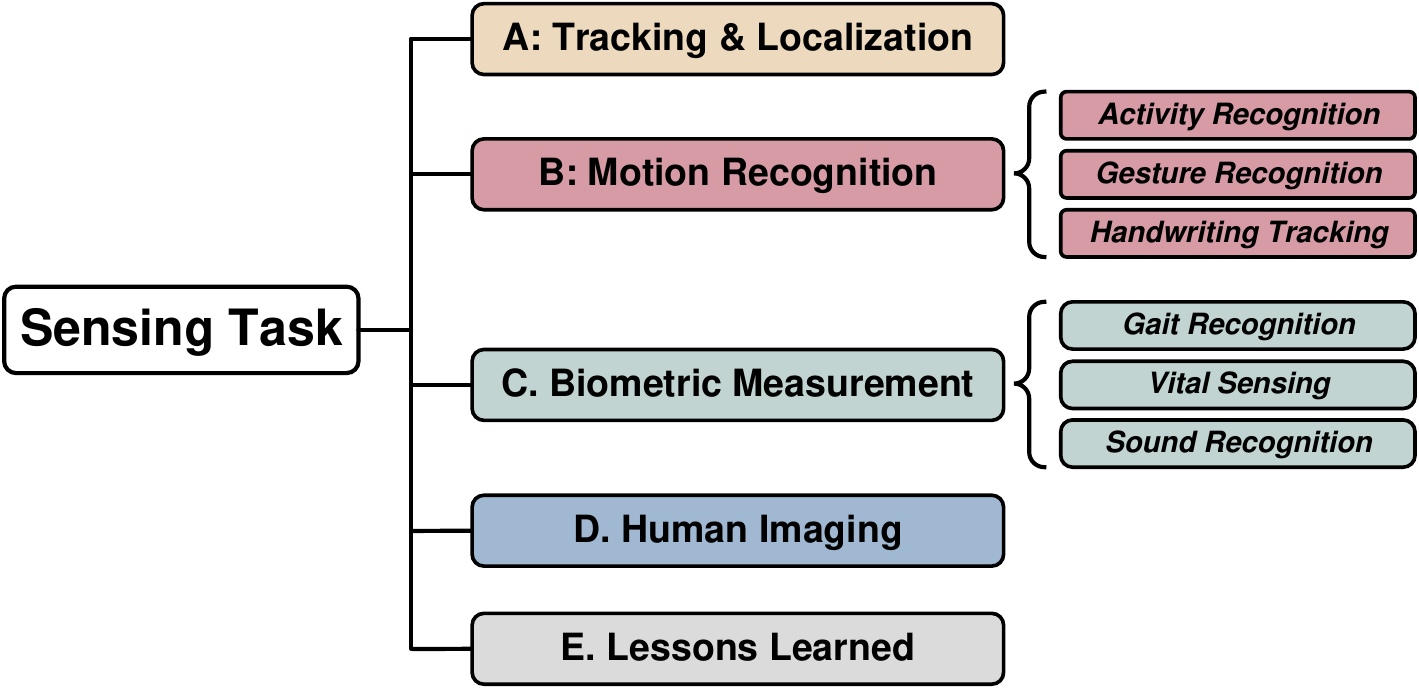}
\caption{\review{The structure of sensing task.}}
\label{fig:task_structure}
\end{figure}

According to the sensing granularity of the human sensing task, we divide the mmWave-based human sensing works into four categories: human tracking and localization, motion recognition, biometric measurement and human imaging. \review{As shown in Fig.~\ref{fig:task_structure}, the motion recognition works are further divided into three categories, including activity recognition, gesture recognition and handwriting tracking for better summary and comparison. The biometric measurement works are also divided into gait recognition, vital sensing and sound recognition according to the sensing task. We will introduce these works in the following. Finally, we summarize the lessons learned to help readers conduct their research more smoothly.}

\begin{table*}[h]
\renewcommand{\arraystretch}{1.3}
  \centering
  %\captionof{table}{THE SIMULATION PARAMETERS.}
\caption{\review{Comparison of Tracking \& Localization Works}}

\begin{tabular}{lccccc} 
\toprule

\textbf{Method}  & \textbf{Signal Form}      & \textbf{Algorithm}                                                                                                            & \textbf{Hardware Platform}               & \textbf{Identify Accuracy} & \textbf{Localization Error}                                                                          \\ \hline
\textbf{mmSense} \cite{mmsense} & \begin{tabular}[c]{@{}c@{}}RSS \\ Fingerprint\end{tabular} & LSTM                                                                                                                 & \makecell[c]{Customized\\ 60GHz Platform}       & 93\%              & -                                                                                           \\ \hline
\textbf{mmTrack} \cite{mmtrack} & CSI & \begin{tabular}[c]{@{}c@{}}Beamforming, K-means\\ Clustering, Weighted Bipartite\\ Graph Matching\end{tabular}         & \makecell[c]{Qualcomm 60GHz\\ 802.11ad Chipset} & 97.8\%            & 16.24 cm                                                                                     \\ \hline
\textbf{mID} \cite{mid}    & Point Cloud     & \begin{tabular}[c]{@{}c@{}}DBSCAN Clustering,\\ Hungarian Algorithm,\\ BiLSTM\end{tabular}                             & TI IWR1443boost                 & 89\%              & 16 cm                                                                                        \\ \hline
\textbf{PALMAR} \cite{palmar} & Point Cloud     & \begin{tabular}[c]{@{}c@{}}Adaptive Order Hidden\\ Markov Model, Crossover Path\\ Disambiguation\end{tabular} & \makecell[c]{TI 79-81GHz\\ mmWave Sensor}       & -                 & \begin{tabular}[c]{@{}c@{}}6.9 cm of Single Person\\ 11.5 cm of Multiple Persons\end{tabular} \\

\bottomrule
\end{tabular}
\label{tab:tracking}
\end{table*}

\subsection{Tracking \& Localization}

Human tracking and localization is the critical task for ubiquitous human sensing. With human position and trajectory, various applications can be developed. Besides, tracking and localization are cornerstones of further human sensing, including motion recognition, biometric measurement and so on. \review{We summarize the related works in TABLE \ref{tab:tracking}.}

The core challenge of human tracking and localization lies in how to distinguish the human from surrounding static objects. As the reflected signal of the human body is likely to be confused with that of other objects, specific human-related features need to be mined to distinguish people. Furthermore, when multi-person localization is required, continuous tracking is also a fundamental problem.

Most of the existing works mainly detect humans by mining the dynamics of the human body. Some works focus on human motion dynamics and utilize the signal variance to detect humans, while other works achieve human recognition by observing the dynamics of the vital signs. \todo{There are also some works that utilize the human body contour to enable human detection.} For multi-person tracking and localization, researchers usually consider it as the association between the detection results and existing trajectories. It can be translated into a bipartite graph-matching problem and can be well-solved by many classical algorithms, such as the Hungarian algorithm.

\textbf{mmSense} \cite{mmsense} enables a multi-person detection and identification system with a single 60GHz mmWave radio. It is implemented based on a customized 60GHz
 sensing platform, which contains a Keysight signal generator equipped with the Vubiq 60GHz front-end \todo{\cite{vubiq}}. mmSense first segments the detection region into multiple areas and constructs the RSS fingerprints with and without human presence. It then utilizes a human detection model based on \todo{long short term memory (LSTM)} to determine whether there are people in each area. In this way, mmSense can detect the presence of humans and localize their positions. mmSense further utilizes multiple features to identify humans, including body surface boundary, body surface curvature and vital signs. Results show that mmSense achieves a human presence classification accuracy of 98.75\% and a human identification accuracy of 93\%.

\begin{figure}[!tb]
\centering
\includegraphics[width=\linewidth]{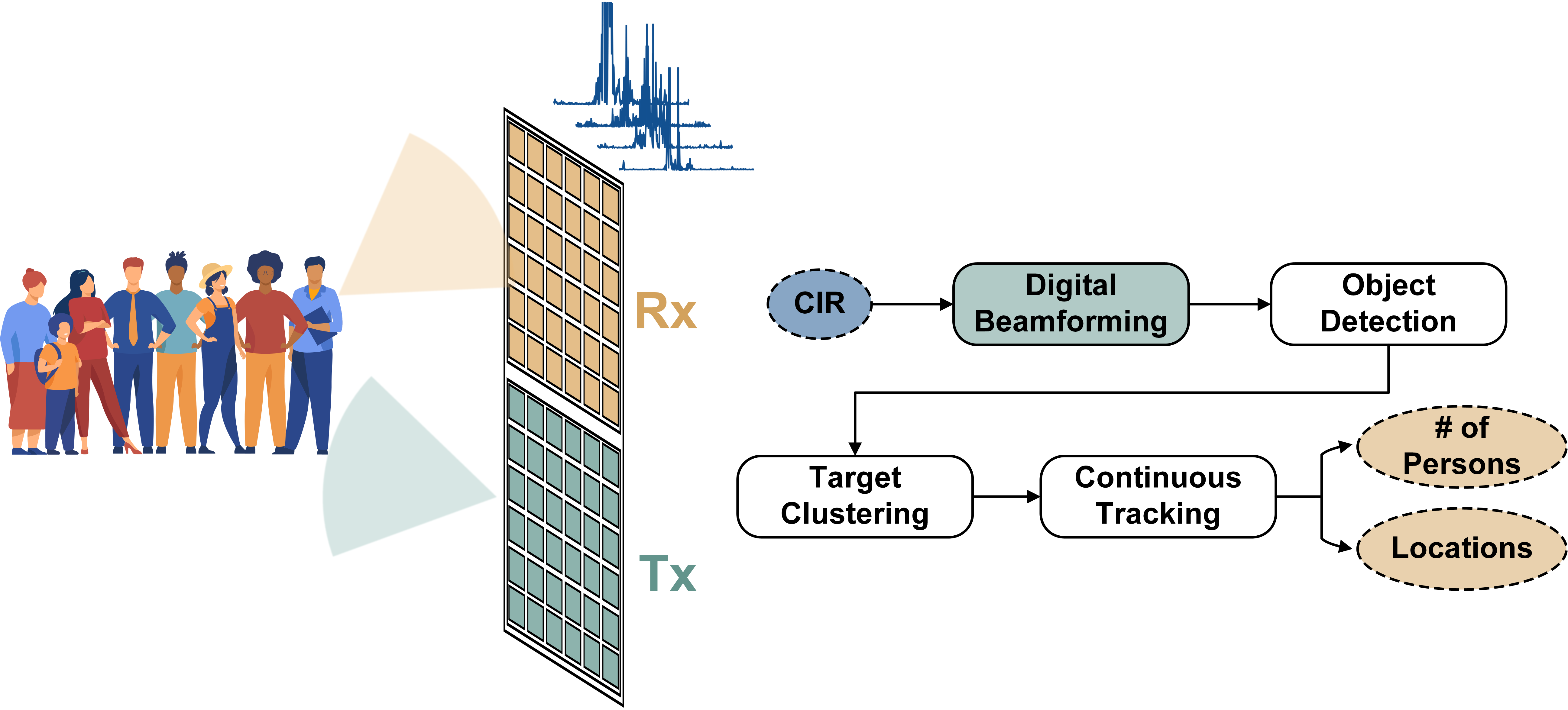}
\caption{The overview of mmTrack.}
\label{fig:mmtrack}
\end{figure}

 Different from mmSense, \textbf{mmTrack} \cite{mmtrack} utilizes the spatial spectrum rather than the RSS fingerprints to achieve multi-person localization. mmTrack is implemented based on a Qualcomm 60GHz 802.11ad chipset with an additional antenna array. The co-located transmitter and receiver arrays are both equipped with 32 elements. As illustrated in Fig.~\ref{fig:mmtrack}, it first extracts the CIR measurements from the received signal and employs the beamforming algorithm to obtain the spatial spectrum. Considering the human motion dynamics, mmTrack calculates the variation of the energy distribution of the spatial spectrum at each range to detect targets. After transforming the points in the Spherical coordinate system to 3D locations in Euclidean space, mmTrack clusters these points via \todo{K-means clustering algorithms \cite{k-means}} to estimate the human locations. mmTrack further leverages the geometric properties of the detected targets to design post-validation techniques for clustering robustness improvement. Finally, mmTrack models continuous human tracking as a weighted bipartite graph matching problem to perform the association between the detection results and existing trajectories. It achieves a median location error of 16.24 cm with moving and static users.

The spatial distribution of point clouds indicates the contour of the human body, which can also be used for human detection. \textbf{mID} \cite{mid} demonstrates the feasibility of human tracking and identifying with mmWave radars. It utilizes the sparse point clouds generated from mmWave radars to track and identify multiple people. The sparse point clouds show the 3D location of the reflection points and indicate the people's information. mID first merges the point clouds into clusters by DBSCAN \cite{dbscan} clustering algorithm. Considering that the distribution of the people's point cloud on the z-axis is more dispersed than that on the x-axis and y-axis, mID modifies the Euclidean distance to place less weight on the contribution from the z-axis. After that, mID employs the Hungarian algorithm \cite{Hungarian} to associate between each cluster at the current timestamp and the track obtained before, which is a classic many-to-many assignment problem. mID further utilizes a Kalman filter \cite{kalmanfilter} to correct tracking errors. Finally, the points of potential human objects are voxelized by a fixed-size bounding box to form an occupancy grid. mID passes it into a bi-directional LSTM network \cite{lstm} for final people identification. mID is developed on top of a COTS mmWave radar, TI IWR1443boost \cite{TI1443}. The experimental results show that mID achieves a median position error of 0.16 m and identification accuracy of 89\% for 12 people.

When people are too close to be distinguished by mmWave radar, the trajectory crossover problem arises and degrades the tracking accuracy. Some works have attempted to handle this problem. \textbf{PALMAR} \cite{palmar} applies the Adaptive Order Hidden Markov Model (AO-HMM) and the Crossover Path Disambiguation Algorithm (CPDA) to handle multiperson path ambiguity and trajectory crossover \cite{findinghumo}. It achieves an overall error of 11.5 cm in tracking multiple persons and outperforms mID with an overall 57.4\% improvement.

In addition to tracking and locating the human body, there are also some works to locate the human parts directly. \textbf{WaveEar} \cite{waveear} utilizes a customized 24GHz mmWave probe with 16 antennas to scan in all directions. It extracts all the time domain features to detect the throat's location. \textbf{RadioMic} \cite{radiomic} leverages the symmetry of the Doppler shift caused by sound vibration to locate the sound source directly. \textbf{RF-SCG} \cite{rf-scg} estimates the optimal 3D location of the heart by combining the estimated heart rate with the time-domain spectral properties extracted by FFT and beamforming techniques.

\subsection{Motion Recognition}

\begin{table*}[h]
\renewcommand{\arraystretch}{1.3}
  \centering
  %\captionof{table}{THE SIMULATION PARAMETERS.}
\caption{\review{Comparison of Motion Recognition Works}}

\begin{tabular}{clcccc} 
\toprule
\textbf{Category}                              & \textbf{Method}     & \textbf{Signal Form}            & \textbf{Algorithm}                                                                                                                          & \textbf{Hardware Platform}                         & \textbf{Performance}                                                                                                       \\ \hline
\multirow{10}{*}{\hfil \makecell[c]{Activity \\ Recognition}} & \textbf{EI} \cite{EI}         & CIR       & \begin{tabular}[c]{@{}c@{}}CNN, Unsupervised Domain \\ Adversarial Training\end{tabular}                                             & \makecell[c]{COTS 60GHz \\mmWave Transceiver}            & 65\% Accuracy                                                                                                          \\ \cline{2-6} 
                                      & \textbf{SPARCS} \cite{sparcs}     & CIR       & \begin{tabular}[c]{@{}c@{}}Joint Probabilistic Data\\ Association Filter, \\Iterative Hard Thresholding\end{tabular}                  & mm-FLEX Platform                          & 90\% Accuracy                                                                                                          \\ \cline{2-6} 
                                      & \textbf{RadHAR} \cite{radhar}     & Point Cloud           & \begin{tabular}[c]{@{}c@{}}Time-distributed CNN,\\ BiLSTM\end{tabular}                                                                                                        & TI IWR1443BOOST                           & 90.47\% Accuracy                                                                                                       \\ \cline{2-6} 
                                      & \textbf{m-Activity} \cite{m-activity} & Point Cloud           & \begin{tabular}[c]{@{}c@{}}Fast-calculation CNN,\\ Less-parameter RNN\end{tabular}                                                  & TI IWR1443BOOST                           & \begin{tabular}[c]{@{}c@{}}93.25\% of Off-line\\ 91.52\% of Real-time\end{tabular}                             \\ \cline{2-6} 
                                      & \textbf{PALMAR} \cite{palmar}     & Point Cloud           & \begin{tabular}[c]{@{}c@{}}DBSCAN and BIRCH\\ Clustering, Adaptive Order \\Hidden Markov Model,\\ Variational Autoencoder\end{tabular} & 79GHz mmWave Radar                        & 91.88\% Accuracy                                                                                                       \\ \hline
\multirow{12}{*}{\begin{tabular}[c]{@{}c@{}}Gesture \\ Recognition\end{tabular}}  & \textbf{Soli} \cite{soli}       & \begin{tabular}[c]{@{}c@{}}Range-Doppler,\\ Spectrum\end{tabular} & Bayesian Filter                                                                                                                    & Soli Chip                                 & 92.1\% Accuracy                                                                                                        \\ \cline{2-6} 
                                      & \textbf{RFWash} \cite{rfwash}     & \begin{tabular}[c]{@{}c@{}}Range-Doppler, \\ Spectrum\end{tabular} & BiLSTM                                                                                                                             & TI IWR1443BOOST                           & 92.59\% Accuracy                                                                                                       \\ \cline{2-6} 
                                      & \textbf{mmASL} \cite{mmasl}      & Doppler Spectrum       & \begin{tabular}[c]{@{}c@{}}STFT and Log Normalization,\\ CNN, Multitask Learning \end{tabular}                              & \makecell[c]{NI+SiBeam 60GHz\\FPGA Software Radio} & 87\% Accuracy                                                                                                          \\ \cline{2-6} 
                                      & \textbf{mmPose} \cite{mmpose}     & Point Cloud           & \begin{tabular}[c]{@{}c@{}}3D Heatmap Conversion\\ Forked CNN\end{tabular}                                                         & TI AWR1642BOOST                           & \begin{tabular}[c]{@{}c@{}}3.2 cm,7.5 cm,2.7 cm  Error in\\ Depth, Elevation and Azimuth\end{tabular} \\ \cline{2-6} 
                                      & \textbf{mHomeGes} \cite{mhomeges}   & Point Cloud           & \begin{tabular}[c]{@{}c@{}}CNN, Ghost Image Separation, \\ Hidden Markov Model-based\\ Voting Mechanism\end{tabular}        & TI IWR1443BOOST                           & 97.96\% Accuracy                                                                                                       \\ \cline{2-6} 
                                      & \textbf{Pantomime} \cite{pantomime}  & Point Cloud           & \begin{tabular}[c]{@{}c@{}}Pointnet++, LSTM\end{tabular}                                                                         & TI IWR1443BOOST                           & 95\% Accuracy                                                                                                          \\ \hline
\multirow{6}{*}{\begin{tabular}[c]{@{}c@{}}Handwriting \\ Tracking\end{tabular}} & \textbf{mTrack} \cite{mtrack}     & RSS/Phase & \begin{tabular}[c]{@{}c@{}}Discrete Beam Steering, \\ Dual-differential Background \\Removal\end{tabular}                              & \makecell[c]{Customized 60GHz \\Software-radio Testbed}   & \makecell[c]{90-percentile Tracking\\ Error below 8 mm}                                                                         \\ \cline{2-6} 
                                      & \textbf{mmWrite} \cite{mmwrite}    & CIR       & \begin{tabular}[c]{@{}c@{}}Background Subtraction, \\Discrete Cosine Transform\end{tabular}                               & \makecell[c]{Qualcomm 60GHz \\802.11ad Chipset}           & 2.8 mm  Median Error                                                                                 \\ \cline{2-6} 
                                      & \textbf{mmKey} \cite{mmkey}      & CIR       & \begin{tabular}[c]{@{}c@{}}Adaptive Background\\ Cancellation, MUSIC\end{tabular}                                                    & \makecell[c]{Qualcomm 60GHz \\802.11ad Chipset}           & \begin{tabular}[c]{@{}c@{}}95\% Accuracy for Single-key \\ 90\% Accuracy for Multi-key\end{tabular}                              \\

\bottomrule
\end{tabular}
\label{tab:motion_recognition}
\end{table*}

Human motion recognition plays an important role in a wide range of real-world applications. As the reflected mmWave signals usually carry substantial information about the human object, human motion can be well recognized with mmWave sensing. Not only can human activities or gestures be distinguished, but the handwriting can be accurately tracked. \review{We summarize the related works in TABLE \ref{tab:motion_recognition}.}

\subsubsection{Activity recognition}
Human activity recognition (HAR) has received continuous attention due to its wide applicability in actual scenes. Based on the high spatial resolution of mmWave signals, many researchers have begun to explore mmWave-based human activity recognition.

The core challenge of activity recognition is to find appropriate and environmental-independent features to represent activity-related information in reflected signals. The CIR frequency response, micro-Doppler spectrum and voxelized point clouds have been exploited as activity-related features in the existing works. Moreover, learning-based techniques such as domain adversarial technique and domain adaptation technique have been applied to resist the impact of the environment in feature extraction and classification.

\textbf{EI} \cite{EI} proposes an environment-independent activity recognition framework. It utilizes a COTS 60GHz mmWave transceiver system equipped with a 24-element phased antenna array to collect multi-user activity data from different environments. EI first obtains the CIR measurements and transforms each CIR measurement into a frequency response sample. These frequency responses in a period of time then form an input matrix. EI employs CNN to extract activity features from the input matrix. After that, a fully-connected layer and a softmax layer are used to obtain the activity probability vector. In particular, EI adopts the unsupervised domain adversarial training technique \cite{EI_adversarial} with unlabeled data to produce environmental-independent activity features. It further proposes three constraints to tackle the overfitting problem and improve the recognition performance, including confident control constraint, smoothing constraint and balance constraint. The experimental results show that EI achieves a recognition accuracy of about 65\% in different environments.

When people move, the different movements of the body parts cause different frequency modulation on the reflected signal, called the micro-Doppler effect. This feature is highly related to human movement and can be exploited to recognize the human activity. \textbf{SPARCS} \cite{sparcs} explores the micro-Doppler spectrum extracted from CIR samples for activity recognition. It reuses existing communication traffic and provides an \todo{integrated human sensing and communication (ISAC)} solution. 

After obtaining the CIR estimation from communication traffic, SPARCS first tracks each person's distance and angular position by performing peak detection, \todo{angle of arrival (AoA)} estimation and Joint Probabilistic Data Association Filter (JPDAF) \cite{jpdaf}. Due to the bursty and irregular traffic pattern, the CIR samples are sparse and irregular. In this case, the micro-Doppler spectrum can not be directly obtained by performing STFT, which needs uniform samples. To recover the micro-Doppler spectrum from such irregular and sparse CIR samples, SPARCS resamples the random CIR samples to obtain the regularly spaced samples and missing values with a fixed interval. Then the recovery of the micro-Doppler spectrum from such incomplete CIR samples can be formulated as a sparse recovery problem and solved with the \todo{Iterative Hard Thresholding (IHT)} algorithm \cite{iht}. Furthermore, when the communication traffic is too sparse to recover the accurate micro-Doppler spectrum, SPARSE injects short sensing units to provide enough CIR measurement. SPARSE is developed on top of the open-source mm-FLEX platform \cite{mmflex} with 60GHz mmWave RF front-ends. The experimental results show that SPARCS reaches over 0.9 F1 scores on all four activities.

\begin{figure}[!tb]
\centering
\includegraphics[width=\linewidth]{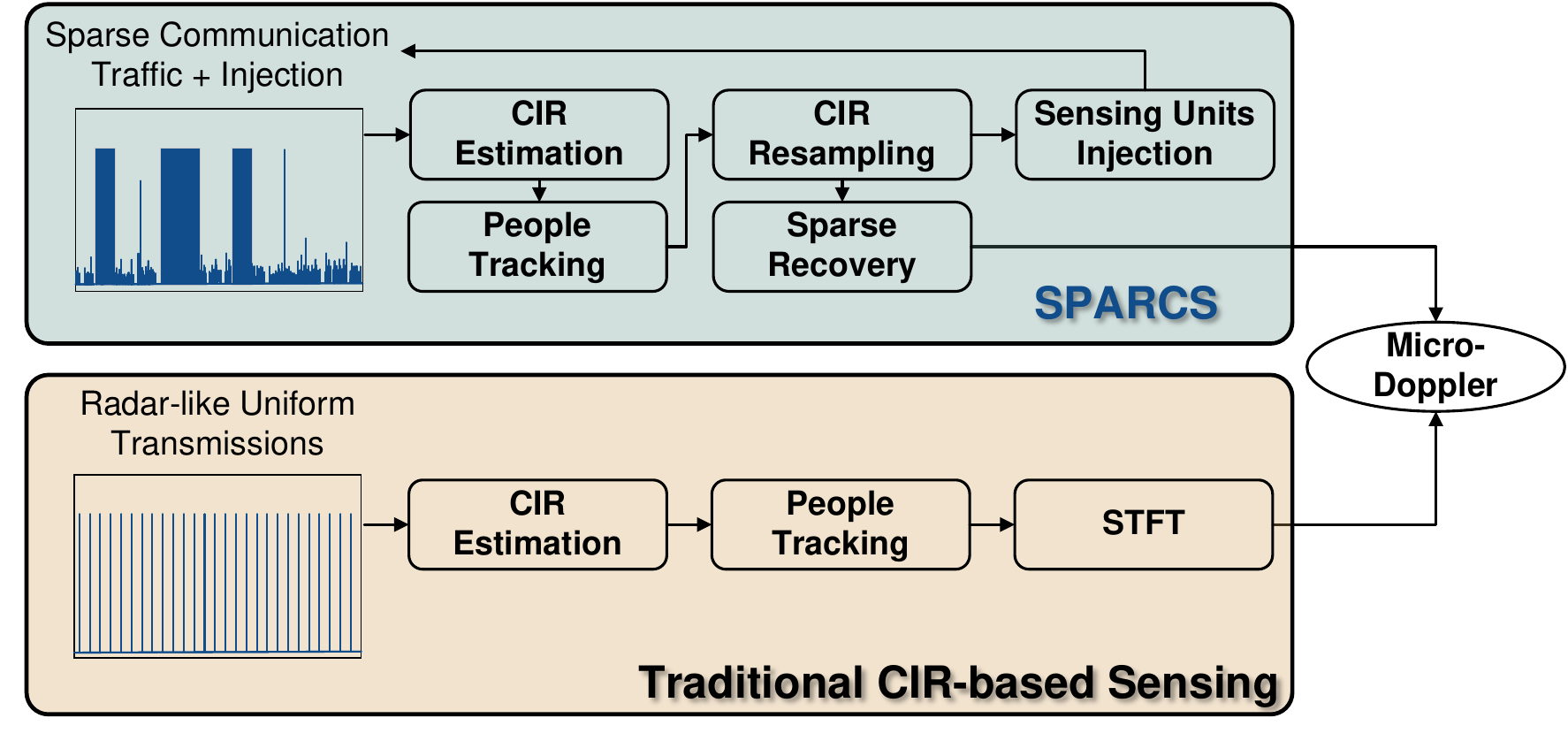}
\caption{Comparison between the traditional CIR-based human sensing and SPARCS.}
\label{fig:sparcs}
\end{figure}

In addition to the frequency response and the micro-Doppler spectrum, point clouds are also exploited to achieve activity recognition. \textbf{RadHAR} \cite{radhar} explores how to perform accurate HAR using point clouds generated through a mmWave radar. RadHAR first captures the point clouds which contain spatial coordinates, velocities, ranges, intensities and angles. The point clouds are then converted into 3D voxels to tackle the non-uniformity of points number in each frame. The value of each voxel is the number of data points within its boundaries. After that, RadHAR utilizes a sliding time window to accumulate point clouds to overcome the sparsity of points number. Finally, RadHAR evaluates different classifiers and selects the time-distributed \todo{convolutional neural network (CNN)} + bi-directional LSTM classifier as the best-performing classifier. The experimental results show that RadHAR achieves a recognition accuracy of 90.47\% with five different activities.

Considering the applicability of activity recognition in the real world, \textbf{m-Activity} \cite{m-activity} enables a real-time activity recognition system in noisy environments. It removes the noise points based on the density of point clouds and extracts the point clouds of human activity. Then the 3D data is accumulated as an integral temporal flow to reduce recognition cost. m-Activity further combines a fast-calculation CNN with a less-parameter \todo{recurrent neural network (RNN)} to classify different activities. It achieves an off-line activity accuracy of 93.25\% and real-time activity accuracy of 91.52\%. 

\textbf{PALMAR} \cite{palmar} further exploits the point clouds to track and recognize multi-person activities simultaneously. It first collects the point clouds generated from a 79GHz mmWave radar and represents them in voxel format. Then PALMAR employs DBSCAN and BIRCH \cite{birch} clustering algorithm to cluster these voxels of each people. PALMAR further applies the Adaptive Order Hidden Markov Model (AO-HMM) to track multiple targets and the Crossover Path Disambiguation Algorithm (CPDA) to handle multiperson path ambiguity and trajectory crossover \cite{findinghumo}. Finally, PALMAR proposes a deep domain adaptation model based on a variational autoencoder to improve activity recognition accuracy and domain adaptation. It achieves a recognition accuracy of 91.88\% with domain adaptation in multi-user scenarios.

\subsubsection{Gesture recognition}
``In air'' gesture recognition using mmWave signals has shown its potential in human-computer interaction and health monitoring. Various gesture recognition-based applications have been developed, including fine gesture interaction, hand hygiene monitoring, sign language communication, skeletal posture estimation and so on.

Similar to activity recognition, gesture recognition requires extracting appropriate features from the received signals. To do so, researchers have explored the potential of Range-Doppler spectrums, Doppler spreads, point clouds and concentrated position-Doppler profiles. Furthermore, as gesture recognition requires more fine-grained analysis and sensing, compared with activity recognition, various feature extraction algorithms and classification approaches have been proposed, including bidirectional LSTM, forked CNN, hidden Markov model-based voting mechanism and so on.

\begin{figure}[!tb]
\centering
\includegraphics[width=\linewidth]{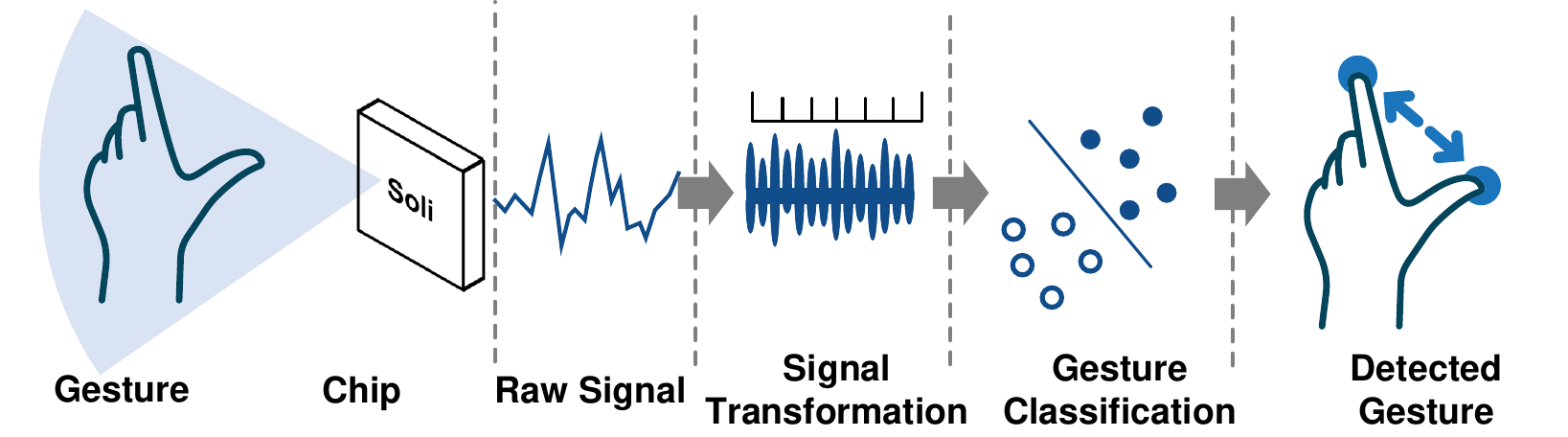}
\caption{Soli: the first mmWave radar gesture sensing system.}
\label{fig:soli}
\end{figure}

\textbf{Soli} \cite{soli} is the first mmWave radar gesture recognition system developed by Google. As shown in Fig.~\ref{fig:soli}, it achieves ubiquitous fine gesture interaction and has been integrated into smart devices such as Google Pixel 4 \cite{soli_home}. Soli customizes its own radar chips in $12 \times 12$ mm and $9 \times 9$ mm scales. Each chip operates at the 60 GHz band with 7 GHz bandwidth. It also has two transmit and four receive antenna elements to support digital beamforming. 

To enable ubiquitous gesture recognition, Soli first proposes the scattering center model of the human hand. It models the RF response of the hand as a superposition of the response of discrete scattering centers on the hand. Then Soli performs slow-time processing and fast-time processing on the received signal to produce the Range-Doppler spectrum. It further projects the Range-Doppler spectrum into the range profile and the Doppler profile for feature extraction. In the feature extraction phase, Soli extracts a variety of features, including explicit scattering center tracking features, low-level descriptors of physical RF measurement and data-centric machine learning features. Finally, different machine learning classifiers can be used to identify the user's gesture. The experimental results demonstrate that Soli can achieve a recognition accuracy of 92.1\% with a Bayesian filter.  

\begin{figure}[!tb]
\centering
\includegraphics[width=\linewidth]{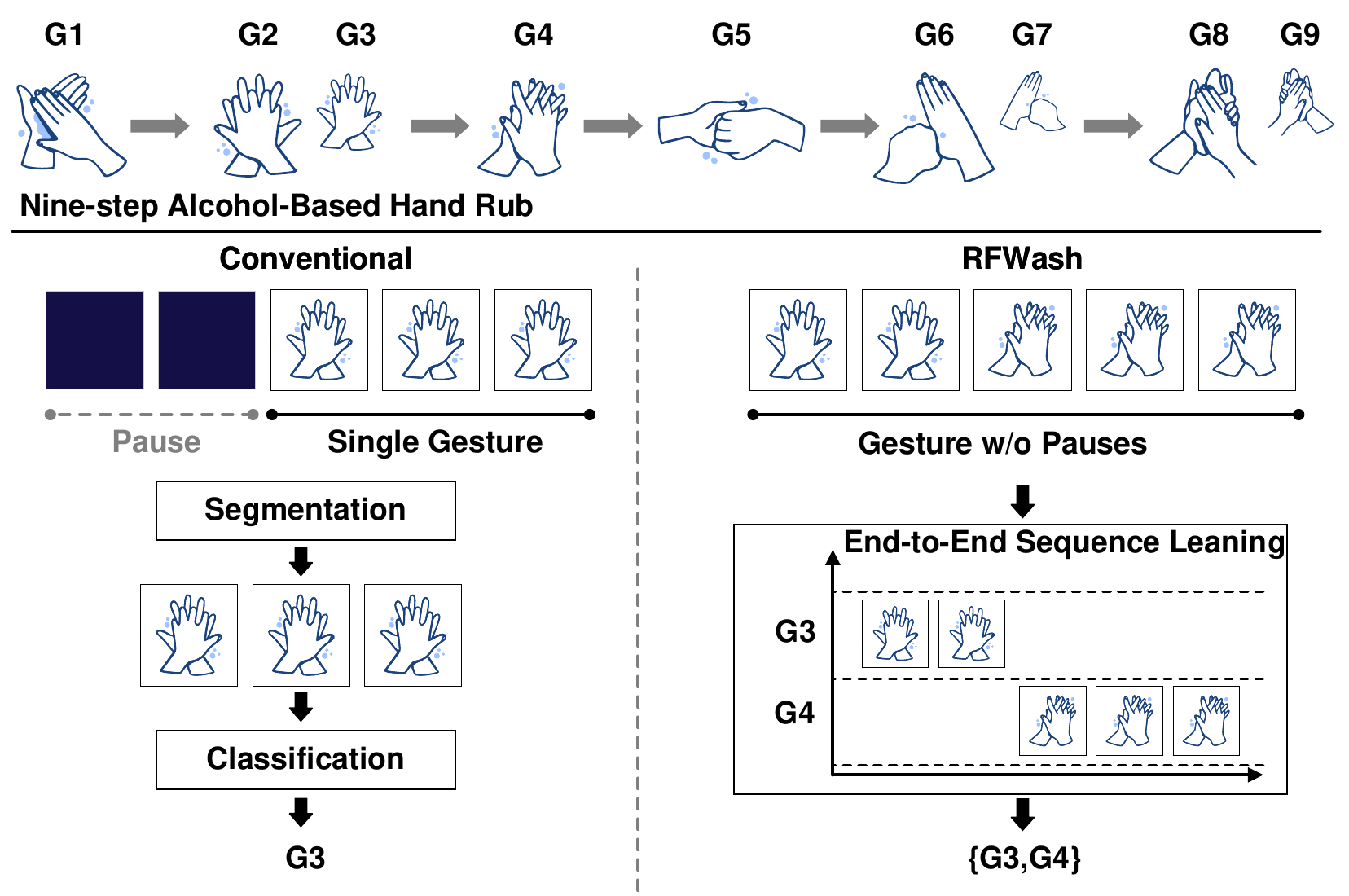}
\caption{Gesture sequence recognition of RFWash.}
\label{fig:rfwash}
\end{figure}

\textbf{RFWash} \cite{rfwash} proposes a mmWave-based gesture sensing approach to monitor the nine-step Alcohol-Based Hand Rub technique. As shown in Fig.~\ref{fig:rfwash}, unlike popular gesture recognition approaches which contain the detection/segmentation step and the recognition step, RFWash demonstrates that data segmentation is a tough task and seriously affects gesture classification accuracy. Inspired by speech and handwriting recognition approaches, RFWash proposes a segmentation-free approach to accurately recognize hand hygiene gestures without segmentation.

RFWash mounts the mmWave radar on a soap dispenser and extracts the Range-Doppler spectrum at each time instant. It further limits the range to less than 1m to remove interference from other people. To predict the gesture sequence from the Range-Doppler spectrums, RFWash first employs bidirectional recurrent layers with LSTM cell type (BiLSTM) \cite{bilstm} to extract the gesture-related spatiotemporal features. Then it adopts temporal alignment learning \cite{ctc} to map the output of BiLSTM to the corresponding gesture sequence. RFWash further employs order-preserving concatenation to augment training data and extends the data size quadratically. The results show that RFWash achieves a gesture error rate of less than 8\% and reduces manual labeling overhead by about 67\%.

Sign language communication is another major application of gesture recognition. To serve the Deaf and Hard-of-Hearing (DHH) community, \textbf{mmASL} \cite{mmasl} proposes a home assistant system that can recognize American Sign Language (ASL) using 60GHz mmWave signals. mmASL utilizes a sinusoid of 1MHz as the baseband signal and extracts the corresponding Doppler spread of the received signal, which is caused by the movements of the body parts. After obtaining the received signal, a low-pass 1KHz filter and a high-pass 10Hz filter are first applied to the received signal for denoising. Then the spectrums are plotted by STFT and enhanced by log normalization. mmASL then utilizes these spectrums to achieve wake-word recognition and ASL sign recognition. Specifically, mmASL employs beam scanning through a set of beam sectors and performs STFT for each sector's data. These results are concatenated to form the spatial spectrograms. Then mmASL employs a CNN-based machine learning model to recognize the wake word quickly. It further leverages a multitask learning network to extract and learn the ASL-related features to achieve ASL recognition. mmASL is implemented on NI+SiBeam 60GHz multi-FPGA software radio platform \cite{sibeam} with a phased antenna array. The experimental results demonstrate that mmASL can detect wake-word with an average accuracy of 94\% and achieves an average sign recognition accuracy of 87\%. 

\textbf{mmPose} \cite{mmpose} exploits the point clouds generated from the mmWave signals to achieve real-time human skeletal posture estimation. It uses TI AWR1642BOOST board \todo{\cite{AWR1642}} to collect the reflected signal and build the 3D point clouds. Then mmPose assigns an RGB weighted pixel value to each point based on its reflection intensity, resulting in a 3D heatmap. mmPose further compresses this heatmap into two low-resolution planes, the depth-azimuth plane and the depth-elevation plane, to overcome the sparsity of the heatmap and reduce unnecessary computing costs. Finally, a forked CNN architecture is used to output the skeletal joint coordinates of the human. The experimental results show that it accurately predicts human motion with four different gestures.

It is desirable to perform gesture recognition in smart home scenarios. However, the surrounding interference can prevent the mmWave radar from discovering users and thus make gesture recognition unavailable. \textbf{mHomeGes} \cite{mhomeges} attempts to overcome this challenge and achieves a real-time arm gesture recognition system. mHomeGes first captures a series of point clouds with a fixed-length sliding window. Then it extracts the concentrated position-Doppler profile (CPDP) that compresses the intensity of each point into the distance dimension and the Doppler dimension. It can be regarded as the sum in the time dimension of the Range-Doppler spectrums. Then CPDP is used as the input of a customized CNN recognition model to recognize fine-grained gestures. To eliminate the multipath effect caused by the surrounding reflectors, mHomeGes proposes a novel ghost image separation algorithm using the velocity synchrony between the user and the corresponding ghost image. Finally, mHomeGes employs a hidden Markov model-based voting mechanism (HMM-VM) to achieve continuous gesture recognition. mHomeGes is implemented based on a commercial mmWave radar, TI IWR1443 board \cite{IWR1642}. Results show that mHomeGes achieves a recognition accuracy of 97.96\% for 25 volunteers across five home scenes.

%\begin{figure}[!tb]
%%\centering
%\includegraphics[width=\linewidth]{figure/pantomime.png}
%\caption{The point cloud classification architecture of Pantomime}
%\label{fig:pantomime}
%\end{figure}

Another challenge of gesture recognition using point clouds is the sparsity of the point clouds. It makes gesture-related signals challenging to be distinguished. \textbf{Pantomime} \cite{pantomime} proposes a novel hybrid model to achieve accurate gesture recognition with sparse point clouds. Pantomime first aggregates the point clouds of a gesture segment to obtain more points in a single frame. Then the aggregated point cloud is resampled to meet a fixed point number with upsampling and downsampling algorithms. To further extract the spatio-temporal features from the aggregated point clouds, Pantomime employs a hybrid architecture combining the Pointnet++ architecture \cite{pointnet++} and the LSTM modules. Pointnet++ is designed to extract spatial features from 3D point clouds, and the LSTM modules continue to extract temporal features from obtained features. Finally, these two feature vectors are concatenated to form the final feature vector and classified by a fully connected layer. The experimental results show that Pantomime achieves 95\% accuracy and 99\% the area under the ROC curve (AUC) with 21 gestures.

\subsubsection{Handwriting tracking}
Passive handwriting tracking is a critical task in performing ubiquitous human-computer interaction. Benefiting from the short wavelength and large bandwidth of mmWave signals, researchers have proposed mmWave-based handwriting tracking approaches that can achieve mm-level tracking accuracy.

Different from activity recognition and gesture recognition, handwriting tracking requires accurate quantitative tracking of the writing object, such as a finger or a pen. The micro displacement measurement based on mmWave signals relies heavily on the phase changes of the received signal. However, the mmWave signals also suffer from background interference that distorts the phases of the mmWave signals. How to extract the phase changes caused by micro displacement becomes the key to achieving accurate handwriting tracking. Researchers have exploited beamforming algorithms and signal-denoising techniques to eliminate background interference and extract handwriting-related phase changes.

\begin{figure}[!tb]
\centering
\includegraphics[width=0.9\linewidth]{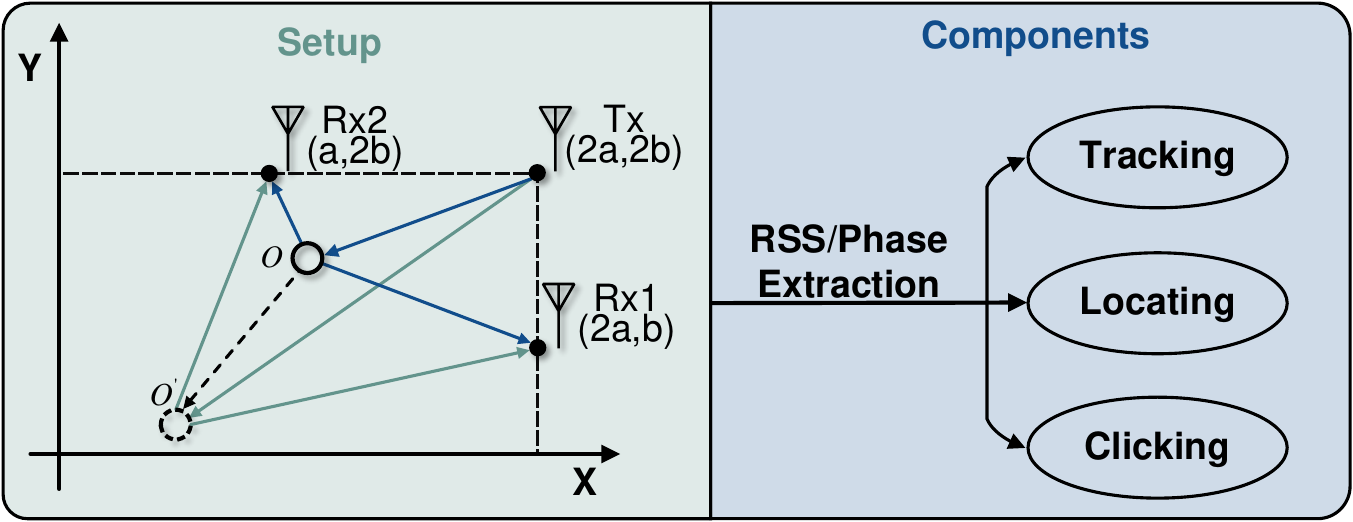}
\caption{mTrack setup and system components.}
\label{fig:mtrack}
\end{figure}

\textbf{mTrack} \cite{mtrack} proposes a high-precision handwriting tracking with 60GHz mmWave radios. The mTrack setup and system components are shown in Fig.~\ref{fig:mtrack}. The transmitter is equipped with a quasi-omni beam pattern to illuminate the tracking region, and the receiver adopts highly directional antennas to boost the target-reflected signals. With three well-designed modules, namely the anchor point acquisition module, the phase tracking module and the touch detection module, mTrack can achieve handwriting detection and tracking.

mTrack extracts the \todo{received signal strength (RSS)} and phase changes from each receiver antenna to locate and track the writing object. It first performs discrete beam steering techniques to obtain the discrete RSS samples and captures the relative angle to each receiver antenna with the interpolated maximum RSS. Then mTrack continuously tracks the writing object by extracting the phase changes. As the phase changes of the received signals are distorted by background interference, directing the conversion of the phase changes to moving distance will result in incorrect tracking results. mTrack proposes a dual-differential background removal (DDBR) algorithm to remove such an impact. Specifically, it differentiates the phase of three consecutive samples to obtain a single-phase shift sample. mTrack also designs a phase counting and regeneration algorithm to complement the DDBR algorithm.
Furthermore, mTrack utilizes the RSS measurement and the phase variance to detect the touch events. mTrack is implemented on a customized 60GHz software-radio testbed with the Vubiq 60GHz RF front-end. Results show a 90-percentile tracking error below 8 mm.

Instead of using multiple separate receiver antennas, \textbf{mmWrite} \cite{mmwrite} enables a high-precision handwriting tracking system with a single mmWave radio. mmWrite is implemented based on a Qualcomm 60GHz 802.11ad chipset with an additional antenna array. mmWrite exploits the CIR measurement from the receiver for handwriting tracking. Specifically, mmWrite first employs the background subtraction algorithm to the recorded CIRs to eliminate the impact of other static objects. Then the digital beamforming technique is performed to determine the range and direction of the reflected signals. mmWrite further extracts the Doppler velocity by STFT and performs 3D \todo{constant false alarm rate (CFAR) \cite{cfar}} target detection based on the Doppler power. This series of target locations construct an estimated trajectory of the target. Finally, mmWrite proposes a smoothing technique based on Discrete Cosine Transform  (DCT) to enhance the estimated trajectory. The experimental results show that it achieves a median tracking error of 2.8 mm. 

Similarly to mmWrite, \textbf{mmKey} \cite{mmkey} enables a universal virtual keyboard with a single mmWave radio. mmKey first utilizes the differential amplitude of the CIR to detect the presence of motions. Then mmKey employs a novel motion filter and adaptive background cancelation to extract only the finger-related reflections. It further performs the \todo{multiple signal classification algorithm (MUSIC) \cite{music}} to enable precise keystroke localization. Results show that mmKey can achieve a keystroke recognition accuracy of more than 95\% for single-key cases and more than 90\% for multi-key scenarios.

\subsection{Biometric Measurement}

\begin{table*}[th]
\renewcommand{\arraystretch}{1}
  \centering
  %\captionof{table}{THE SIMULATION PARAMETERS.}
\caption{\review{Comparison of Biometric Measurement Works}}

\begin{tabular}{clcccc} 
\toprule

\textbf{Category}                                                                     & \textbf{Method}      & \textbf{Signal Form}                                                      & \textbf{Algorithm}                                                                                 & \textbf{Hardware Platform}                                                         & \textbf{Performance}                                                                                             \\ \hline
\multirow{6}{*}{\begin{tabular}[c]{@{}c@{}}Gait\\ Recognition\end{tabular}}  & \textbf{mmGait} \cite{mmgaitnet}      & Point Cloud                                                      & \begin{tabular}[c]{@{}c@{}}DBSCAN Clustering\\ Hungarian Algorithm, DNN\end{tabular}        & \begin{tabular}[c]{@{}c@{}}TI IWR1443BOOST\\ TI IWR6843BOOST\end{tabular} & \begin{tabular}[c]{@{}c@{}}90\% Accuracy for Single-person\\ 88\% Accuracy for Five-person\end{tabular} \\ \cline{2-6} 
                                                                             & \textbf{MU-ID} \cite{muid}       & \begin{tabular}[c]{@{}c@{}}range-Doppler\\ Spectrum\end{tabular} & CNN                                                                                       & TI AWR1642BOOST                                                           & \begin{tabular}[c]{@{}c@{}}97\% Accuracy for Single-person\\ 92\% Accuracy for Four-person\end{tabular} \\ \cline{2-6} 
                                                                             & \textbf{GaiCube} \cite{gaitcube}     & \begin{tabular}[c]{@{}c@{}}micro-Doppler\\ Spectrum\end{tabular} & \begin{tabular}[c]{@{}c@{}}Log Spectrum Analysis\\ CNN\end{tabular}                       & TI IWR1443BOOST                                                           & 98.3\% Accuracy                                                                                         \\ \midrule
\multirow{24}{*}{\begin{tabular}[c]{@{}c@{}}Vital \\ Sensing\end{tabular}}                                              & \textbf{mmVital} \cite{mmvital}     & RSS                                                              & Band-pass Filters                                                                         & Vubiq Platform                                                            & \begin{tabular}[c]{@{}c@{}}0.43 Bpm Error\\ 2.15 bpm Error\end{tabular}                                   \\ \cline{2-6} 
                                                                             & \textbf{ViMo} \cite{vimo}        & CIR                                                              & \begin{tabular}[c]{@{}c@{}}2D-CFAR\\ Autocorrelation\end{tabular}                         & \makecell[c]{Commercial \\ 60-GHz WiFi}                                                    & \begin{tabular}[c]{@{}c@{}}0.19 Bpm Error\\ 0.92 bpm Error\end{tabular}                                   \\ \cline{2-6} 
                                                                             & \textbf{mBeats} \cite{mBeats}      & \begin{tabular}[c]{@{}c@{}}Phase\\ Waveform\end{tabular}         & \begin{tabular}[c]{@{}c@{}}Biquad Cascade IIR Filter\\ DNN\end{tabular}                   & TI IWR6843BOOST                                                           & 95.26\% Accuracy                                                                                        \\ \cline{2-6} 
                                                                             & \textbf{RF-SCG} \cite{rf-scg}      & \begin{tabular}[c]{@{}c@{}}Phase\\ Waveform\end{tabular}         & \begin{tabular}[c]{@{}c@{}}4D Cardiac Beamformer\\ CNN\\ Translator\end{tabular}          & TI IWR1443BOOST                                                           & 0.72 Correlation Coefficient                                                                            \\ \cline{2-6} 
                                                                             & \textbf{mmHRV} \cite{mmhrv}       & \begin{tabular}[c]{@{}c@{}}Phase\\ Waveform\end{tabular}         & Modified VMD                                                                              & TI IWR1443BOOST                                                           & 97.96\% Accuracy                                                                                        \\ \cline{2-6} 
                                                                             & \textbf{mmECG} \cite{mmecg}       & \begin{tabular}[c]{@{}c@{}}Phase\\ Waveform\end{tabular}         & \begin{tabular}[c]{@{}c@{}}Hierarchy VMD\\ Template-based Optimization\end{tabular}       & TI AWR1642BOOST                                                           & 0.37 bpm Error                                                                                           \\ \cline{2-6} 
                                                                             & \textbf{HeartPrint} \cite{heartprint}  & \begin{tabular}[c]{@{}c@{}}Phase\\ Waveform\end{tabular}         & WPT                                                                                       & TI IWR1443BOOST                                                           & 96.16\% Accuracy                                                                                        \\ \cline{2-6} 
                                                                             & \cite{rf-vital}    & \begin{tabular}[c]{@{}c@{}}Phase\\ Waveform\end{tabular}         & LSTM                                                                                      & TI AWR1243Boost                                                           & \begin{tabular}[c]{@{}c@{}}5.57 Bpm Error for Moving\\ 3.32 bpm Error for Moving\end{tabular}             \\ \cline{2-6} 
                                                                             & \textbf{Movi-Fi} \cite{movifi}     & \begin{tabular}[c]{@{}c@{}}Phase\\ Waveform\end{tabular}         & \begin{tabular}[c]{@{}c@{}}Deep Contrastive Learning\\ Encoder-decoder Model\end{tabular} & TI IWR1443BOOST                                                           & \begin{tabular}[c]{@{}c@{}}about 2\% Bpm Error\\ about 1\% Bpm Error\end{tabular}                         \\ \cline{2-6} 
                                                                             & \textbf{VED} \cite{ved}         & \begin{tabular}[c]{@{}c@{}}Phase\\ Waveform\end{tabular}         & Variational Encoder-decoder                                                               & TI IWR1843BOOST                                                           & above 0.92 Similarity                                                                                   \\ \cline{2-6} 
                                                                             & \textbf{CardiacWave} \cite{cardiacwave} & \begin{tabular}[c]{@{}c@{}}Phase\\ Waveform\end{tabular}         & \begin{tabular}[c]{@{}c@{}}Mask Filter\\ DNN\end{tabular}                                 & TI IWR1443BOOST                                                           & 97.96\%                                                                                                 \\ \cline{2-6} 
                                                                             & \textbf{mmBP} \cite{shi22sensys}       & \begin{tabular}[c]{@{}c@{}}Phase\\ Waveform\end{tabular}         & Functional Link Adaptive Filter                                                           & TI IWR1843BOOST                                                           & \begin{tabular}[c]{@{}c@{}}0.87 mmHg SBP Error\\ 1.55 mmHg DBP Error\end{tabular}                         \\ \midrule
\multirow{14}{*}{\begin{tabular}[c]{@{}c@{}}Sound\\ Recognition\end{tabular}} & \textbf{WaveEar} \cite{waveear}     & \begin{tabular}[c]{@{}c@{}}Phase\\ Waveform\end{tabular}         & \begin{tabular}[c]{@{}c@{}}DNN\\ Griffin-Lim Phase Reconstruction\end{tabular}            & \makecell[c]{Customized 24GHz \\ mmWave Probe}                                             & Lower than 6\% WER                                                                                      \\ \cline{2-6} 
                                                                             & \textbf{VocalPrint} \cite{vocalprint}  & \begin{tabular}[c]{@{}c@{}}Phase\\ Waveform\end{tabular}         & Resilience-aware Clutter Removal                                                          & TI AWR1642BOOST                                                           & 96\% Accuracy                                                                                           \\ \cline{2-6} 
                                                                             & \textbf{mmPhone} \cite{mmphone}     & \begin{tabular}[c]{@{}c@{}}Phase\\ Waveform\end{tabular}         & \begin{tabular}[c]{@{}c@{}}DNN\\ Training-free Harmonic Extension\end{tabular}            & TI AWR1843BOOST                                                           & 93\% Accuracy                                                                                           \\ \cline{2-6} 
                                                                             & \textbf{Wavedropper} \cite{wavesdropper} & \begin{tabular}[c]{@{}c@{}}Phase\\ Waveform\end{tabular}         & \begin{tabular}[c]{@{}c@{}}Wavelet-based Analysis\\ RNN\end{tabular}                      & TI IWR1642BOOST                                                           & 91.3\% Accuracy                                                                                         \\ \cline{2-6} 
                                                                             & \textbf{MILLIEAR} \cite{milliear}    & \begin{tabular}[c]{@{}c@{}}Phase\\ Waveform\end{tabular}         & Conditional GAN                                                                           & TI IWR1642BOOST                                                           & 3.68 MCD                                                                                                \\ \cline{2-6} 
                                                                             & \textbf{Wavoice} \cite{wavoice}     & \begin{tabular}[c]{@{}c@{}}Phase\\ Waveform\end{tabular}         & Multi-modal Signals Fusion                                                                & TI IWR1642BOOST                                                           & \begin{tabular}[c]{@{}c@{}}0.69\% CER\\ 1.72 WER\end{tabular}                                           \\ \cline{2-6} 
                                                                             & \textbf{AmbiEar} \cite{ambiear}     & \begin{tabular}[c]{@{}c@{}}Phase\\ Waveform\end{tabular}         & \begin{tabular}[c]{@{}c@{}}Modified MVDR\\ End-to-end Network\end{tabular}                & TI IWR1642BOOST                                                           & 87.21\% Accuracy for NLoS                                                                               \\

\bottomrule
\end{tabular}
\label{tab:biometric_measurement}
\end{table*}

% Please add the following required packages to your document preamble:
% \usepackage{multirow}

Human biometric measurement is an important step to achieve ubiquitous human sensing. Through mmWave-based sensing of gait, vital signs or sounds, people can achieve health monitoring, human-computer interaction and other applications. \review{We summarize the related works in TABLE \ref{tab:biometric_measurement}.} 
\subsubsection{Gait recognition}
Gait-based human identification is a key requirement for secure and smart applications. Compared with visual-based solutions, mmWave-based identification systems are more acceptable, as the concerns about privacy security can be eliminated. With the strong directivity of mmWave signal, they are capable of tracking and identifying multiple targets simultaneously.

The key problem of gait recognition is to define appropriate feature forms to effectively extract gait-related features from the raw data. Considering that gait-related features include the temporal, spatial and velocity variation from the reflected signal. Point clouds and Range-Doppler spectrums are widely used because they maintain rich gait-related information. Recent works further propose to use the micro-Doppler spectrum to detailedly describe the human gait.

\begin{figure}[!tb]
\centering
\includegraphics[width=\linewidth]{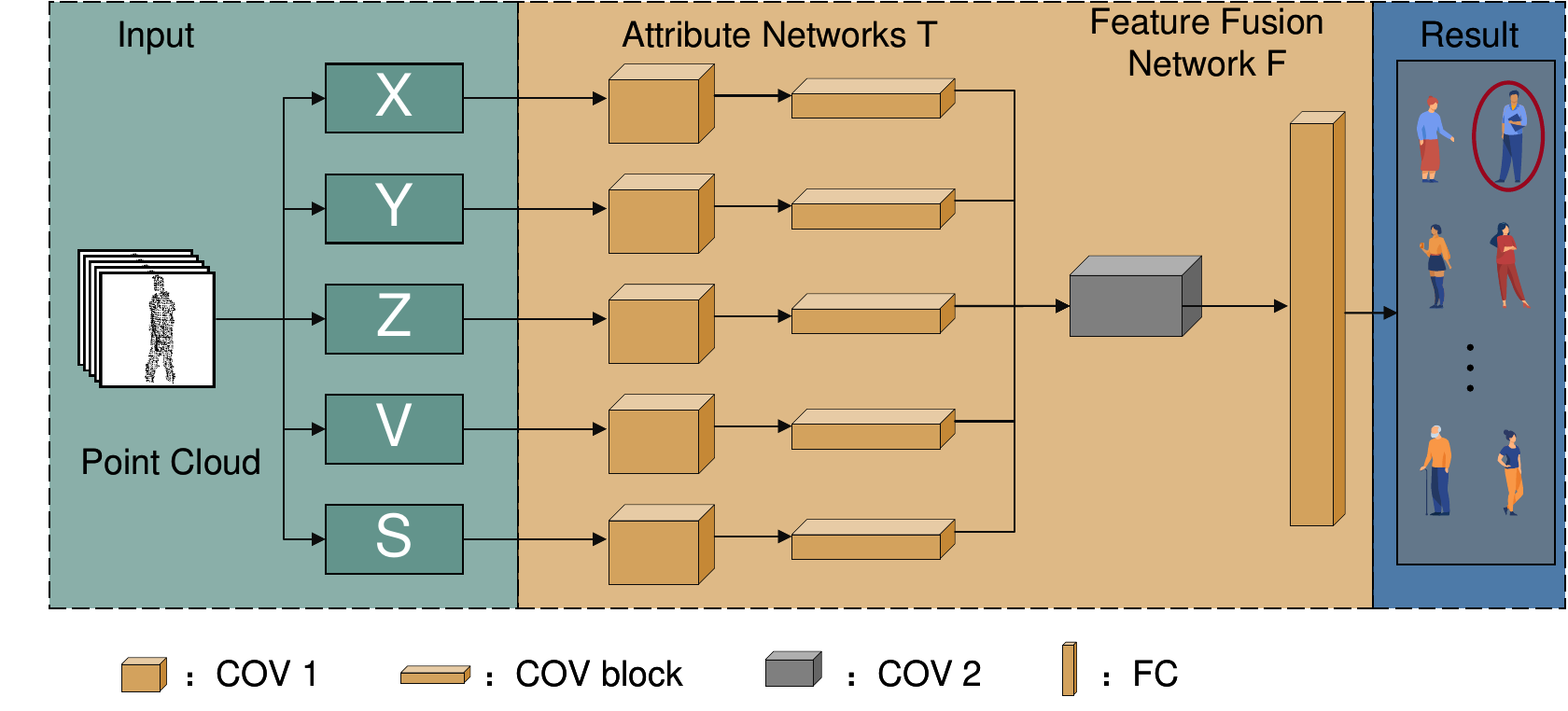}
\caption{The overview of mmGaitNet.}
\label{fig:mmgait}
\end{figure}

\textbf{mmGait} \cite{mmgaitnet} proposes a deep-learning driven method to achieve mmWave gait recognition. mmGait uses two mmWave devices, TI IWR1443 and IWR6843, to build a mmWave gait dataset from 95 volunteers. It first employs DBSCAN algorithm and Hungarian algorithm to obtain the point clouds corresponding to every single human. it further proposes a customized neural network structure, called mmGaitNet, to distinguish the point cloud. The network structure is shown in Fig.~\ref{fig:mmgait}. The inputs of mmGaitNet are point clouds' five attributes $X$, $Y$, $Z$, $V$ and $S$, where $X$, $Y$, $Z$ represent the spatial location of each point, $V$ and $S$ denote the corresponding radial speed and signal strength, respectively. The attribute features are first extracted by the attribute network, then they are input into the feature fusion network to fuse the features. Finally, the final fully connected layer output the class score. The experimental results show that mmGait achieves 90\% identification accuracy for single-person scenarios and 88\% accuracy for five co-existing persons. 

In addition to using the distribution of point clouds to recognize different humans, researchers also capture the signal frequency changes caused by human motion to achieve gait recognition. \textbf{MU-ID} \cite{muid} focuses on the gait feature from the lower limb motion in the spatial-temporal domain to achieve multi-user identification. MU-ID first converts the raw data into the Range-Doppler spectrum and removes the stationary interference. Then the Range-Doppler spectrum is compressed along the velocity dimension. A sequence of the compressed result is further arranged in the order of time to extract the lower limb motion features. When multiple users appear simultaneously, their features are separated based on AoA difference and segmented based on silhouette analysis. Finally, MU-ID develops a deep CNN-based classifier to output the classification probability distribution. The experimental results show that MU-ID achieves 97\% single-person identification accuracy and over 92\% accuracy for four people.

Compared with velocity analysis, micro-Doppler spectrum indicates more detail local motion information and can be used for gait recognition. Inspired by this observation, \textbf{GaitCube} \cite{gaitcube} proposes to extract a joint-feature representation of micro-Doppler and micro-range signatures over time, called gait data cube to embody the physical relative features of human gait. Then GaitCube can achieve accurate gait classification with the gait cubes. Specifically, GaitCube first obtains the time-range spectrum from raw data and chooses the maximum-variance trace to detect a subject. Then the human's walking phase can be detected with speed extraction and threshold comparison. After that, GaitCube estimates the gait cycle by extracting the Time-Doppler log spectrum energy and segments each gait cycle. The micro-Doppler spectrums over the range in each gait cycle are further downsampled and aligned to form the gait cubes. Assisted by several additional features such as trace length and cycle duration, the gait cube is fed into a CNN-based classification for gait recognition. The experimental results show that GaitCube achieves an accuracy of 96.1\% with a single gait cycle with one RX antenna and the accuracy increases to 98.3\% with all RX antennas.

\subsubsection{Vital sensing}
Non-contact continuous monitoring of human vital signs is one of the important applications of mmWave sensing. Through the continuous in-depth analysis of the vital information contained in the mmWave reflected signals, not only the vital feature can be extracted but also the vital waveform can be recovered.

The core challenge of vital sensing lies in the fine-grained characterization of the time-domain vital-related signal changes and exact vital signal extraction from such signals. When the sensing target is vital features such as breathing and heart rate, frequency domain analysis or learning-based feature extraction methods are widely adopted. When the sensing target spreads to the vital waveform, learning-based waveform extraction methods or signal decomposition technologies begin to play an important role.

\textbf{mmVital} \cite{mmvital} first explores how to enable continuous monitoring of human breathing and heart rates with 60GHz mmWave signals. mmVital locates the users based on the reflection loss of the human body, which is different from other objects. Then mmVital adjusts Tx and Rx angles to transmit the signal to the human body and extract the \todo{receiver signal strength (RSS)} of the reflected signal. The RSS signal is sensitive to periodic movement of the human body, resulting in a peak in the frequency domain. To achieve better accuracy, mmVital detects the corresponding frequency peaks with band-pass filters to estimate the breathing and heart rates. mmVital utilizes a mmWave development platform provided by Vubiq \cite{vubiq} for transceiver angle adjustment and RSS value collection. The experimental results show that mmVital provides reliable vital sensing with a mean estimation error of 0.43 breaths per minute (Bpm) and 2.15 beats per minute (bpm).

As the \todo{channel impulse response (CIR)} can obtain more detailed vital-related information than the RSS, \textbf{ViMo} \cite{vimo} try to extract the vital features with the CIR time-series. It enables multiperson vital sign monitoring using a commercial 60GHz WiFi. Due to the extremely high frequency and large bandwidth of the mmWave signal, 60GHz WiFi can offer high directionality with large phased arrays in small size. When the mmWave signal is reflected by the human body, the phase of the CIR measurement changes periodically due to periodic human motions, i.e. breath and heartbeat. ViMo first employs 1D-CFAR in range dimension and 2D-CFAR in angle dimension to detect reflecting objects. Then humans can be distinguished from the static reflecting objects according to the variation of their corresponding phases. ViMo estimates the breathing rate by applying the autocorrelation function to the CIR phase and finding the peak location. After that, ViMo adopts a smooth spline to estimate and eliminate the breathing signal. Finally, the enhanced heartbeat signal is utilized to continuously estimate the heart rate with dynamic programming. ViMo is embedded in a commodity 60GHz WiFi \todo{\cite{80211ad}}, which has 32 antennas assembled in a $6 \times 6$ layout. The median errors are 0.19 Bpm and 0.92 bpm in the case of a single user, and the mean accuracy of both breathing rate and heart rate is over 92.8\% in the case of 3 users.

Besides frequency domain analysis, learning-based feature extraction methods are also applied to extract vital features. \textbf{mBeats} \cite{mBeats} provides a heart rate monitoring system that is insensitive to user orientation and position. mBeats utilizes a domestic service robot equipped with a mmWave radar to track the human's orientation and position and perform vital sensing. As the service robots are typically low in height, mBeats measures the heart rate in a user's leg. The measured phase variation is caused by the micro displacement of the user's skin but filled with noise. To accurately estimate the heart rate, mBeats first leverages a biquad cascade IIR filter to extract heartbeat waveforms. Then the heart rate estimation problem is formulated as a regression problem and solved by a customized \todo{deep neural network (DNN)} predictor. mBeats utilize a commercial radar, TI IWR6843ISK board \todo{\cite{TI6843}}, to send mmWave signals and extract the phase variation. The experimental results show that mBeats provides a heart rate estimation accuracy of 95.26\% under 8 different user poses.

mmWave sensing can capture the subtle displacement of the human chest, which is mainly caused by the superposition of the heartbeat signal and respiration signal. Inspired by this observation, researchers begin to explore how to recover the vital waveform rather than just extract vital features based on mmWave sensing. 

\begin{figure}[!tb]
\centering
\includegraphics[width=\linewidth]{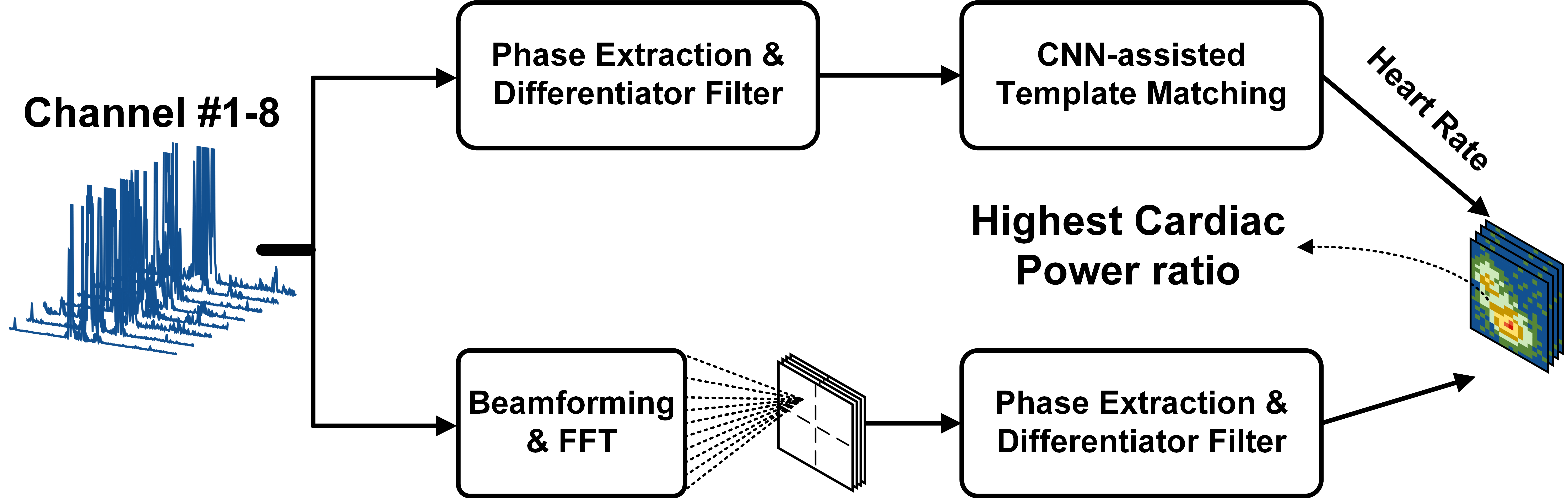}
\caption{The architecture of 4D Cardiac Beamformer.}
\label{fig:rf-scg}
\end{figure}

\textbf{RF-SCG} \cite{rf-scg} introduces how to utilize the reflection of the mmWave radar signal to reconstruct the seimocardiogram (SCG) waveform, which captures fine-grained cardiovascular events. To sense the cardiac micro-vibrations, RF-SCG first needs to exactly capture the reflected mmWave signal from the apex of the heart. RF-SCG designs a 4D Cardiac Beamformer component to discover the 3D location of the heart. The detailed architecture is shown in Fig. ~\ref{fig:rf-scg}. RF-SCG first formulates the heart rate estimation problem as a 1D matching problem and solves it with a CNN-assisted template matching technical. Then RF-SCG applies FFT and beamforming techniques to the received signal to extract the corresponding time-domain signals in 3D bins. Combining the estimated heart rate with the time-domain spectral properties, RF-SCG can estimate the optimal 3D location of the heart.

After obtaining the reflection signal from the heart, RF-SCG learns a CNN-based RF-to-SCG translator to recover SCG waveforms from the mmWave reflections. Moreover, RF-SCG modifies the Unet architecture \todo{\cite{unet}} to automatically label five fiducial points, including aortic valve opening, aortic valve closing, mitral valve opening, mitral valve closing, and isovolumetric contraction. RF-SCG is implemented based on a commercial mmWave radar, TI IWR1443BOOST board \todo{\cite{TI1443}}. Results show that RF-SCG has more than 0.72 correlation coefficient between each subject's ground truth and the recovered waveform. It also can robustly time five key cardiovascular events with a median error between 0.26\%-1.29\%.

In addition to directly recovering the vital waveform from the reflected signal, researchers explore extracting the vital signals with different signal decomposition technologies. \textbf{mmHRV} \cite{mmhrv} utilizes a modified VMD algorithm to extract the heartbeat signal from the reflected signal and monitor the heart rate variability (HRV). Similarly, \textbf{mmECG} \cite{mmecg} designs a hierarchy VMD approach to estimate the heart movement in mmWave signals. It further leverages a template-based optimization method to reconstruct the cardiac cycle in driving environments. \textbf{HeartPrint} \cite{heartprint} analyzes the minor changes in heartbeat waveform with a 3-level \todo{wavelet packet transform (WPT)}, which can conduct multi-resolution analysis in multiple frequency domains. \cite{rf-vital} designs customized LSTM models to estimate a baseline of heart rate and respiration rate separately only when the human is static. It further predicts these features on top of the baseline when the human moves.

The fundamental assumption of the signal decomposition algorithm is the linear superposition of signals. However, \textbf{MoVi-Fi} \cite{movifi} states a different point of view: The reflected signals caused by vital signs are composited with other motion-incurred reflections in a nonlinear manner. To solve such a nonlinear blind source separation problem, MoVi-Fi employs deep contrastive learning to reverse the nonlinear composition between body movement and vital signs. Then an encoder-decoder model is utilized to recover the heartbeat waveform and the breathing waveform. The experimental results show that MoVi-Fi can recover fine-grained vital signs waveforms under severe body movements. \textbf{VED} \cite{ved} also claims that the composition between respiration and heartbeat can be highly nonlinear. VED explores the ability of the variational encoder-decoder, a novel deep neural network, to convert the raw data to the heartbeat waveform. It achieves a heart rate estimation median error less than 2.4\% and a median waveform cosine similarity higher than 0.92.

Different from these works, \textbf{CardiacWave} \cite{cardiacwave} recovers vital signs by analyzing the electromagnetic (EM) field changes rather than the chest movements brought by the heart. CardiacWave presents a novel and interesting observation: the electromagnetic field induced by cardiac electrical activity will modulate the chest-scattered mmWave signals, called Cardiac-mmWave scattering effect (CaSE). Specifically, the frequency spectrums of the intermediate frequency (IF) signal are different when the heat activity changes. CardiacWave employs a mask filter with learnable coefficients to extract the CaSE feature from the IF signal. Then a DNN-based profiler is developed to obtain the ECG-like signals and perceive the cardiac events. Results show that the recovered waveforms have a high positive correlation with the ground truth and contain high-fidelity heart clinical characteristics.

Apart from the aforementioned applications, mmWave sensing also has great potential in blood pressure (BP) monitoring. To measure BP values, mmWave reflections are processed to reconstruct fine-grained pulse waves which contain unique BP-related features, e.g., minimum value, peak value, and first inflection. However, due to high frequency and short wavelength, mmWave signals in the time domain are susceptible to noise, degrading the performance of pulse waveform reconstruction and BP measurement. Moreover, the tiny body motion is another concern for mmWave-based BP monitoring, which would cause severe non-linear signal distortions and lead to a huge performance drop in BP monitoring.

To address the above challenges, {\textbf{mmBP}} \cite{shi22sensys} develops a contact-free BP monitoring system to achieve high accuracy and tiny-motion robustness. To reduce noise, mmBP first converts the received mmWave signals from time domain to the delay-Doppler (DD) domain, and then retains the pulse-related information but filters out noise by leveraging on their different properties in the DD domain. To alleviate the impact of tiny motion, mmBP develops a functional link adaptive filter (FLAF) by exploiting the periodic property and correlation character of the pulse signals. mmBP is implemented with an off-the-shelf mmWave radar (TI IWR1843BOOST), and results demonstrate that mmBP achieves the mean errors of 0.87 mmHg and 1.55 mmHg for systolic blood pressure (SBP) and diastolic blood pressure (DBP), respectively, and the standard deviation errors of 5.01 mmHg and 5.27 mmHg for SBP and DBP, respectively. Results meet the acceptable error range specified in the Association for the Advancement of Medical Instruments (AAMI) standard and rank Grade A in the Britain Hypertension Society (BHS) standard.

\subsubsection{Sound recognition}
Sound recognition is another important application field of mmWave sensing. Conventional solutions based on microphone(s) often fail in noisy environments or outside soundproof scenes. Thanks to the penetrability and high directivity of mmWave signals, people can directionally capture the voice-related reflected signals by mmWave sensing to achieve sound recognition or eavesdropping.

Different from gait recognition or vital sensing, sound recognition mainly extracts the time-frequency spectrum from the sound-related signal to reconstruct the sound.
This means that sound recognition is more sensitive to the complex noise in the frequency domain, which may come from the surrounding acoustic field, electromagnetic interference or even imperfect hardware. Moreover, as the sensing target is the tiny near-throat region,  how accurately locate this vocal part becomes a key problem.

For denoising in the frequency domain, learning-based denoising methods are often applied as the complex noise overlaps with the speech frequency band. These methods can also help recover the severely distorted high-frequency part. To accurately locate the speaker's throat, the signal variance brought by throat vibration are often considered. Some works also solve this problem with the help of microphones or the surrounding environment.

\begin{figure}[!tb]
\centering
\includegraphics[width=0.85\linewidth]{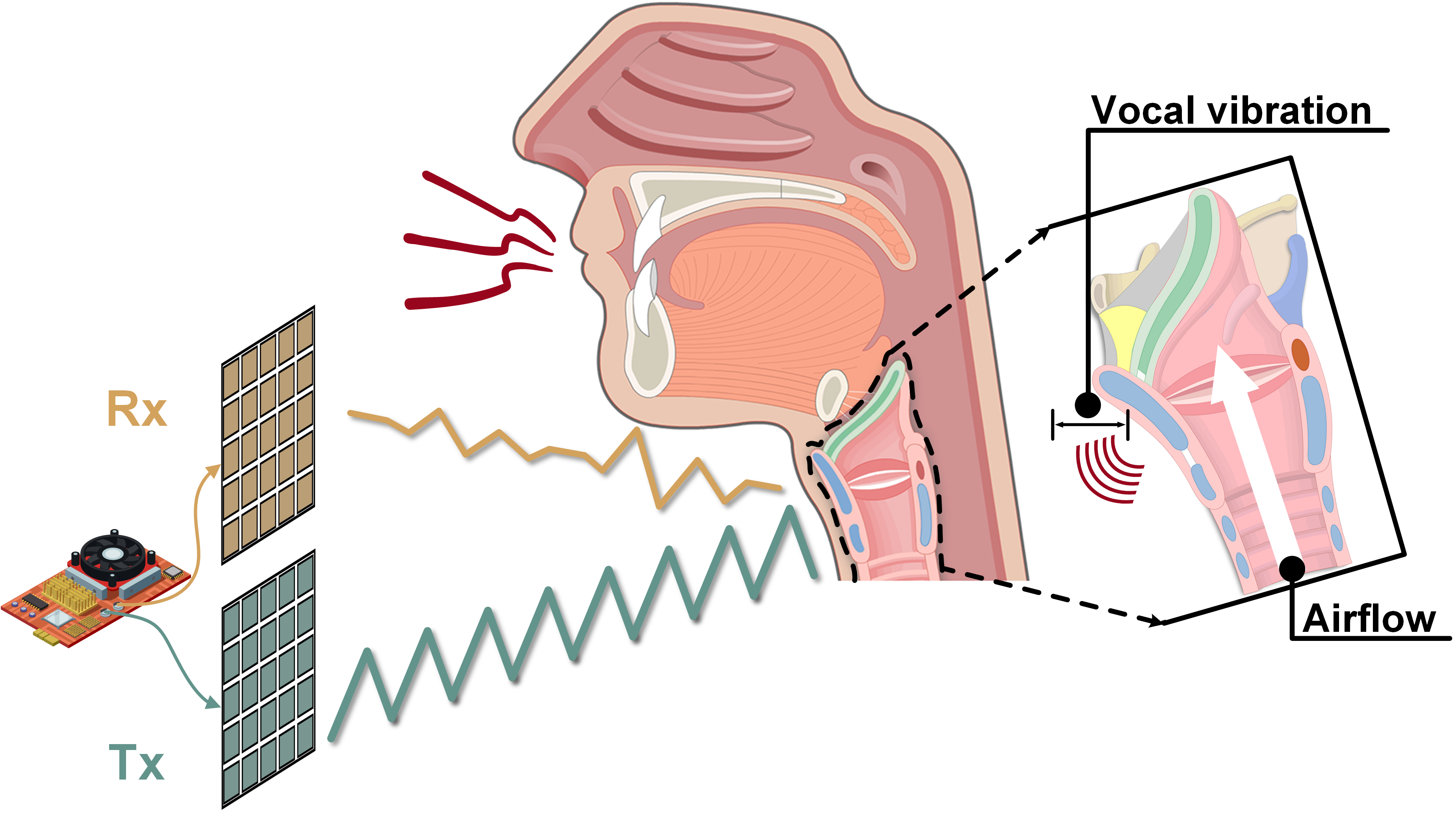}
\caption{The reflected mmWave signal indicates speech information.}
\label{fig:waveear}
\end{figure}

\textbf{WaveEar} \cite{waveear} first explores how to utilize the mmWave signal to achieve noise-resistant speech sensing. Its observation is that the vibration of the speaker's vocal cord contains speech information and can be measured on the throat's surface. As shown in Fig. ~\ref{fig:waveear}, the skin-reflected mmWave signal has a strong correlation with human speech. WaveEar designs a customized 24GHz mmWave probe, where both Tx and Rx consist of 16 antennas following the $4 \times 4$ layout. With the help of such a powerful probe, WaveEar can easily scan in all directions and extract the time domain features to detect the throat's location. After that, the mmWave reflected signal is converted to a series of spectrograms and fed into a novel DNN to obtain the corresponding sound spectrograms. Finally, the Griffin-Lim phase reconstruction approach \todo{\cite{GL}} is employed to recover the sound. Results show that WaveEar can achieve a mean \todo{Mel-Cepstral distortion (MCD) \cite{MCD}} lower than 1.5 and a \todo{word error rate (WER)} lower than 6\%.

\begin{figure*}[t]
\centering
\includegraphics[width=\linewidth]{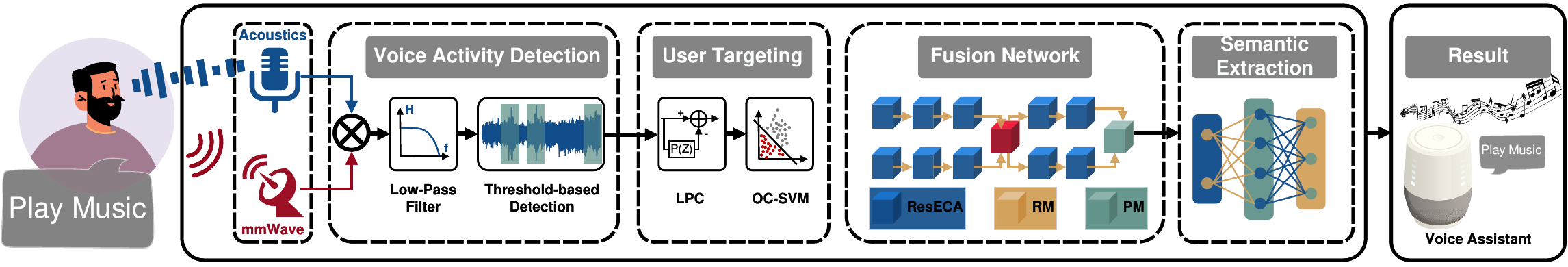}
\caption{The multi-model speech recognition system.}
\label{fig:wavoice}
\end{figure*}

With the directivity and small range resolution of mmWave sensing, Many noises can be eliminated. \textbf{VocalPrint} \cite{vocalprint} enables a secure voice authentication system by exploiting the unique characteristics of the vibrating near-throat region. As the vocal cords and vocal tract vary from person to person, the voice pronunciation process can be regarded as a unique feature, just like the voice itself. VocalPrint first applies a resilience-aware clutter removal scheme on the \todo{Range-Doppler matrix (RDM)} calculated from the reflected mmWave signals. In this way, the complex clutters can be moved from RDM and the vocal biometric properties can be preserved. Furthermore, VocalPrint extracts a series of text-independent features (e.g.,\todo{ residual phase Cepstrum coefficients (RPCC) \cite{RPCC} and Mel frequency cepstral coefficient (MFCC) \cite{MFCC}}) and feeds them into a classifier for user authentication. VocalPrint is implemented based on a commercial mmWave radar, TI AWR1642BOOST \todo{\cite{AWR1642}}. The experimental results show that it achieves over 96\% authentication accuracy under different conditions.

The work on \cite{vcc+lm} also proposes a speaker verification system based on mmWave sensing. It utilizes the mmWave radar to capture both vocal cord vibration (VCV) and lip motion (LM) as multimodel biometrics for user verification. MFCC features and fuzzy WPT techniques are separately applied to extract VCV features and LM features. These features are further fed into a CNN for user verification. 

Sound recognition based on mmWave sensing can be utilized not only for authentication but also for illegal eavesdropping. \textbf{mmPhone} \cite{mmphone} utilizes piezoelectric films and a mmWave radar to eavesdrop on the loudspeaker speech protected by soundproof environments. The properties of piezoelectric films can change with sound pressure and can be captured by the mmWave radar. In this case, the films can be converted to a passive ``microphone'' as long as they are in the sound field. To improve the \todo{signal-to-noise ratio (SNR)} of the received signal, mmPhone employs a DNN-based denoising method to handle non-stationary noise. Finally, a training-free harmonic extension scheme is utilized to improve speech intelligibility. Different from mmPhone, \textbf{Wavedropper} \cite{wavesdropper} eavesdrops on speech contents by directly sensing the speaker's throat vibration in the obstructed condition. It applies wavelet-based analysis to extract the voice-related signal from the hybrid signal and an RNN-based classifier for speech recognition. The experimental results show that Wavedropper can achieve 91.3\% accuracy for 57-word recognition.

As the reflected mmWave signal is seriously affected by various noises, such as environmental acoustic noise and electromagnetic noise, the SNR of the received signal constrains the recovered sound quality. To resist the impact of complex noises, \textbf{RadioMic} \cite{radiomic} first utilizes a line fitting algorithm to extract the phase change caused by the sound. Then RadioMic employs a selection combining scheme on these signals from multiple antennas and multipath components. Finally, a sound enhancement scheme via deep learning is proposed for neural bandwidth expansion and denoising. \textbf{MILLIEAR} \cite{milliear} directly
employs a \todo{conditional generative adversarial network (cGAN) \cite{cgan}} to enhance the audio components and reduce noise. It achieves a mean MCD of less than 4, which implies that the reconstructed audio is highly similar to the original speech.

Furthermore, mmWave-based sound recognition is constrained by another thorny problem: Due to the dynamics of the speaker's position and posture, the speaker's throat is hard to detect and locate in real-world scenarios, making the voice-related signal can not be accurately extracted and analyzed. \textbf{Wavoice} \cite{wavoice} explores the inherent correlation between the reflected signal from a mmWave radar and the audio signals collected from a microphone to achieve noise-resistant speech recognition. The collected audio signals, although noisy, can guide mmWave radars to detect the vibrating throat and compensate for information loss in mmWave signals.

The architecture of Wavoice is shown in Fig. ~\ref{fig:wavoice}. Wavoice first detects the voice activity by comparing the audio signals and mmWave signals in each range bin. It multiplies these two signals and inputs the result into a low-pass filter. If these two signals have the same or similar frequency components, there will be an energy peak in the output of the low-pass filter. Then Wavoice proposes two learning-based modules for multi-modal signals fusion. One module exchanges valid features while the other module projects respective information into a joint feature space. Finally, the semantic information can be extracted from the joint feature space with a typical speech-to-text translation system. Results show that Wavoice outperforms existing methods with the \todo{character recognition error rate (CER)} below 1\% in a range of 7 m.

\begin{table*}[h]
\renewcommand{\arraystretch}{1}
  \centering
  %\captionof{table}{THE SIMULATION PARAMETERS.}
\caption{\review{Comparison of Human Imaging Works}}

\begin{tabular}{lcccc} 
\toprule

\textbf{Method} & \textbf{Signal Form} & \textbf{Algorithm} & \textbf{Hardware Platform} & \textbf{Performance} \\ \hline
\textbf{mmFace} \cite{mmface} & \begin{tabular}[c]{@{}c@{}}Amplitude,\\ Phase\end{tabular} & \begin{tabular}[c]{@{}c@{}}Range Migration Algorithm,\\ Similarity-based Matching, Ray-tracing\end{tabular} & TI IWR1642BOOST & \begin{tabular}[c]{@{}c@{}}96\% ASR\\ lower than 5\% EER\end{tabular} \\ \hline
\textbf{SquiggleMilli} \cite{squigglemilli} & \begin{tabular}[c]{@{}c@{}}Phase,\\ Waveform\end{tabular} & \begin{tabular}[c]{@{}c@{}}Non-linear Motion Compensation,\\ Compressed Sensing,\\ Unconstrained Basis Pursuit De-noising,\\ cGAN\end{tabular} & TI IWR1443BOOST & \begin{tabular}[c]{@{}c@{}}0.85-0.95 Similarity\\ 90\% Classification Accuracy\end{tabular} \\ \hline
\textbf{milliCam} \cite{millicam} & \begin{tabular}[c]{@{}c@{}}Phase,\\ Waveform\end{tabular} & \begin{tabular}[c]{@{}c@{}}SAR,\\ Motion Compensation Algorithm\end{tabular} & Customized 60GHz Testbed & \begin{tabular}[c]{@{}c@{}}20\% Error for $5 \times 5$ cm$^{2}$ \\ 4\% Error for $20 \times 20$ cm$^{2}$ \end{tabular} \\ \hline
\textbf{MILLIPOINT} \cite{millipoint} & Point Cloud & \begin{tabular}[c]{@{}c@{}}Dynamic Programming,\\ Automatic Multi-focusing Mechanism,\\ Capon Algorithm\end{tabular} & TI MMWCAS-DSP-EVM & \begin{tabular}[c]{@{}c@{}}1.2\% Cumulative\\ Tracking Error\end{tabular} \\ \hline
\textbf{mmMesh} \cite{mmmesh} & Point Cloud & \begin{tabular}[c]{@{}c@{}}Attention Mechanism,\\ Skinned Multi-Person Linear Model\end{tabular} & TI AWR1843BOOST & 2.47 cm Error \\ \hline
\textbf{mmFER} \cite{zhang23mobicom} & \begin{tabular}[c]{@{}c@{}}Phase,\\ Waveform\end{tabular} & \begin{tabular}[c]{@{}c@{}}Gaussian Mixture Model,\\ Cross-domain Transfer,\\ Autoencoder\end{tabular} & TI IWR1843BOOST & 80.57\% Accuracy \\ \hline
\textbf{RPM} \cite{rpm} & Point Cloud & \begin{tabular}[c]{@{}c@{}}Attention Module, Deformable\\ Multi-stage Convolution Module\end{tabular} & Two FMCW Radars & 5.71 cm Error \\ \hline
\textbf{Hawkeye} \cite{hawkeye} & \begin{tabular}[c]{@{}c@{}}Phase,\\ Waveform\end{tabular} & \begin{tabular}[c]{@{}c@{}}cGAN, Ray-Tracing,\\ SAR\end{tabular} & Customized 60GHz Testbed & \begin{tabular}[c]{@{}c@{}}30 cm Ranging Error\\ $27^{o}$ Orientation Error\end{tabular} \\ \hline
\textbf{mmEye} \cite{mmeye} & Point Cloud & \begin{tabular}[c]{@{}c@{}}MUSIC,\\ Joint Transmitter Smoothing,\\ Background and Noise Cancellation\end{tabular} & \makecell[c]{Qualcomm 60GHz\\ 802.11ad Chipset} & 7.6 cm Keypoint Precision \\ \hline
\textbf{m3Track} \cite{m3track} & \begin{tabular}[c]{@{}c@{}}Doppler\\ Spectrum\end{tabular} & \begin{tabular}[c]{@{}c@{}}MVDR,\\ Two-steam Deep Learning Architecture,\\ K-means Algorithm\end{tabular} & TI AWR1443BOOST & 42.4 mm Tracking Error \\ 

\bottomrule
\end{tabular}
\label{table:human_imaging}
\end{table*}

However, the voice detection problem still exists in \todo{non-line-of-sight (NLoS)} scenarios, where the \todo{line-of-sight (LoS)} path between the throat and the radar is hard to find or even does not exist. \textbf{AmbiEar} \cite{ambiear} pinpoints that although the near-throat region may not be sensed, surrounding objects around the speaker can provide voice-related information. Its insight is that sound propagates as a mechanical wave. The speaker's voice causes similar vibrations of the surrounding objects, which are highly correlated with the voice. To detect the surrounding objects around the speaker, AmbiEar proposes a variance-based human trajectory detection algorithm to detect the human's trajectory and locate surrounding objects. Then a modified \todo{minimum variance distortionless response (MVDR) \cite{mvdr}} algorithm is applied to these signals from surrounding objects to extract their common components and enhance each signal. Finally, these signals are superimposed and fed into an end-to-end network for voice recognition. AmbiEar provides an effective voice recognition approach in NLoS scenarios with a word recognition accuracy of 87.21\%.

\subsection{Human Imaging}

mmWave imaging system emits artificially generated signals to illuminate the target object, and 2D/3D images can be reconstructed by the reflected signals from the target according to the reflectance distribution. Different from the existing vision, IR, and X-ray imaging systems, mmWave signals can penetrate clothes and work in low visibility conditions. Due to its higher privacy and millimeter-scale ranging resolution, mmWave-based imaging has been widely pursued for pose/posture tracking\cite{mmmesh}\cite{squigglemilli}, automatic driving\cite{millipoint}, security concerns\cite{mmface}\cite{millicam}\cite{3dried}, etc. \review{We summarize the related works in TABLE \ref{table:human_imaging}.}

% \todo{table of method comparison}

The biggest challenge of mmWave imaging system is the data sparsity. The small size of the antenna array of mmWave radar device and its specularity lead to a  poor spatial resolution. In general, the imaging resolution of a radar system is defined by $resolution \propto wavelength \times distance/aperture$. For an antenna array size of 1.8 cm $\times$ 1.8 cm, its imaging resolution can only reach up to 28 cm at 1 m distance. To enhance resolution, a universal technique is the \todo{Synthetic Aperture Radar (SAR)\cite{sar}}, in which the mmWave device needs to move a pre-determined trajectory and collect reflections from uniformly and densely grid locations. Recent works further propose to use the deep learning algorithm to generate images from mmWave signals.

\begin{figure}[t]
\centering
    \includegraphics[width=0.8\linewidth]{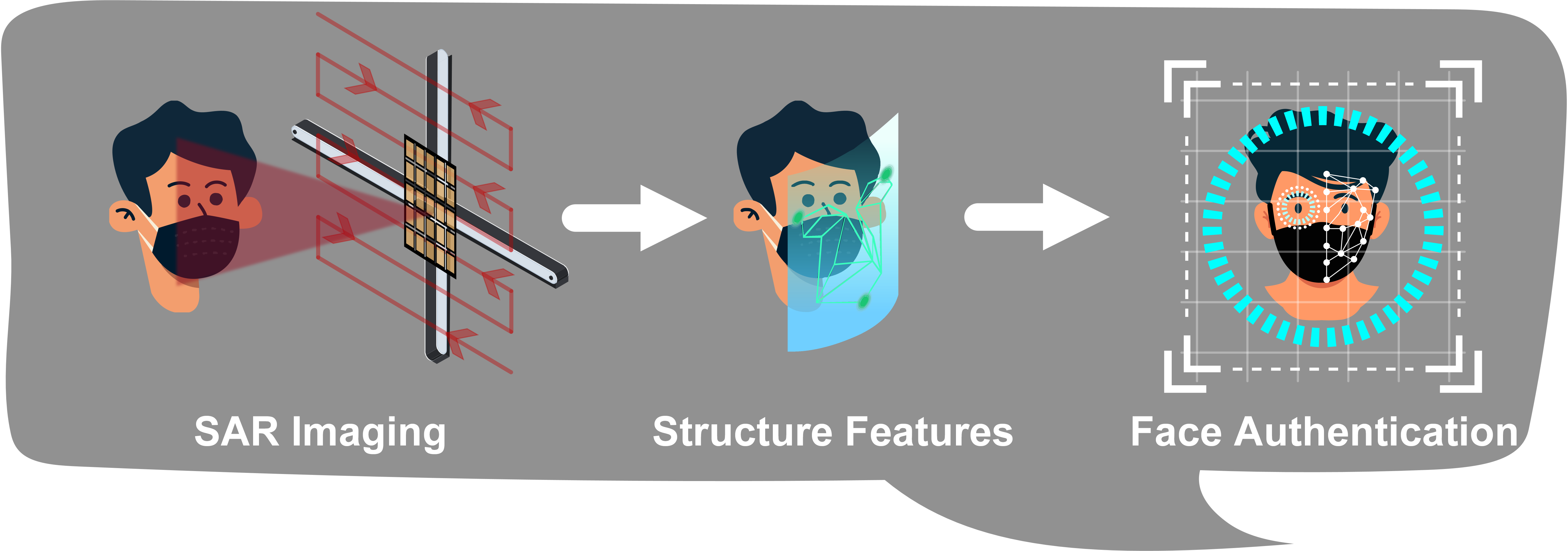}
    \caption{The flowchart of mmFace.}
    \label{fig:mmface}
\end{figure}

\textbf{mmFace}\cite{mmface} proposes a anti-spoofing face authentication (FA) system using a COTS mmWave radar (TI IWR1642-BOOST~\cite{IWR1642}). As Fig.~\ref{fig:mmface} illustrates, mmFace emulates a large aperture planar antenna array by moving a radar along a specific 2D slide rail, and collects the reflected signals from the human face. Firstly, as the reflection coefficient only depends on the material of the facial surface \cite{msense}, mmFace designs a novel algorithm to extract biometric features using the amplitudes of the reflected signals for liveness detection. With the reflected signals, a Range Migration algorithm (RMA) \cite{rma} is applied to construct a 3D facial image. However, it is distance-dependent and requires the authentication distance and registration distance to be identical, which is impractical. To further obtain distance-resistant features, mmFace utilizes the contour of the bright area on the facial image, which corresponds to the user's facial region and its surface curvature is relatively stable. This feature is utilized to calculate Fourier descriptors as the facial structure features for similarity-based user matching. Furthermore, to mitigate the registration overhead, mmFace also proposes a VRS (Virtual Registration Signals) generation method rather than directly collecting mmWave signals. Given three 2D facial photos taken from different perspectives, mmFace can reconstruct a virtual scene and generate the reflected mmWave signals by building a theoretical model of mmWave signal propagation. With the generated VRSes, the user's corresponding facial structure features are stored as templates. The experimental results that mmFace can achieve an average \todo{authentication success rate (ASR)} of 96\% and an average \todo{equal error rate (EER)} lower than 5\%. 

Unfortunately, the pre-determined trajectory and uniform sampling make it extremely challenging to emulate SAR on a hand-held device. To this end, \textbf{SquiggleMilli}\cite{squigglemilli} uses a non-linear motion compensation and a compressed sensing-based framework to generate uniformly and densely spaced grid locations, as the traditional SAR imaging system does. As a user freely moves the device in the air, SquiggleMilli maps the measured sample to the nearest point on the uniform grid according to its phase change.  To avoid shape aliasing from the missing grid samples, SquiggleMilli estimates the missing by combining several measurements around its corresponding location. The recovery procedure then can be represented as an L1-norm minimization problem and solved by the unconstrained basis pursuit de-noising method \cite{yang2011alternating}. Then it applies voxel segmentation to extract the 3D shape of objects at different depths. To enhance the perceptibility of the reconstructed 3D shape, SquiggleMilli  leverages a pre-trained cGAN\cite{cgan} model to recover the high spatial frequencies in the object to generate an accurate 2D shape with the partial 3D mmWave shape. It further designs two networks to predict its 3D features and category. The experimental results show that SquiggleMilli can reconstruct 2D shapes with a similarity score ranging from 0.85 to 0.95. For the 3D feature, it can predict the mean depth and rotation angle with less than 1\% error and 1.5$^{\circ}$ error for the 90-th percentile. Meanwhile, it achieves more than 90\% accuracy for classifying the objects. 

Different from SquiggleMilli, \textbf{milliCam}~\cite{millicam} proposes a mmWave-based scanning system for capturing a high-resolution shape of a small metallic object by emulating the SAR imaging principle and swiping the hand-held mmWave device over the air. As SAR requires stringent linear motion and mm-scale localization accuracy, milliCam first leverages the co-located camera to compute the position and tractory of the device by measuring the translation and rotation. The camera has sub-mm pixel resolution and can achieve a 7mm localization error. Considering the target-scene is very close-by, such localization errors still cause significant errors. milliCam observes that most of the target-scene in the mmWave signals are sparse, due to the specular reflection. It re-designs a novel motion compensation algorithm based on the airborne SAR compensation by fusing a clustering algorithm and a bicubic interpolation. Furthermore, to resist the varied reflectivity and noise reflections, milliCam iteratively segments and corrects the defocused image using Otsu's method and squint correction, separately. After that, in order to ensure the phase coherency across the emulated apertures, milliCam calibrates the incoherent phases relying on the line-of-sight signals from Tx to Rx. Results show that the measurement errors are less than 20\% and 4\% for $5 \times 5$ cm$^{2}$ and $20 \times 20$ cm$^{2}$, respectively.

\begin{figure*}
\centering
    \includegraphics[width=0.8\linewidth]{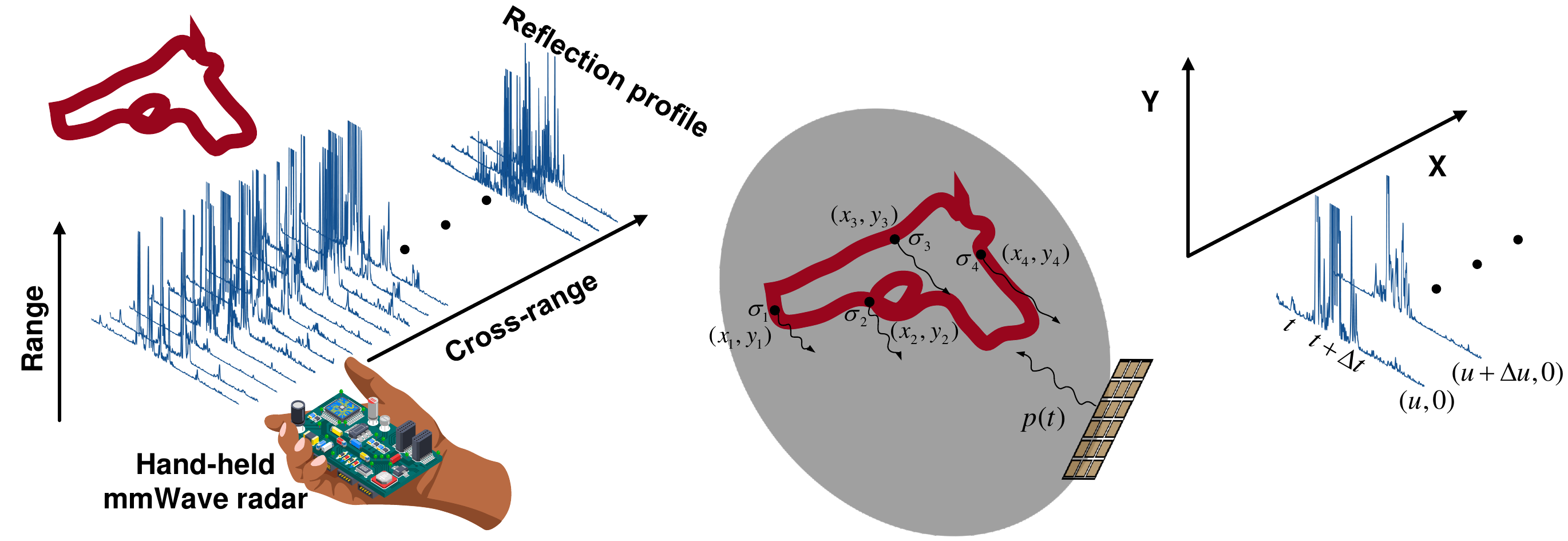}
    \caption{The flowchart of milliCam.}
    \label{fig:millicam}
\end{figure*}

However, autonomous driving cars move fast and their trajectories are uncertain, making it very challenging for precise localization at the scale of the signal wavelength and uniform sampling, which are the requirements of emulating the SAR technique. To this end, \textbf{MILLIPOINT} \cite{millipoint} enables SAR imaging on low-cost commodity radar to generate 3D point clouds. First, MILLIPOINT observes that different Tx/Rx pairs may experience similar channel responses with a lagging effect, and the delay depends on the spacing between antenna pairs. It characterizes the radar's cross-movement by correlating the received signals. The correlation reaches the maximum at the range of half a spacing of two antenna pairs. In addition, MILLIPOINT uses multiple antenna pairs with different spacing to overcome the influence of side peaks and achieves robust self-tracking by applying dynamic programming. 

Furthermore, due to the existence of specular reflection, only reflections around the normal direction of the target can be received by the radar, resulting in a very small effective aperture. To synthesize a full image of the 3D environment, MILLIPOINT proposes an automatic multi-focusing mechanism to focus on each target separately and then synthesize them. To achieve it, MILLIPOINT first compensates for the antenna gain and the attenuation in the cross-range spectrums. It further identifies targets as the peaks of frequency spectrums along the range direction by an empirical threshold. The corresponding peak widths can be regarded as the target's effective aperture lengths. Then the classical SAR imaging algorithm is applied and synthesizes a 2D image by selecting pixels with the maximum value across all images. To extend it with height information, MILLIPOINT models the relationship between height and phase response of pixel, according to the idea of AOA estimation with antenna array \cite{aoaest}. The height is estimated with the Capon algorithm \cite{capon}. Through evaluation of the commercial mmWave Radar with 6 Tx and 8 Rx antennas, results show that MILLIPOINT achieves the median cumulative tracking error of 1.2\% and has a precision rating of sub-mm.

As the SAR-based imaging methods require the target to be stationary or relatively stationary, they are not suitable for tracking the target's posture that dynamically changes. However, the spatial relationships and temporal dependencies of the joint skeletons are concealed in the sequence of mmWave reflected signals. This information can be utilized for pose tracking. \textbf{MilliPose}~ \cite{millipose} proposes a machine-learning model for pose tracking and solves the problems of poor resolution, specular reflection, and varied reflectivity. Firstly, MilliPose leverages cGAN to generate a high-resolution 2D full-body silhouette image from the low-resolution 3D mmWave signals. Meanwhile, Considering that the \todo{degrees of freedom (DoF)} and the ranges of body joints are very limited, MilliPose designs an RNN, consisting of a two-layered GRU network~\cite{gru} and a Structured Prediction Layer (SPL) \cite{spl}, to learn these rules and predict skeleton joint of the next pose. Then it feedbacks the predicted pose to cGAN and can generate a high-quality body shape. Results show that the median error of joint location prediction is only 2.1 cm.

Instead of applying SAR imaging, \textbf{mmFER}\cite{zhang23mobicom} develops a mmWave radar-based system to recognize facial expressions by extracting and exploring subtle facial muscle movements from raw mmWave signals. To this end, mmFER first leverages the MIMO technique to localize the facial areas of the target and eliminate ambient noise. Then it utilizes Gaussian Mixture Model (GMM) to mitigate the impact of body motions on facial localization. mmFER further proposes a cross-domain transfer pipeline to guarantee an effective and reliable model knowledge transformation from image to mmWave. In particular, a hybrid learning loss function is designed to effectively address the over-fitting issue caused by small-scale training datasets. Moreover, an autoencoder is explored to learn the transition of latent features to reshape mmWave data, thereby eliminating the influence of data heterogeneity. mmFER is fully implemented using a commercial mmWave radar (TI IWR1843BOOST), and results show that mmFER obtains an accuracy of at least $80.57\%$ when the distance between the subject and radar is within the range of 0.3-2.5 m.

Different from generating 2D silhouette images, 
%Due to the limited number of antennas, the low spatial resolution and sparsity of point cloud make the commercial mmWave radar difficult to accurately estimate a complex 3D human mesh. To this end, 
\textbf{mmMesh}\cite{mmmesh} proposes a deep learning framework to reconstruct the dynamic 3D human mesh from the mmWave signals. To alleviate the influence of ambient noise and multipath effects, mmMesh uses an attention mechanism to discriminate the point cloud and the points reflected by the subject tend to have higher qualities. Then a rough estimation of the shape and pose can be generated. To correctly align the points and make the human mesh more accurate, mmMesh dynamically chooses some "virtual locations" near the subject as anchor points that are each related to a part of the human body. According to the anchor locations, mmMesh dynamically groups the 3D point cloud into several subsets, and each one corresponds to a different body segment. Meanwhile, the local structure and associations can be learned. Finally, mmMesh incorporates the Skinned Multi-Person Linear (SMPL)\cite{smpl} model as an additional constraint. It allows us to represent the whole 3D human mesh with 86 parameters and does not need to calculate the positions of thousands of vertices. The experimental results show that mmMesh can achieve an average vertex error of 2.47 cm and a mesh localization error of 1.27 cm. Besides, the average joint localization and rotation errors are 2.18 cm and 3.8 degrees, respectively. 

Besides, \textbf{RPM}~\cite{rpm} designs a multi-dimensional feature fusion backbone network for human pose estimation. It consists of two blocks: one is a channel fusion block that can effectively fuse horizontal and vertical RF features based on the channels' correlation. The other one is a deformable multi-stage convolution module that maintains scale-insensitive feature representations. Due to the high directivity of mmWave radar, it is possible to receive the reflected signal of a subset of limbs at each RF snapshot. Therefore, a spatial attention module is applied to recover the remaining body parts from a single RF snapshot by modeling non-local skeleton relationships. A temporal attention module is further applied to model the temporal dependencies across the 3D skeleton sequences. Finally, the complete 3D skeleton can be obtained. The experimental results reveal that RPM can achieve a Mean Per Joint Position Error (MPJPE) of 5.71 cm.

\begin{figure*}[t]
\centering
    \includegraphics[width=\linewidth]{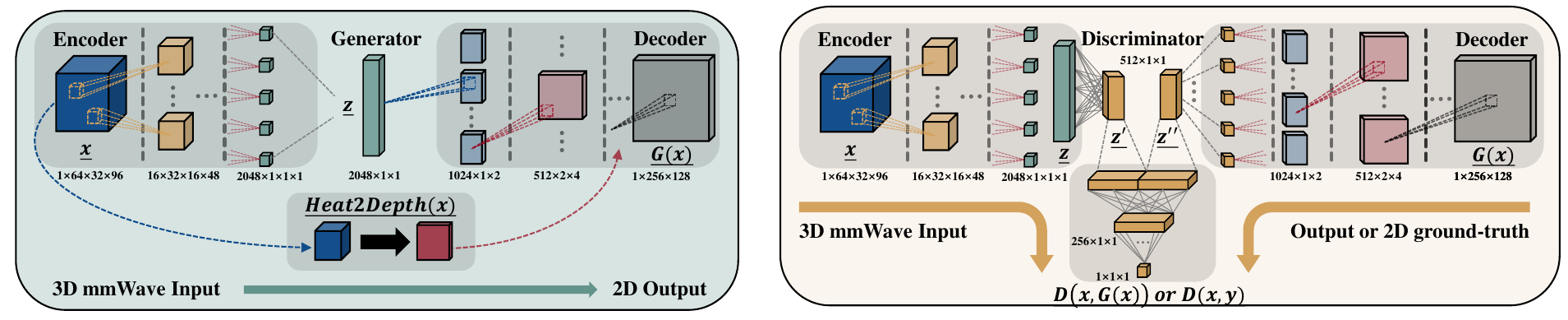}
    \caption{The architecture of HawkEye.}
    \label{fig:hawkeye}
\end{figure*}

Although deep learning-based methods are proven to be effective, the procedure of collecting real-world signals for training is very time-consuming and labor-intensive. Benefiting from the success of the GAN technique and its robust resistance to noise, \textbf{HawkEye} \cite{hawkeye} uses the GAN technique to generate the mmWave signals for training. In this work, a cGAN-based architecture is designed to predict high-resolution depth maps with a low-resolution mmWave heatmap input. HawkEye utilizes an encode-decode framework as the generator network to generate a 2D depth image. Meanwhile, the skip connection mechanism \cite{skipcon} is applied to retain the high-frequency details in depth. The discriminator employs a two-stream architecture to classify real vs. generated samples. Furthermore, to avoid collecting real-world mmWave data, HawkEye proposes a data synthesizer to simulate the propagation of the mmWave signals in the real world using ray-tracing \cite{raytracing} and generate realistic mmWave heatmaps.

Different from the use of radar devices, \textbf{mmEye}\cite{mmeye} is the pioneering work of attempting to achieve mmWave imaging on a single commodity 60GHz WiFi device. To break through the resolution limited by the physical aperture, mmEye proposes to reconstruct the spatial spectrum by applying the \todo{Multiple Signal Classification (MUSIC)\cite{music} with Akaike information criterion (AIC)\cite{aic}}, a well-known information-theoretic approach for model order selection. However, The rank of MUSIC is likely much less than the number of incoming signals, making its performance greatly deteriorate or even fail. To overcome the rank deficiency problem without loss in precision, mmEye designs a novel 2D spatial smoothing termed \textit{Joint transmitter smoothing (JIT)}. This technique reuses all the Rxs for each Tx rather than dividing the receive array. mmEye further proposes a \todo{background and noise cancellation (BANC)} algorithm to remove interfering signals and a \todo{minimum mean square error (MMSE)\cite{mmse}} estimator is applied to compute the background CIR. Additionally, mmEye searches the point of interest (PoI) based on the variance of the energy distribution of the spatial spectrum. Finally, mmEye transforms the PoIs into plain images by solving a weighted least absolute deviation problem. The experimental results show that it achieves a median silhouette (shape) difference of 27.2\% and a median boundary keypoint precision of 7.6 cm.

Although commodity RF signals have been utilized to realize 3D human posture tracking, these studies cannot handle cases where multiple users are in the same space. Towards this end, \textbf{$m^3$Track}\cite{m3track} proposes a 3D posture tracking system. Based on the varying  Doppler responses along with the speed of movement, $m^3$Track performs a convolution operation on the Range-Doppler-Profile to detect all users with their corresponding ranges. In the meanwhile, it utilizes \todo{minimum variance distortionless response (MVDR)\cite{mvdr2}} to separate all users from different angles. Given the range and the angle, $m^3$Track extracts the mmWave profiles of each user based on a 3-cylinder model, relating to the head-neck, chest-arm and lap-leg, respectively. Each profile contains two Range-Angle-Profiles (spatial features) and one Range-Doppler-Profile (temporal features). Then, $m^3$Track designs a two-steam deep learning architecture. It takes the spatial features and temporal features as inputs to extract the global body shape features and local body motion features. By concatenating these features, a regression model is applied to predict the 3D coordinates of skeleton joints and reconstruct the 3D body posture. For tracking posture in multi-user scenarios, it is necessary to map the reconstructed posture into real-world 3D space accurately. $m^3$Track leverages the K-means algorithm \cite{kmeans} to separate the point clouds into multiple clusters. Each cluster center corresponds to a posture joint, and the optimal mapping relation can be obtained by searching the minimal mapping error. Considering that the above computed 3D coordinate is in the polar coordinate system, $m^3$Track further designs a coordinate-correlated extended Kalman filter to calculate the user's position in the Cartesian coordinate system. With the position trajectories, $m^3$Track realizes continuous 3D posture tracking in real-world multi-user scenarios. With TI AWR1443~\cite{AWR1443}, the experimental results show that the overall joint tracking error is 42.4 mm in a 4-user scenario.

\review{
\subsection{Lessons Learned}
%目前的毫米波人体感知已经涵盖了多种多样的任务以及对应的多种解决方法。当人们希望利用毫米波感知解决某一个感知任务时，人们应该首先选择合适的硬件平台和处理信号形式。例如，点云数据更适合用于活动识别或者人体成像等轮廓相关的感知任务，而相位波形更适合用于手写追踪或者生物特征测量等细粒度的量化感知任务。在此基础上，人们被鼓励去发掘对应感知任务的核心挑战，这决定了所要采用算法的技术目标。例如在活动识别、姿势识别、声音识别等任务中，考虑到不同个体带来的数据差异（时长、强度、变化趋势等），learning-based classification手段往往表现出很好的识别效果；而在手写追踪、生理特征感知等任务中，由于需要对感知目标进行细粒度的量化刻画，信号中的噪声会对感知结果产生显著的影响。因此，合适的降噪手段需要被设计。此外，感知场景应该被更仔细的观察和考虑，以获得额外的信息来帮助感知。例如，人体与环境反射物的多次反射、人体对周围环境的影响等，这些间接信息都蕴含着与人体相关的信息，能够实现更通用或者更准确的感知。
The current mmWave-based human sensing has covered a variety of sensing tasks and corresponding solutions. When people try to utilize mmWave sensing to achieve a certain sensing task, they should first choose the appropriate hardware platform and processing signal form. For example, the point cloud is more suitable for contour-related sensing tasks such as activity recognition or human imaging, while the phase waveform is more suitable for fine-grained quantitative sensing tasks such as handwriting tracking or biometric measurement. 

\reviewII{As we mentioned before, different sensing tasks have different technical frameworks and thus different challenges. For example, tracking and localization works are still plagued by trajectory crossing and multipath effect, which leave significant room for improvement in indoor multi-person localization. In terms of motion recognition, continuous motion recognition and dynamic human orientation changes are still open problems. Considering the unique requirements of biometric measurements for localization and sensing of human body parts (chest, throat, etc.), precise body parts localization and enhancement of the weak biometric signals are worthy of research, due to the uncertainty of human body position and posture. In human imaging, since SAR technique and deep learning models are widely used, how to improve the human imaging speed and model generalization is worthy of further study.} 

People then are encouraged to explore the core technical challenges of the corresponding sensing task, which determine the detailed technical targets of the algorithms to be adopted. For example, in recognition tasks such as activity recognition, gesture recognition, and voice recognition, considering the data differences (duration, intensity, variation, etc.) brought by different individuals, the learning-based classification methods often show better recognition abilities. Moreover, in measurement tasks such as handwriting tracking and vital sensing, the complex noise has a significant impact on sensing results, due to the requirement for fine-grained quantitative measurements. Therefore, appropriate noise reduction methods need to be carefully designed. 

Furthermore, the sensing scenes should be carefully observed to assist sensing methods and achieve better sensing results. For example, human orientation and radar deployment need to be carefully considered to reduce potential signal distortion and environmental noise. On the other hand, multiple reflections between human and environmental reflectors and the impact of the human body on the environment may contain additional human-related information, which can help to achieve more general or accurate sensing.

Finally, people are encouraged to deploy their systems in various scenarios to verify their performance in detail. Considering the increasing demand for fine-grained sensing in complex scenarios, such as in smart cockpits and smart homes, multi-scenario experimental verification can not only explore the potential capabilities but also promote the deployment of mmWave-based human sensing in actual scenarios.
}
\review{
\section{Application Scenarios}
\label{sec:application scenarios}

Various human sensing tasks can be applied to many actual scenarios to realize attractive applications. In this section, we briefly introduce some application scenarios of mmWave-based human sensing, including smart home, smart health, smart vehicle and security.

\subsection{Smart Home}
Smart home is very attractive in our daily scenarios as people can directly interact with devices through gestures or voices to achieve various applications. Moreover, smart home can automatically monitor our daily activities to provide customized services and protect our safety \cite{weiguo2022motorbeat, wang2021mavl, weiguo2020symphony, shen2020voice, 9808346, weiguo2020chordmics, aranzazu2017distributed, wang2010, gu2011, gai2022digital}. As mmWave-based human sensing techniques can capture human-related activity information in real-time, they can be well adapted to smart homes.

As the mmWave radar can capture people's commands through activity recognition \cite{EI, radhar, palmar}, gesture recognition \cite{mmasl, mhomeges, pantomime} and sound recognition \cite{waveear, wavoice, ambiear}, people can only give preset actions or voices to directly interact with devices, such as opening curtains and turning off lights. \reviewII{For example, mHomeGes \cite{mhomeges} enables real-time arm gesture recognition in smart home scenarios to provide natural user experiences. Furthermore, mmWave-based gesture recognition can also help Deaf and Hard of Hearing (DHH) people communicate with others. mmASL \cite{mmasl} provides a home assistant system for DHH users that can perform American Sign Language (ASL) recognition with a single mmWave radar.} Considering complex home scenarios, signal denoising techniques need to be adopted and deep learning-based sensing models are highly recommended. 

Smart home is also expected to automatically monitor people's daily activities to provide more intelligent services. The tracking and localization of people can help to identify the location of people to provide more accurate services \cite{mid, mmtrack}. \reviewII{mmTrack \cite{mmtrack} proposes a passive indoor localization system and enables various applications that demand ubiquitous human localization, such as intelligently rotating the TV's viewing angle with the user's location and turning on/off the TV when users are coming/leaving. Furthermore,} the mmWave-based vital sensing \cite{vimo, rf-scg} can monitor the sleep state of people, so as to choose the appropriate light intensity and temperature. \reviewII{Considering the non-invasive sensing capability of mmWave radar, it is more comfortable and suitable for long-term use, compared with wearable sensor-based sensing.}

Furthermore, mmWave-based human sensing can help smart home achieve intrusion detection. by locating the intruder's location \cite{palmar, mid} and identifying its gait \cite{gaitcube, muid} or voiceprint \cite{vcc+lm, vocalprint}, smart home can judge whether the intruder's identity is legal and take corresponding warning measures. \reviewII{Compared with existing WiFi-based intrusion detection systems \cite{li2020wiborder, wang2019wi}, mmWave sensing is expected to achieve more accurate detection results, thanks to its high spatial resolution and fine-grained sensing capabilities.}

\subsection{Smart Health}
Smart health is another attractive application scenario \cite{6177188, singh2020multi, kebe2020human, wang2019contactless, yu2021, yang2017vital}. Through the continuous non-intrusive sensing of the user's physical state, the mmWave devices can provide the user with daily health monitoring and timely anomaly detection.

When the user suffers from certain chronic diseases and needs real-time monitoring, mmWave-based vital sensing can provide the users with continuous health monitoring and assist doctors by providing sufficient auxiliary data, such as heart rate \cite{vimo, rf-scg}, breathing rate \cite{movifi, mBeats}, blood pressure changes \cite{shi22sensys}, etc. \reviewII{Especially, thanks to the contactless fine-grained sensing capability of mmWave radar, users do not have to remain still to obtain the vital sensing results. Movi-Fi \cite{movifi} enables a motion-robust vital signs monitoring system and can recover vital signs waveform under severe body movements. mmECG \cite{mmecg} utilizes the mmWave radar to monitor the heart movements of drivers in moving vehicles.}

When some special groups (e.g., children, the elderly, patients in convalescence, etc.) cannot guarantee their own safety during activities, the mmWave devices can monitor their activities \cite{sparcs} and timely detect abnormal activities \cite{wifall}, such as falls, bumps, chokes, etc. In this case, body-harmful actions can be detected and dealt with in time. \reviewII{Furthermore, mmWave-based activity recognition can also help people adopt recommended habits and stay healthy. For example, RF-Wash \cite{rfwash} attempts to track hand hygiene by monitoring hand washing activities and is expected to be applied in future healthcare facilities.}

\subsection{Smart Vehicle}
With the rapid development of autonomous driving techniques and the Internet of Vehicles, smart vehicles are moving from a beautiful vision toward reality \cite{weiguo2023micnest, RoS, yimiao2023aim, wang2017, Millimetro}. Smart vehicles can monitor road conditions and provide drivers with comprehensive assisted driving services.

When driving in scenes with limited vision such as heavy fog and rainy days, it is difficult for the driver to judge pedestrians and cyclists beside the car, which leads to potential driving accidents. Thanks to the penetrability of mmWave in these scenarios, the mmWave devices can identify potential pedestrians and cyclists through human imaging \cite{millipoint, hawkeye} or activity recognition\cite{radhar, m-activity} to avoid driving accidents. Moreover, smart vehicles can predict the trajectories of pedestrians by tracking and localizing pedestrians \cite{m3track, palmar}, thereby providing safer assisted driving information. \reviewII{Researchers have emphasized that being able to recognize, detect and track objects from afar is essential in smart vehicles to prevent traffic accidents \cite{venon2022millimeter, deep-learning-survey}. Moreover, the radar manufacturers, such as Texas Instruments, have provided detailed guidance \cite{ramasubramanian2017mmwave} on the application of mmWave radar in automotive applications, which can help to further explore the application potential of mmWave radar in smart vehicles.}

\subsection{Security}
Security has been widely concerned by people as they are closely related to our property and personal information \cite{8854247, 9060970, cross-sensing-survey}. On the one hand, mmWave-based identity authentication can provide people with high security and protect people from troubles brought by attackers. On the other hand, the sniffing of private information by mmWave devices may also bring new challenges to existing security services.

mmWave-based identity authentication is being gradually developed and adopted. Through sensing tasks such as gait recognition \cite{mmgaitnet, gaitcube}, sound sensing \cite{vcc+lm, vocalprint} and human imaging \cite{mmface}, mmWave devices can identify and authenticate the human's physical characteristics, including gait, voiceprint and facial features. As mmWave-based identity authentication relies on changes in mmWave signals, it is hard to attack and can work with other authentication methods to further enhance authentication security. \reviewII{mmFace \cite{mmface} points out that the camera-based face authentication approaches are vulnerable to emerging 3D spoofing attacks. With the material sensitivity of RF signals, the mmWave radar can be utilized to distinguish real human faces from other materials and defend against 3D spoofing attacks. In this way, combining cameras and radars appears to be a more secure authentication method.}

Conversely, mmWave-based human sensing can capture subtle changes in human physiological signs, such as throat vibrations \cite{waveear, wavoice} and gesture variance \cite{mmkey, mmwrite}. This means that mmWave devices can sniff sensitive private information, such as private voice and manually entered passwords, thereby threatening existing security services. \reviewII{For example, mmKey \cite{mmkey} enables a virtual keyboard to sense the position and action of human fingers. This approach, in turn, can be utilized by attackers for illegal eavesdropping. Moreover, AmbiEar \cite{ambiear} validates the feasibility of recovering human voice from surrounding objects. This approach may inspire rethinking the security issues of acoustic systems.}
}
\section{Challenges \& Future Directions}
\label{sec:challenge}
%Although mmWave-based human sensing has been greatly developed, it still faces many challenges due to the limited hardware capabilities and complex deployment environment, resulting in limited sensing capabilities and applications. On the other hand, with the deepening of sensing tasks and the emergence of novel technologies, there is still a lot of research space for mmWave-based human sensing.

According to the discussion in the previous section, we can see that there is still a big gap between the state of the arts and the ideal vision of mmWave-based human sensing, with respect to the capacity, applicability, accuracy, and ubiquity. Filling this gap requires continuous innovations in all the dimensions around this technology, such as hardware, algorithm, sensing mediums, and application designs. Therefore, our discussion in this section will accordingly focus on four main research directions, namely hardware and platforms, enhancing the applicability, novel sensing schemes, and integration with new mediums.

%powerful hardware, complex deployment, enhanced and ubiquitous sensing algorithm. 

\subsection{Hardware and Platforms}
\label{sec:harddev}
\subsubsection{Current limitations}
The current mmWave-based sensing technology still deeply suffers from limited hardware capabilities. The insufficient number of antennas, limited transmission power and hardware denoise capability hinder the further development of mmWave-based sensing. 

Taking mmWave radar based sensing as an example, the COTS mmWave radar only has several antennas (e.g., 3 TX antennas and 4 RX antennas in IWR1443 and IWR6843ISK-ODS), which leads to very limited angle resolution (15$^\circ$). \review{The reason is that these radars can only provide $3*4=12$ virtual antennas by letting 3 Tx antennas take turns transmitting signals and 4 Rx antennas receive signals simultaneously. Then the Angle-FFT operation can obtain an angle resolution of $\frac{180^\circ}{3*4}=15^\circ$ \cite{gwaltz}.} Such angle resolution is insufficient to enable many fine-grained sensing tasks, such as face imaging. Researchers have attempted to perform Rx beamforming \cite{mmvib, gwaltz} and synthetic aperture radar (SAR) technique \cite{mmface, millipoint} to expand the radar's angle resolution. However, they either have limited improvements or need a customized slide rail, causing the current solutions still be unsatisfactory. To further enhance the angle resolution, it's necessary to increase the number of antennas and exploit novel signal enhancement techniques. Texas Instruments has launched a powerful radar, TI MMWCAS-RF-EVM \cite{MMWCAS-RF-EVM}, which has 12 Tx antennas and 16 Rx antennas and has a much better angle resolution. \review{However, its four-radar cascaded array implementation makes its form factor multiply. Moreover, its power consumption (higher than 20 W \cite{MMWCAS-RF-EVM}) is much higher than that of those radars mentioned before (less than 2 W \cite{TI1443, TI6843}). These factors limit the practical deployment of this radar, especially in mobile scenarios where energy consumption needs to be carefully considered. In fact, the impact of its high energy consumption has been discussed in the autonomous driving community.} Moreover, some signal local resolution techniques, such as Chirp-Z transform techniques \cite{rabiner1969chirp}, can be explored.

The sensing range is another important technical indicator. Although the mmWave radar can sense vehicle targets from tens of meters, the range of human sensing is very limited. For example, mmTrack \cite{mmtrack} can only track users within 5 m range. WaveEar \cite{waveear} localizes the position of the speaker and performs speech recognition within 2 m range. The reason is that the extremely high frequency of the mmWave causes its rapid attenuation, which makes the human-related signal features blurred a few meters away. To overcome the limitation of sensing range, more powerful equipment is needed. They can be radars with higher transmission power or some auxiliary equipment that can enhance the reflected signals, such as retro-reflective tags \cite{RoS, Millimetro, mmTag}.

To obtain more accurate sensing results, careful signal noise reduction is necessary to resist hardware deficiency and environmental interference. However, the existing hardware denoise capability cannot fully meet our needs. For example, mmSpy \cite{mmspy} models the noise from the ramp/settle times in the radar's oscillator. It further employs statistical error correction techniques to subtract it from the phase waveform. These imperfect hardware features may affect our sensing results and must be carefully modeled and handled.

\subsubsection{Potential directions}
Considering that the current limited hardware capabilities seriously constrain the development of sensing techniques, we hope to discuss the development trend of the future mmWave hardware to provide some ideas and reasonable imagination for researchers in industry and academia.

We think the future mmWave hardware will have the characteristics of miniaturization, an increased number of antennas and flexible customization, etc. With the continuous development of the radio-frequency integrated circuit (RFIC) technique, the RF front-end and computing units of the radar have been able to be integrated into a small-size chip. In this way, the mmWave hardware can be easily integrated into mobile phones, smartwatches and other devices to achieve more ubiquitous sensing. In the past few years, Soli \cite{soli} has made this attempt but most radars still have not been integrated. Nevertheless, we believe that miniaturization will be a major trend of mmWave hardware.

Furthermore, since the number of antennas determines the angle resolution and directional sensing capability of the mmWave hardware, more antennas will definitely be welcomed in industry and academia. Texas Instruments has noticed this extensive demand and launched a more powerful radar, TI MMWCAS-RF-EVM \cite{MMWCAS-RF-EVM}, which forms a $12 \times 16$ virtual antenna array. Compared with mmWave radars, mmWave probes usually have more antennas, such as the 32 Tx antennas and 32 Rx antennas of Qualcomm 802.11ad 60GHz WiFi. However, these probes cannot easily form a larger virtual antenna array. In addition, the sensing pipeline of transmitting and receiving in turn also limits their sensing frequency.

Finally, we believe that more flexible customization is attractive and fascinating. The current mmWave hardware platform has achieved a lot of signal collection and processing functions. However, due to the high integration of hardware, it is difficult for developers to modify the hardware to make their attempts, such as antenna layout and polarization direction modification. We believe that more flexible customization can certainly stimulate the community creatively.

\subsection{Enhancing the Applicability}
\subsubsection{Challenges in real deployments}
In addition to limited hardware capabilities, complex deployment environments also bring various critical challenges to mmWave-based human sensing. NLoS scenarios and multipath effects are the most common challenge in these complex environments.

Due to the human's position and posture dynamics and the blockage between the radar and the target, the human-related signal sometimes cannot be obtained directly, especially in NLoS scenarios. In this case, we can obtain the human-related signal with environmental reflection \cite{gwaltz} or realize indirect sensing through the influence of the human on surrounding objects \cite{ambiear}. Some works also try to utilize the penetrability of the mmWave signal to achieve human sensing in NLoS scenarios \cite{mmphone, wavesdropper, milton}. However, their application scenarios are limited due to insufficient penetrability.

Multipath effects also challenge the actual deployment of mmWave-based human sensing, especially in dense environments. Due to the reflection of mmWave signals on the surface of environmental objects, the received signal is the superposition of multiple signals with different time delays and attenuation. Researchers have attempted to eliminate multipath interference by either deploying multiple radars \cite{pointillism} or exploiting the consistency of the dynamic between the target and the ghost images \cite{mhomeges}. Some works begin to utilize rather than eliminate the multipath effect. For example, mmReliable \cite{mmreliable} utilizes multiple beams to achieve more reliable communication links. However, the exploitation of multipath effects in human sensing remains to be explored.

In addition, large-scale ground truth data is indispensable to the actual deployment of mmWave-based human sensing. Although the existing works have provided a lot of datasets, as mentioned in Sec.~\ref{sec:dataset}. Datasets that contain quantitative sensing results are lacking. For example, the actual measurement dataset of human respiration and heartbeat is significant to develop the relevant mmWave-based sensing techniques.

\subsubsection{Promising directions}

To resist the complex deployment environment and provide accurate sensing results, researchers have made many various efforts \cite{ambiear, Millimetro, mhomeges, mmvib, gwaltz}. Based on those works, we discuss the promising directions that appear to be effective in enhancing the applicability of mmWave-based human sensing. Retracing the continuous enrichment of mmWave-based human sensing applications in the past few years, we think that the future mmWave sensing applications will be more quantitative, ubiquitous and fine-grained, etc.

Although mmWave-based human sensing has realized a variety of sensing applications, varying from tracking and localization to vital sign sensing, the quantitative sensing of the human body still needs to be developed \cite{mmvib, gwaltz}. Taking gesture recognition as an example, the existing works can well distinguish the gesture types. However, the quantitative indicators such as the amplitude of each gesture are difficult to analyze. To further promote the development of mmWave sensing techniques, accurate quantization results are needed. Moreover, quantitative sensing is a key step to realizing the practical application of sensing techniques. Without quantitative indicators that meet the error standard, it is difficult for the public to pay for these sensing techniques.

Another development trend of mmWave sensing applications lies in ubiquitous sensing, including wider sensing ranges, more universal sensing scenarios and richer sensing capabilities. As we mentioned before, due to the complex deployment environment, most of the existing works are the realization of single sensing tasks within the limited sensing range of specific scenarios. How to achieve more ubiquitous sensing is still an open problem. In the actual scenes, many factors such as range, occlusion, multipath and so on will lead to the limited application of mmWave sensing. Some works have tried to expand the sensing application scenarios \cite{ambiear, Millimetro, mhomeges}. However, more efforts are needed for this vision.

Finally, with the deepening of mmWave sensing techniques, more and more fine-grained sensing applications have been developed, including vital sensing and sound recognition, etc. However, we believe that the potential of mmWave sensing techniques goes beyond this. With directional and fine-grained sensing capabilities, mmWave sensing techniques are expected to enable more sophisticated sensing applications, such as skin disease detection, blink recognition, etc.

\subsection{Novel Sensing Schemes}

To further enrich the mmWave-based sensing capabilities, researchers have begun to find novel sensing schemes, such as fusion sensing and multitask sensing. Furthermore, researchers have noticed the indirect sensing ability of mmWave sensing and tried to realize enhanced sensing through ``sensing side channels''. 

\subsubsection{Fusion sensing}
After realizing the limitation of existing mmWave-based human sensing techniques, some researchers begin to explore the feasibility of fusion sensing. By combining mmWave signal with other sensing media, such as vision, inertial measurement unit (IMU) and acoustic, fusion sensing is expected to provide more accurate and robust performance.

On the one hand, the sensing efficiency and ability can be increased by fusion sensing. The sensing techniques based on other media can share a part of the sensing task for mmWave sensing. For example, \cite{milton} utilizes a mmWave + camera multimodal sensing system to track the box's position and the relative position of fragile products in the box. The vision-based sensing algorithm is employed to locate the box along the conveyor belt, making the mmWave sensing can focus on computing the product's location.

On the other hand, by mutual correction with the sensing results based on other media, fusion sensing can further improve the sensing accuracy. \cite{SLAM_mmWave_IMU} employs a mmWave radar based simultaneous localization and mapping (SLAM) solution assisted by IMU. The IMU data is utilized to combine continuous radar scanning point clouds into ``multi-scan'' to achieve accurate and robust SLAM results. Wavoice \cite{wavoice} explores the inherent correlation between the audio sensing result from a mmWave radar and a microphone. By combining the noisy audio signals and the aimless mmWave signal, Wavoice can achieve noise-resistant speech recognition. Considering the complexity of human sensing, fusion sensing is an attractive and effective way.

\subsubsection{Multitask sensing}
The complexity of human sensing is also reflected in the interlacing of human sensing tasks. For example, both gesture recognition and vital sensing are important components of health monitoring in smart home. If we can sense these two tasks simultaneously with a single mmWave radar, the deployment and implementation of smart home can be simplified, and the constraints on people will be further released.

However, multitask sensing is not simply the superposition of multiple single-task sensing. For example, human gestures and vital signals are closely related and are difficult to be separated. The relation between multiple human sensing tasks is a challenge to achieve multitask sensing, which makes it difficult to describe and analyze the reflected signals. Conversely, it is also an opportunity as mutual correlation and improvement can be performed. In this case, the multitask learning techniques \cite{multitask_learning}, which have made great progress in the past ten years, may be able to show unique advantages in multitask sensing.

\subsubsection{Side channel for human sensing}
In addition to directly sensing the human body, there are many potential ``sensing side channels'' that can be used to obtain human-related information. By analyzing the influence of the human body on other objects, we can deduce the characteristics of the human body itself.

The current sensing side channel is most commonly used in sound recognition especially in eavesdropping. mmEve \cite{mmeve} and mmSpy \cite{mmspy} attempt to recover speech emitted from smartphone earpieces. They both utilizes the fact that the reflected mmWave signals from the smartphone's rear are highly correlated with the soundwaves emitted from the smartphone's earpiece. Similarly, mmPhone \cite{mmphone} utilizes the changes of piezoelectric films with sound pressure to decode the speech through soundproof obstacles. There are also some works to eavesdrop devices with ``sensing side channels''. For example, SpiralSpy \cite{spiralspy} utilizes a malicious software to encode data into the fan control signals. Then the fan motion status can be sensed and decoded by a mmWave radar.

Considering that the influence of the human body on the surrounding environment does not only exist in voice but also in many other physical signs, researchers can further explore the sensing side channel in many sensing tasks, such as keyboard sensing and gait recognition.

\subsection{Integration with New Mediums}
\label{sec: sensing with communication}
Innovations in the wireless medium is the extra boost of mmWave-based sensing.
Emerging wireless communication techniques are deemed to provide ideal sensing mediums. With the continuous development and innovation of wireless communication techniques, many of them have been applied to sensing tasks to achieve ubiquitous sensing, including integration sensing and communication, backscatter, intelligent reflecting surface and THz sensing.

\subsubsection{Integration sensing and communication}
With the increasing demand for human sensing and the number of mmWave network techniques such as 3GPP 5G-NR and IEEE 802.11ad/ay, Integration sensing and communication (ISAC) has received extensive attention from academia and industry. 

Given the growing popularity of mmWave networks, human sensing based on communication devices seems to be more suitable for life scenes. Chenshu Wu \textit{et al.} propose a series of works based on 60GHz probe, such as mmWrite \cite{mmwrite}, ViMo \cite{vimo} and mmTrack \cite{mmtrack}. However, these works all need to adjust the probe to a radar-like mode to give up the communication function. Therefore, such a design can only support the rotation of sensing and communication but can not achieve ISAC. On the contrary, SPARCS \cite{sparcs} proposes an IEEE 802.11ay based ISAC solution. It utilizes the CIR measurements from sparse communication data packages to analyze human movement without interfering with communication. Considering the popularity of 5G network and WiZig, ISAC has great potential to be applied to daily scenarios.

Moreover, some researchers have begun to endow the mmWave radar, which is customized for sensing, with additional communication capabilities. For example, people have changed the frequency, transmission time and interval of the mmWave signal to encode information. We will not expand as this is beyond the scope of this survey. 

\review{there are several critical technical challenges in implementing ISAC, both communication-assisted sensing and sensing-assisted communication. Firstly, people need to carefully consider the trade-off between information delivery and channel state estimation. The latter usually serves as the basis for sensing. More power allocated to information delivery means less sensing capability, and vice versa. Since the ratio of communication and sensing is different in different scenarios, people should analyze the trade-off by actual measurement and modeling to achieve the desired target. Secondly, due to hardware limitations and environmental interference (e.g., LoS path blockage), the coverage of ISAC is often limited, which hinders the application of ISAC in many scenarios. To solve this problem, new auxiliary devices such as Intelligent Reflecting Surface (IRS) can be considered to be deployed around the main ISAC device to further expand the coverage of ISAC. For example, Mosiac \cite{mosaic} utilizes multiple curved reflectors to provide comprehensive coverage to achieve omnidirectional around-corner automotive radars. Finally, the interference from other devices will be stronger in ISAC compared to pure wireless sensing scenarios. The interfering signals can cause the ISAC signals to be corrupted, delayed or even lost. To improve the robustness of ISAC in scenes full of interference, people can either adopt compensatory operation similar to SPARCS \cite{sparcs} or protect ISAC signals from interference through scheduled measures, such as frequency division multiplexing and time division multiplexing.}

\subsubsection{Sensing over backscatter}
\label{sec:backscatter}

Backscatter technology \cite{Passive-RFID1, leggiero, guo2020aloba, Lora-backscatter, anyscatter} has been developed rapidly in the past few years to provide communication capability for low-end devices. Compared with active radios, backscatter \cite{saiyan, Palantir, Ambient-backscatter, FDLoRa, LScatter, aloba_ton} promises to be extremely low power, smaller and cheaper alternative. As the reflected signal of the backscatter tag is affected by both the modulation information and the channel state, we can sense the target state by attaching the tag to the target and analyzing the change of the reflected signal.

Backscatter technology has been used to improve the sensing range of mmWave sensing. For example, Millimetro \cite{Millimetro} exploits a customized backscatter tag to realize accurate localization in a long range. The tag is formed by Van Atta retro-reflectors, which can reflect the carrier signals back in the direction of arrival regardless of the incident angle and the tag's mobility. With the retroreflected tag, The attached target can be localized at high accuracy (centimeter-level) over extended distances (over 100~m). Such a design can enhance the mmWave reflected signal and expand the sensing range of mmWave sensing.

The mmWave sensing capability can also be enhanced by combining mmWave sensing and backscatter technology. RoS \cite{RoS} utilizes multiple configurable Van Atta arrays to build radar readable road signs. The arrays are formed as customized geometrical layouts and can be distinguished by an automotive radar.
With the continuous development of mmWave backscatter technology \cite{mmTag, mmID}, we believe that there will be more mmWave sensing works combined with backscatter technology adapting to various sensing tasks.

\subsubsection{Sensing with intelligent reflecting surface}
\label{sec:IRS}

As a cost-effective technology, Intelligent Reflecting surface (IRS)~\cite{IRS1, IRS2, IRS3, IRS4, IRS5, xu2023reconfiguring} has been an attractive subject to improve the performance of communication and sensing systems by reconfiguring the wireless propagation environments smartly. Generally, IRS is a metasurface consisting of a large number of passive reflecting elements. Each reflector induces an adjustable phase shift and amplitude variation to the incident signals. With the ability to reconfigure the direction of mmWave signals, IRS is expected to play an important role in enhancing the coverage and energy efficiency of mmWave systems.

Due to the strong directivity and rapid attenuation of mmWave signals, the coverage of mmWave sensing is very limited. As IRS can reconfigure the mmWave signal direction, it can guide the mmWave signal to the originally inaccessible positions to enhance the sensing coverage. For example, MilliMirror~\cite{MilliMirror} designs a passive metasurface prototype which expands the coverage of mmWave radios to blind spots by redirecting and reshaping mmWave signals to any anomalous directions. mmWall~\cite{mmWall} proposes another tunable smart surface made of metamaterial, which enables a fast mmWave beam relay through the wall and redirects the beam power to another direction when a human body blocks a line-of-sight path. In addition, mmWall also supports splitting the incoming signal into multiple beams and concurrently steering the multi-armed beams. In this way, ``through-wall'' sensing or even multi-person sensing can be performed.

With such powerful capability, however, IRS has the potential to be utilized for illegal attacks. MeSS \cite{Mess} has achieved illegal eavesdropping based on mmWave metasurface. By leveraging two degrees of freedom (space and time) in reconfigurable surfaces, MeSS generates and steers a concealed directional sideband toward the eavesdropper while maintaining the direction of the mainband toward the legitimate client. Joonas Kokkoniemi \textit{et al.}~\cite{channel} also discuss the channel modeling of the phased array type reconfigurable IRSs in the mmWave band. 

To sum up, we believe the research of IRS has the opportunity to open a novel direction for solving the pain points of mmWave sensing, such as NLoS, mobility, and so on. Meanwhile, it also brings new challenges to our privacy protection.

\review{
\subsubsection{From mmWave to THz}

With the rapid increase of smart applications and devices, it is foreseeable that 5G network is hard to meet the ever-increasing network traffic. According to ITU-T's estimation \cite{union2015imt}, the global mobile traffic will grow to 5016 EB per month in 2030, compared to 62 EB per month in 2020. To solve this problem, the 6G network has been discussed and researched by industry and academia. mmWave signal and THz signal are expected to be used in 6G because of their wide unexplored spectrum. At present, mmWave communication and THz communication have reached 1 terabyte/s and 206.25 Gbit/s data rate in laboratory environments \cite{6G-THz,6G-mmWave}.

On the other hand, due to the lack of corresponding standards, we make some predictions about the requirement of 6G in human sensing from application scenarios. The 6G network is expected to be human-centric and provide comprehensive human sensing. Therefore, 6G will pay more attention to various human-related applications, such as holographic-type communications, digital twins and tactile internet. Taking the tactile internet as an example, human sensing techniques may be used to recognize human body movements and the 6G network is expected to meet the low transmission delay of 1 millisecond or less reaction time \cite{fettweis2014tactile}. Based on our summary of mmWave sensing, we give some discussions on THz sensing to complement blueprint for 6G.

THz signals share many similarities with mmWave signals, including high frequency, large bandwidth and extremely high attenuation. With the continuous development of wireless sensing, THz signals have also begun to enter the field of vision of researchers and are used to realize some sensing tasks, such as indoor positioning \cite{kludze2022quasi}. Since THz signal frequency (0.3-3 THz) is much higher than that of mmWave signal (0.03-0.3 THz), THz sensing faces many unique challenges and has attractive potential capabilities.

The high-frequency characteristics of THz signals will cause extremely higher attenuation, resulting in very limited sensing coverage and low-SNR received signal. The limited sensing coverage may make it difficult to apply THz sensing to outdoor human sensing. Furthermore, how to reliably extract the human-related phases from the low-SNR received signal needs to be further explored. Novel low-noise electronics may be demanded. In addition, considering the separated transceiver device, tight synchronization is needed to extract human-related information, which will be more challenging in higher frequencies.

Conversely, the higher frequency and larger bandwidth of THz signals enable THz sensing to have finer sensing capability and higher spatial resolution, which means that more fine-grained sensing tasks can be achieved, such as fine-grained medical imaging, fingerprint detection, skin texture detection, etc. Considering the frequency limitation of mmWave signals, THz sensing is more suitable for achieving such fine-grained sensing tasks.
}

\section{Conclusion}
\label{sec:conclusion}

mmWave sensing has become more and more popular with the development of the 5G network and automatic driving. This survey focuses on recent research advances in mmWave-based human sensing. This survey presents various hardware and the key techniques of mmWave sensing, reviews existing mmWave sensing works based on different sensing tasks, i.e., human tracking and localization, motion recognition, biometric measurement and human imaging. We further discuss potential challenges and future directions, including hardware and platforms, enhancing the applicability, novel sensing schemes, and integration with new mediums.

%mmWave sensing has become more and more popular with the development of the 5G network and automatic driving. This survey focuses on recent research advances in mmWave-based human sensing. This survey presents various hardware and the key techniques of mmWave sensing, reviews existing mmWave sensing works based on different sensing tasks, i.e., human tracking and localization, motion recognition, biometric measurement and human imaging. We further discuss potential challenges including hardware and deployment limitations, and present future directions including sensing technologies for enhanced and hidden sensing, and integrated sensing and communication.

\section*{Acknowledgment}
This work is supported by the National Science Fund of China under grant No. U21B2007, National Science Fund of China under grant No. 62202264 and the R\&D Project of Key Core Technology and Generic Technology in Shanxi Province (2020XXX007).

\renewcommand\nomgroup[1]{%
  \item[\bfseries
  \ifstrequal{#1}{A}{}{%
  \ifstrequal{#1}{B}{}{%
  \ifstrequal{#1}{C}{}{%
  \ifstrequal{#1}{D}{}{%
  \ifstrequal{#1}{E}{}{%
  \ifstrequal{#1}{F}{}{%
  \ifstrequal{#1}{G}{}{%
  \ifstrequal{#1}{H}{}{%
  \ifstrequal{#1}{I}{}{%
  \ifstrequal{#1}{J}{}{%
  \ifstrequal{#1}{K}{}{%
  \ifstrequal{#1}{L}{}{%
  \ifstrequal{#1}{M}{}{%
  \ifstrequal{#1}{N}{}{%
  \ifstrequal{#1}{O}{}{%
  \ifstrequal{#1}{P}{}{%
  \ifstrequal{#1}{Q}{}{%
  \ifstrequal{#1}{R}{}{%
  \ifstrequal{#1}{S}{}{%
  \ifstrequal{#1}{T}{}{%
  \ifstrequal{#1}{U}{}{%
  \ifstrequal{#1}{V}{}{%
  \ifstrequal{#1}{W}{}{%
  \ifstrequal{#1}{X}{}{%
  \ifstrequal{#1}{Y}{}{%
  \ifstrequal{#1}{Z}{}{}}}}}}}}}}}}}}}}}}}}}}}}}}%
]}

\nomenclature[A]{\textbf{ADC}}{Analog-to-digital converter}
%\nomenclature[A]{\textbf{AIC}}{Akaike information criterion}
\nomenclature[A]{\textbf{AiP}}{Antenna in package}
\nomenclature[A]{\textbf{AoA}}{Angle of arrival}
%\nomenclature[A{\textbf{AO-HMM}}{Adaptive order hidden markov model}
\nomenclature[A]{\textbf{ASL}}{American sign language}
\nomenclature[A]{\textbf{ASR}}{Authentication success rate}
%\nomenclature[A]{\textbf{ATAP}}{Advanced technologies and project}
\nomenclature[A]{\textbf{AUC}}{Area under the ROC curve}

\nomenclature[B]{\textbf{BANC}}{background and noise cancellation}
%\nomenclature[B]{\textbf{BIRCH}}{Balanced iterative reducing and clustering using hierarchies}

%\nomenclature[C]{\textbf{CaSE}}{Cardiac-mmWave scattering effect}
\nomenclature[C]{\textbf{CER}}{Character error rate}
\nomenclature[C]{\textbf{CFAR}}{Constant false alarm rate}
\nomenclature[C]{\textbf{CGAN}}{Conditional generative adversarial nets}
\nomenclature[C]{\textbf{CIR}}{Channel impulse response}
\nomenclature[C]{\textbf{CNN}}{Convolutional neural network}
\nomenclature[C]{\textbf{COTS}}{Commercial off-the-shelf}
\nomenclature[C]{\textbf{CPDA}}{Crossover path disambiguation algorithm}
\nomenclature[C]{\textbf{CPDP}}{Concentrated position-Doppler profile}

\nomenclature[D]{\textbf{DBSCAN}}{Density-based spatial clustering of applications with noise}
\nomenclature[D]{\textbf{DCT}}{Discrete cosine transformd}
\nomenclature[D]{\textbf{DDBR}}{Dual-differential backgound removal}
\nomenclature[D]{\textbf{DHH}}{Deaf of hard-of-hearing}
\nomenclature[D]{\textbf{DNN}}{Deep neural network}
\nomenclature[D]{\textbf{DoF}}{Degrees of freedom}

\nomenclature[E]{\textbf{EER}}{Equal error rate}
\nomenclature[E]{\textbf{EKF}}{Extended Kalman filter}
\nomenclature[E]{\textbf{EM}}{Electromagnetic}
\nomenclature[E]{\textbf{EMD}}{Empirical mode decomposition algorithm}

\nomenclature[F]{\textbf{FA}}{Face authentication}
\nomenclature[F]{\textbf{FFT}}{Fast fourier transform}
\nomenclature[F]{\textbf{FMCW}}{Frequency modulated continuous wave}
\nomenclature[F]{\textbf{FSK}}{Frequency shift keying}

\nomenclature[G]{\textbf{GAN}}{Generative adversarial network}
\nomenclature[G]{\textbf{GRU}}{Gated recurrent network}

\nomenclature[H]{\textbf{HAR}}{Human activity recognition}
\nomenclature[H]{\textbf{HMM-VM}}{Hidden markov model-based voting mechanism}
\nomenclature[H]{\textbf{HRV}}{Heart rate variability}
\nomenclature[H]{\textbf{HWA}}{Hardware accelerator block}

\nomenclature[I]{\textbf{IF}}{Intermediate frequency}
\nomenclature[I]{\textbf{IHT}}{Iterative hard thresholding}
\nomenclature[I]{\textbf{IMF}}{Intrinsic mode function}
\nomenclature[I]{\textbf{IMU}}{Inertial measurement unit}
\nomenclature[I]{\textbf{IRS}}{Intelligent reflective surface}
\nomenclature[I]{\textbf{ISAC}}{Integrated sensing and communication}

\nomenclature[J]{\textbf{JIT}}{Joint transmitter smoothing}
%\nomenclature[J]{\textbf{JPDAF}}{Joint probabilistic data association filter}

\nomenclature[L]{\textbf{LoS}}{Line-of-sight}
\nomenclature[L]{\textbf{LSTM}}{Long short term memory}
%\nomenclature[L]{\textbf{LM}}{Lip motion}

\nomenclature[M]{\textbf{MCD}}{Mel-Cepstral distortion}
\nomenclature[M]{\textbf{MFCC}}{Mel frequency cepstral coefficient}
%\nomenclature[M]{\textbf{MMIC}}{Monolithic microwave integrated circuit}
\nomenclature[M]{\textbf{MMSE}}{Minimum mean square error}
%\nomenclature[M]{\textbf{MPJPE}}{Mean per joint position error}
\nomenclature[M]{\textbf{MTI}}{Moving target indication}
\nomenclature[M]{\textbf{MUSIC}}{Multiple signal classification}
\nomenclature[M]{\textbf{MVDR}}{Minimum variance distortionless response}

\nomenclature[N]{\textbf{NLoS}}{Non-line-of-sight}

\nomenclature[P]{\textbf{PoI}}{Point of interest}

\nomenclature[R]{\textbf{RAI}}{Range angle image}
\nomenclature[R]{\textbf{RDM}}{Range-Doppler matrix}
\nomenclature[R]{\textbf{RF}}{Radio frequency}
%\nomenclature[R]{\textbf{RMA}}{Range migration algorithm}
\nomenclature[R]{\textbf{RNN}}{Residual neural network}
\nomenclature[R]{\textbf{RPCC}}{Residual phase cepstrum coefficients}
\nomenclature[R]{\textbf{RSS}}{Receiver signal strength}
\nomenclature[R]{\textbf{RFIC}}{Radio-frequency integrated circuit}

\nomenclature[S]{\textbf{SAR}}{Synthetic aperture radar}
\nomenclature[S]{\textbf{SLAM}}{Simultaneous localization and mapping}
%\nomenclature[S]{\textbf{SMPL}}{Skinned multi-person linear}
\nomenclature[S]{\textbf{SNR}}{Signal-to-noise ratio}
%\nomenclature[S]{\textbf{SPL}}{Structured prediction layer}
\nomenclature[S]{\textbf{STFT}}{Short-time Fourier transform}

\nomenclature[V]{\textbf{VCV}}{Vocal cord vibration}
\nomenclature[V]{\textbf{VMD}}{Variational modal decomposition}
%\nomenclature[V]{\textbf{VRS}}{Virtual registration signals}

\nomenclature[W]{\textbf{WPT}}{Wavelet packet transform}
\nomenclature[W]{\textbf{WER}}{Word error rate}
\printnomenclature

%\section{References Section}
\bibliographystyle{IEEEtran}
\bibliography{main.bib}

%\newpage

%\section{Biography Section}
\vspace{-3em}
\begin{IEEEbiography}[{\includegraphics[width=1in,height=1.35in,clip,keepaspectratio]{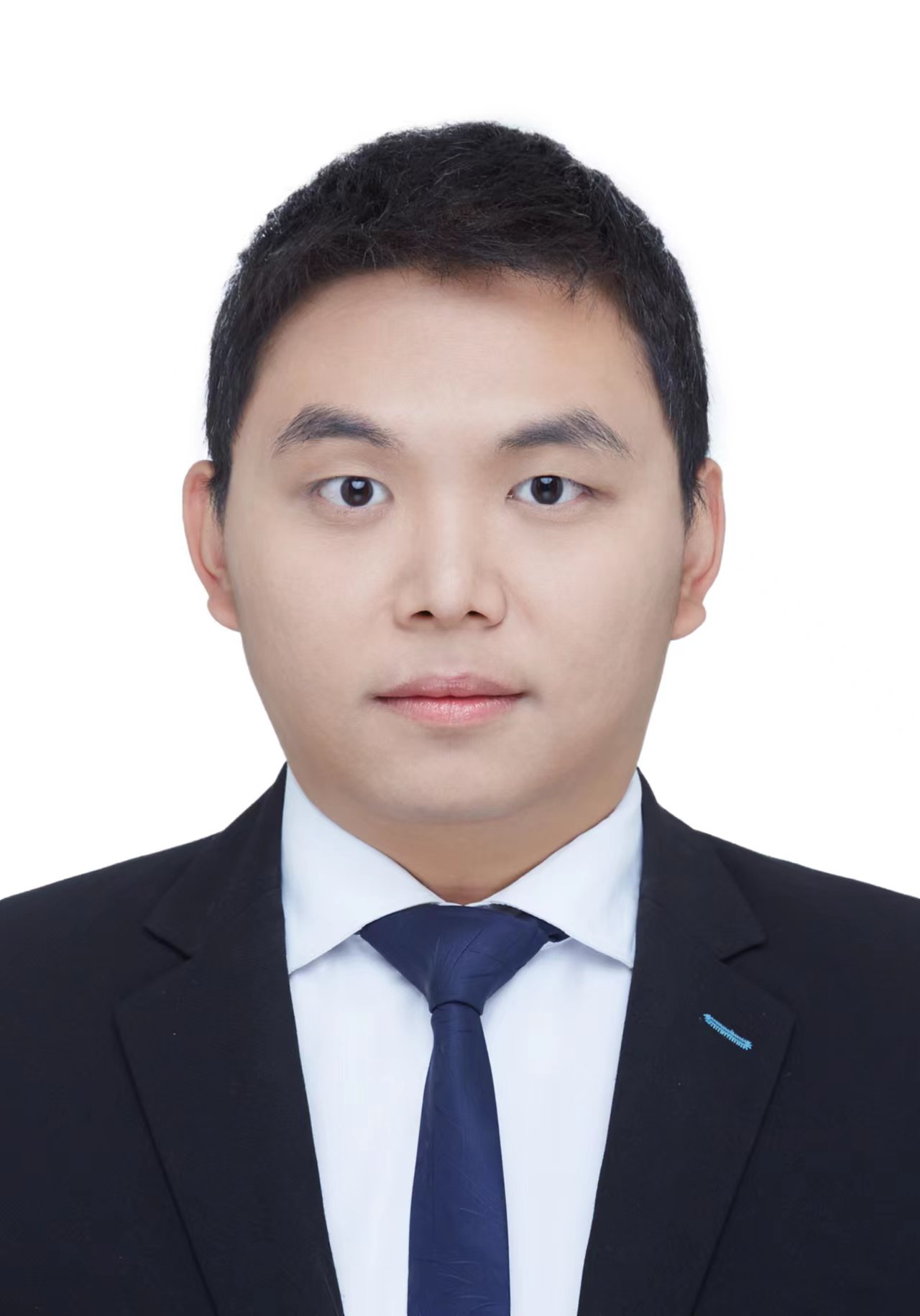}}]{Jia Zhang}
is currently a PhD. student in the School of Software and BNRist of Tsinghua University. He received his B.E. degree in Tsinghua University in 2019. His research interests include Internet of Things and wireless sensing.
\end{IEEEbiography}
\vspace{-3em}
\begin{IEEEbiography}[{\includegraphics[width=1in,height=1.35in,clip,keepaspectratio]{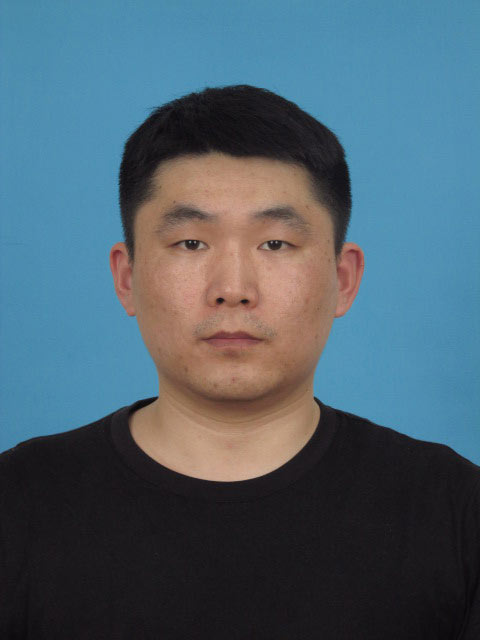}}]{Rui Xi} is currently an Assistant Researcher at School of Computer Science and Engineering in University of Electronic Science Technology of China (UESTC) since August 2022. During 2019.07 - 2022.07, he was a postdoctoral researcher at Tsinghua University, worked with Prof. Yuan He. He obtained the M.S. and Ph.D. degrees in Computer Science from UESTC, Chengdu, China, in 2014 and 2019, respectively. His main research interests include Internet of Things, wireless sensing and AI.
\end{IEEEbiography}
\vspace{-3em}
\begin{IEEEbiography}[{\includegraphics[width=1in, height=1.35in]{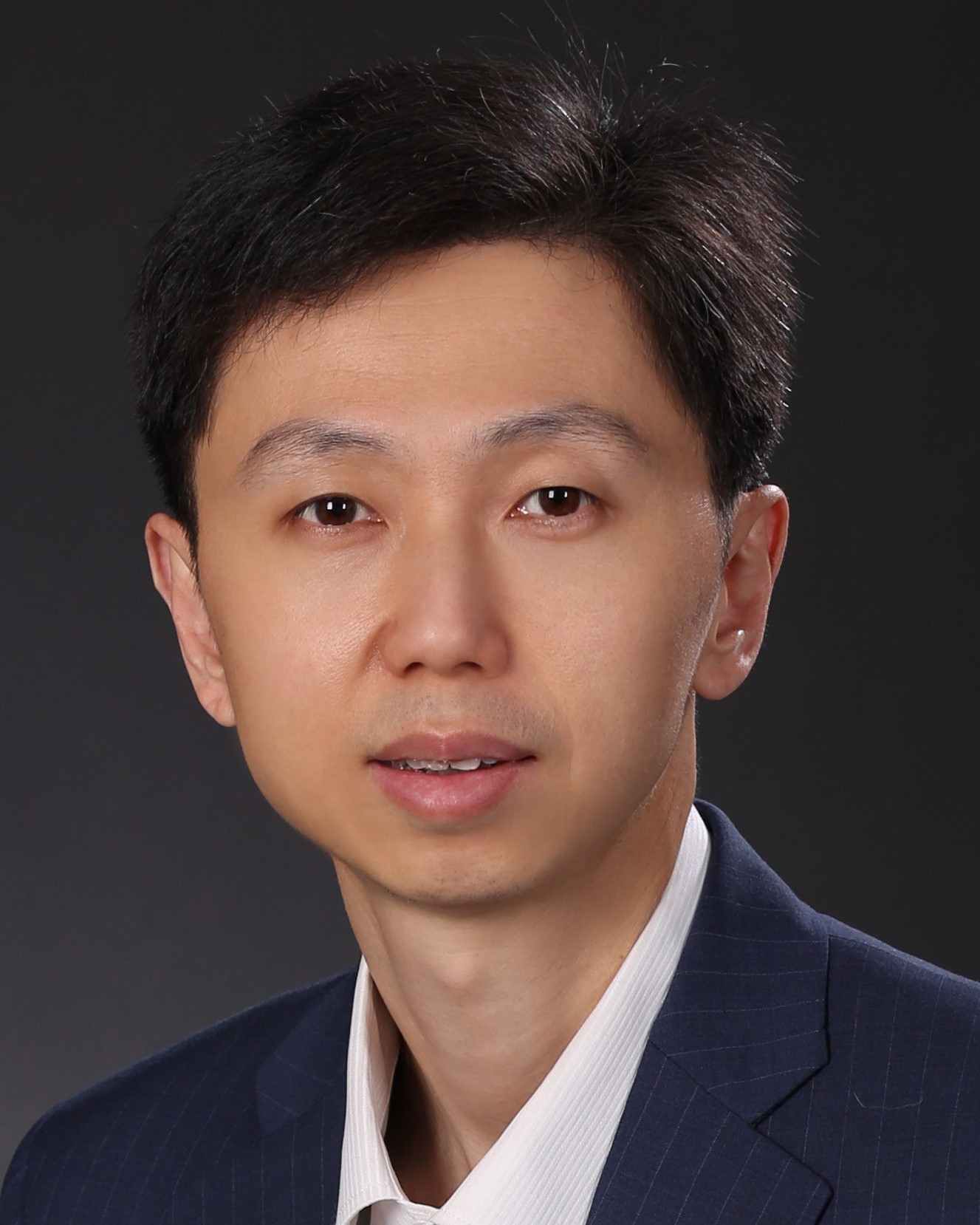}}]{Yuan He} is an associate professor in the School of Software and BNRist of Tsinghua University. He received his B.E. degree in the University of Science and Technology of China, his M.E. degree in the Institute of Software, Chinese Academy of Sciences, and his PhD degree in Hong Kong University of Science and Technology. His research interests include wireless networks, Internet of Things, pervasive and mobile computing. He is a senior member of  IEEE and a member of ACM.\end{IEEEbiography}
\vspace{-3em}
\begin{IEEEbiography}[{\includegraphics[width=1in,height=1.35in,clip,keepaspectratio]{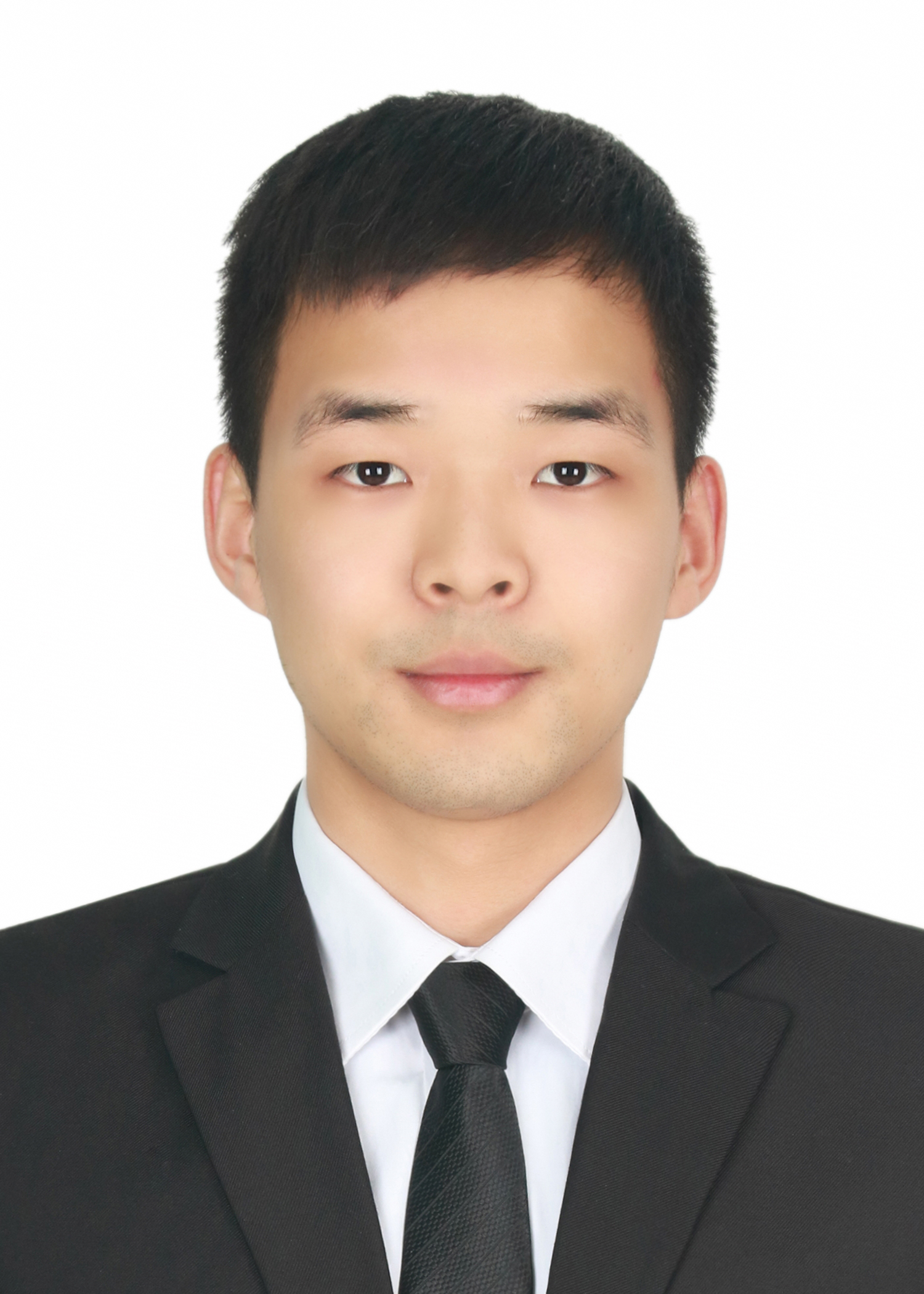}}]{Yimiao Sun}
is currently a PhD. student in Tsinghua University. He received his B.E. degree in the  University of Electronic Science and Technology of China (UESTC) in 2021. His research interests include Internet of Things and wireless sensing.
\end{IEEEbiography}
\vspace{-3em}
\begin{IEEEbiography}[{\includegraphics[width=1in, height=1.35in]{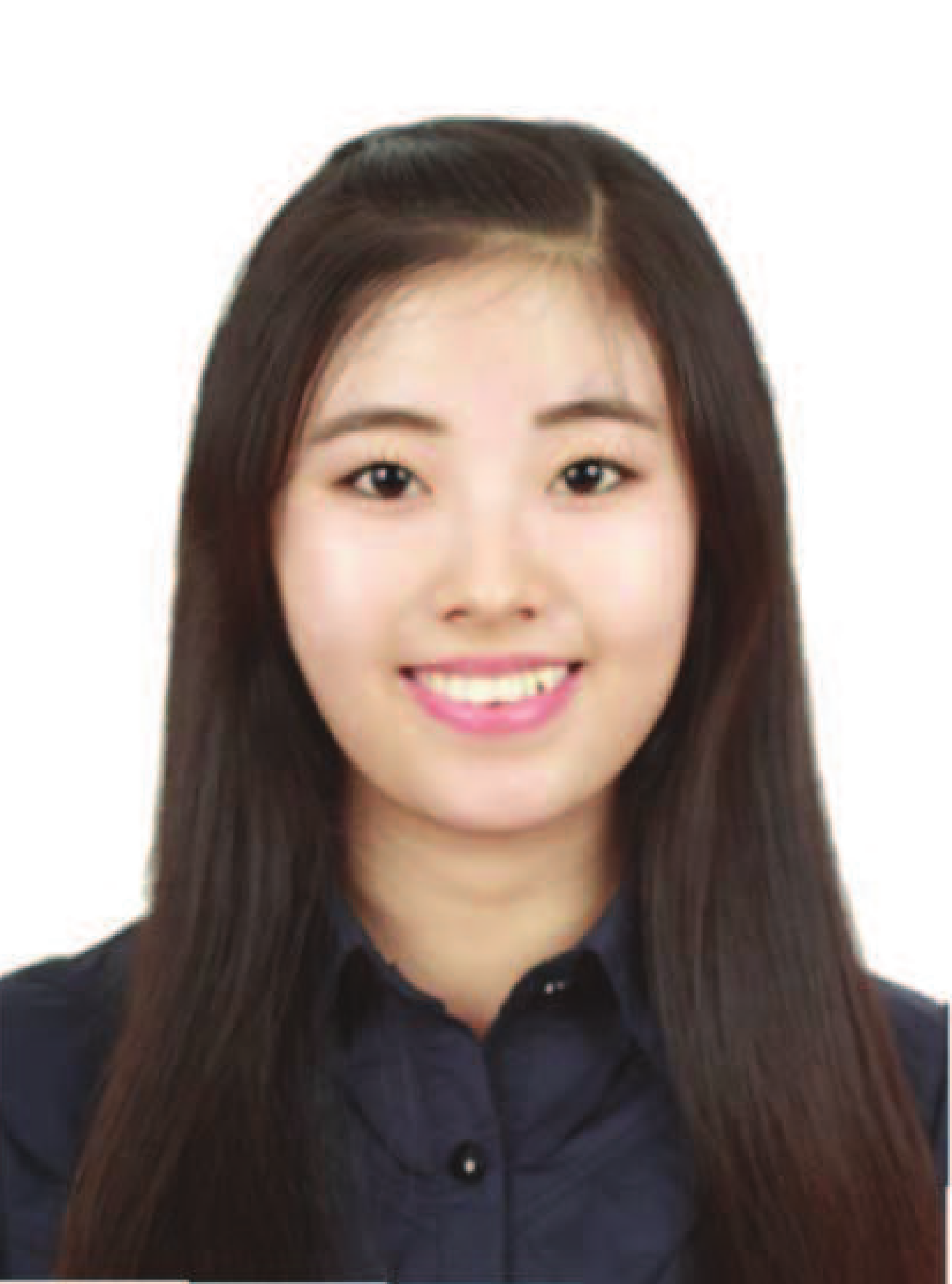}}]{Xiuzhen Guo} is a postdoc in the School of Software and BNRist of Tsinghua University. She received her B.E. degree in Southwest University, and her PhD degree in Tsinghua University. Her research interests include wireless networks and Internet of Things.
\end{IEEEbiography}
\vspace{-3em}
\begin{IEEEbiography}[{\includegraphics[width=1in, height=1.35in]{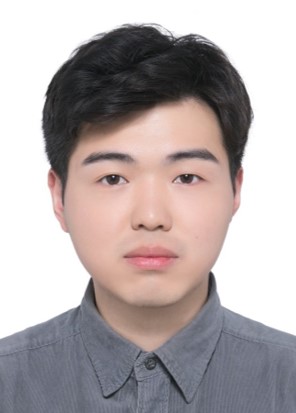}}]{Weiguo Wang} is currently a PhD. student in Tsinghua University. He received his B.E. degree in the University of Electronic Science and Technology of China (UESTC). His research interests include acoustic sensing and mobile computing.  
\end{IEEEbiography}
\vspace{-3em}
\begin{IEEEbiography}[{\includegraphics[width=1in, height=1.35in]{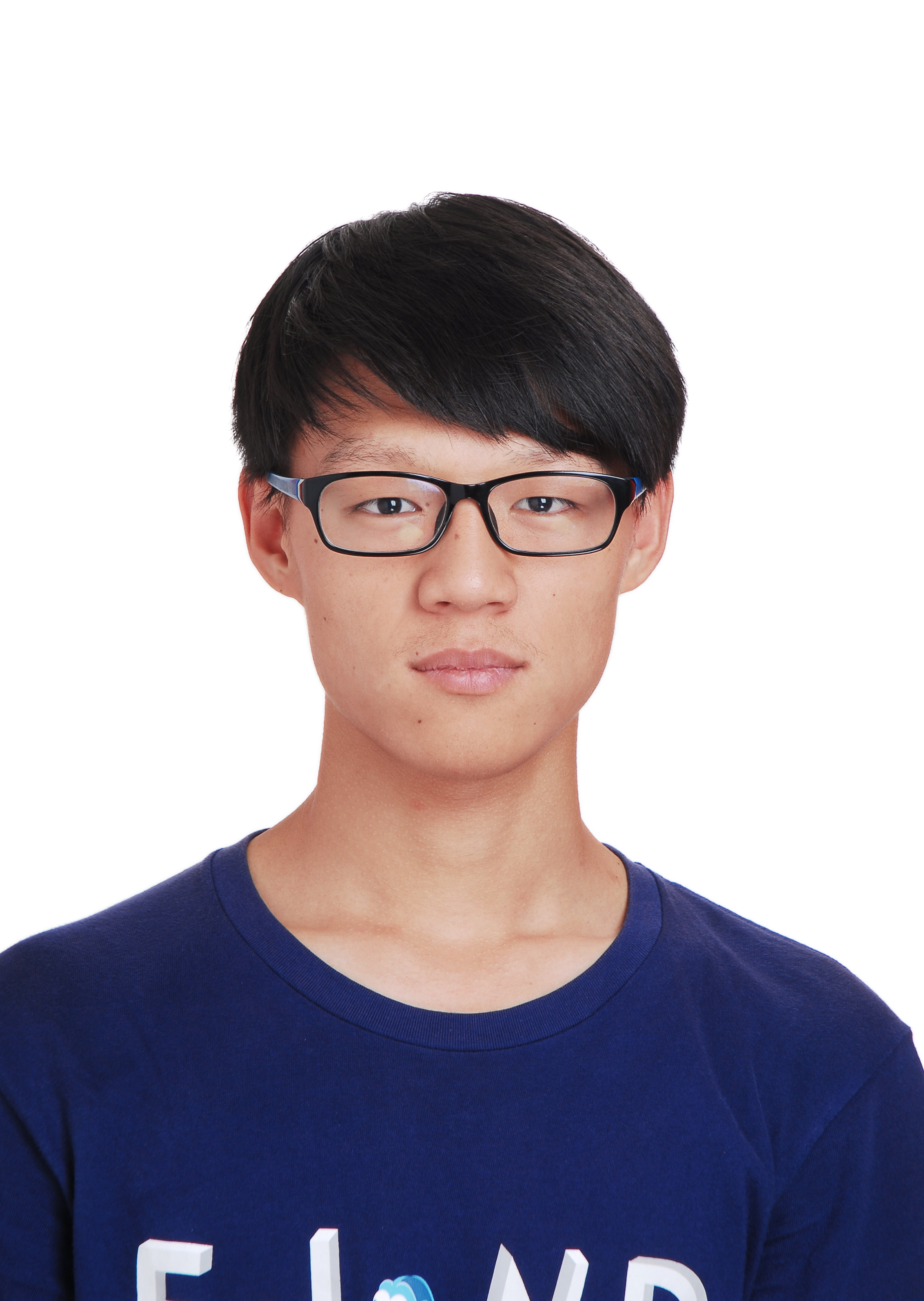}}]{Xin Na} is currently a PhD. student in Tsinghua University. He received his B.E. degree in Tsinghua University in 2020. His research interests include wireless networks and Internet of Things.  
\end{IEEEbiography}
\vspace{-3em}
\begin{IEEEbiography}[{\includegraphics[width=1in, height=1.35in]{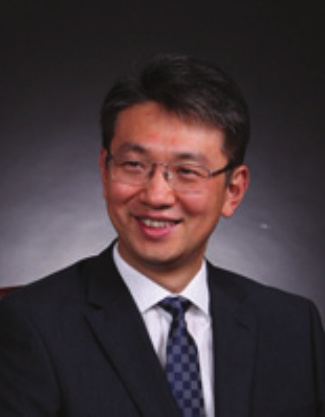}}]{Yunhao Liu} received his B.S. degree in Automation Department from Tsinghua University, and an M.A. degree in Beijing Foreign Studies University, China. He received an M.S. and a Ph.D. degree in Computer Science and Engineering in Michigan State University, USA. He is now a professor at Automation Department and Dean of the GIX in Tsinghua University, China. He is a Fellow of IEEE and ACM.  
\end{IEEEbiography}
\vspace{-3em}
\begin{IEEEbiography}[{\includegraphics[width=1in, height=1.35in]{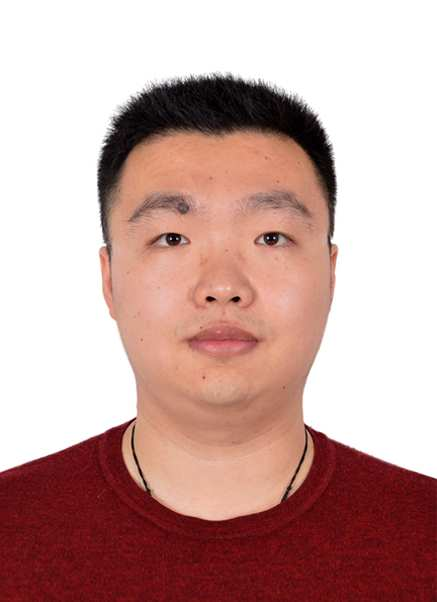}}]{Zhenguo Shi} received his M.S.(in 2011) degree and PhD (in 2016) from Harbin Institute of Technology (HIT), China. He received his second PhD degree (in 2022) from University of Technology Sydney, Australia. He was a visiting student at the University of Leeds (2012-2013) and a visiting scholar at the University of Technology Sydney, Australia (2016- 2017). He is currently working as a Postdoctoral Research Associate in the School of Computing at the Macquarie University, Australia. His research interests include Wireless sensing, Human activity recognition, IoT, Deep learning, Cognitive radio, Interference alignment, Massive MIMO and wireless communication.
\end{IEEEbiography}
\vspace{-3em}
\begin{IEEEbiography}[{\includegraphics[width=1in, height=1.35in]{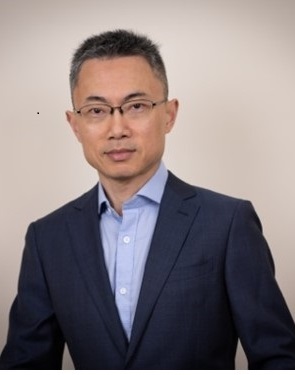}}]{Tao Gu} is currently a Professor at Macquarie University, and he is an expert in the fields of Internet of Things, Embedded/Edge AI, Mobile and Ubiquitous Computing. His publications typically appear in conferences including MobiCom, SenSys, IPSN, UbiComp and INFOCOM. He has also served a leadership role in many conferences, including General Co-Chair of MobiCom 2022 and TPC Co-Chair of IoTDI 2021.
\end{IEEEbiography}
%\vspace{11pt}

%\vfill

\end{document}